\documentclass[12pt]{article}
\pdfoutput=1
\usepackage{amssymb}
\usepackage{amsmath}
\usepackage{latexsym}
\usepackage[usenames]{color}
\usepackage{cite}
\definecolor{darkblue}{cmyk}{0.9,0.9,0,0}
\definecolor{darkred}{rgb}{0.6,0,0.3}
\usepackage{graphicx} 
\usepackage[setpagesize=false,pagebackref=false, linktocpage, bookmarksopen=true, colorlinks=true, linkcolor=darkblue,citecolor=darkblue,urlcolor=darkblue]{hyperref}
\usepackage{hyperref}
\newcommand{\arXiv}[2]{\href{http://arxiv.org/abs/#1}{{\tt arXiv:#2}}}
\newcommand{\hep}[2]{\href{http://arxiv.org/abs/#1}{{\tt #2}}}

\renewcommand{\thefootnote}{\arabic{footnote}}

\def\del{\partial}

\def\nn{\nonumber}
\def\eqref#1{(\ref{#1})}
\def\comma{\,,}
\def\period{\,.}
\def\Tr{{\rm Tr}}
\def\red#1{\textcolor[rgb]{1, 0, 0}{#1}}
\def\blue#1{\textcolor[rgb]{0,0.1,1}{#1}}

\newcommand{\beq}{\begin{equation}}
\newcommand{\eeq}{\end{equation}}

\begin{document}
\thispagestyle{empty}

\renewcommand{\thefootnote}{\fnsymbol{footnote}}
\setcounter{page}{1}
\setcounter{footnote}{0}
\setcounter{figure}{0}
\begin{center}
$$$$
{\Large\textbf{\mathversion{bold}
Hexagonalization of Correlation Functions
}\par}

\vspace{1.3cm}

\textrm{Thiago Fleury$^{\textcolor[rgb]{0,0.6,0}{\blacktriangledown}}$, Shota Komatsu$^{\textcolor[rgb]{1,0.8,0}{\blacktriangleright}}$}
\\ \vspace{1.2cm}
\footnotesize{\textit{
$^{\textcolor[rgb]{0,0.6,0}{\blacktriangledown}}$ 
Instituto de F\'isica Te\'orica, UNESP - Univ. Estadual Paulista, 
ICTP South American Institute for Fundamental Research,
Rua Dr. Bento Teobaldo Ferraz 271, 01140-070, S\~ao Paulo, SP, Brasil
\vspace{1mm} \\
$^{\textcolor[rgb]{1,0.8,0}{\blacktriangleright}}$ Perimeter Institute for Theoretical Physics,
31 Caroline St N Waterloo, Ontario N2L 2Y5, Canada\\
}  
\vspace{4mm}
}

\par\vspace{1.5cm}

\textbf{Abstract}\vspace{2mm}
\end{center}
We propose a nonperturbative framework to study general 
correlation functions of single-trace operators 
in $\mathcal{N}=4$ supersymmetric Yang-Mills theory 
at large $N$. The basic strategy is to decompose them into 
fundamental building blocks called the hexagon form factors, which 
were introduced earlier to study structure constants 
using integrability. The decomposition is akin 
to a triangulation of a Riemann surface, and we 
thus call it {\it hexagonalization}. We propose a set of 
rules to glue 
the hexagons together based on 
symmetry, which 
naturally incorporate the dependence on the conformal and 
the R-symmetry cross ratios. Our method is conceptually different 
from the conventional operator product expansion and automatically 
takes into account multi-trace operators exchanged in 
OPE channels. To illustrate the idea 
in simple set-ups, we compute four-point functions of BPS 
operators of arbitrary lengths and correlation functions 
of one Konishi operator and three short BPS operators, all 
at one loop. In all cases, the results 
are in perfect agreement with  the perturbative data. We also 
suggest that our method can be a useful tool to study 
conformal integrals, and show it explicitly for the case of 
ladder integrals.
\noindent

\setcounter{page}{1}
\renewcommand{\thefootnote}{\arabic{footnote}}
\setcounter{footnote}{0}
\setcounter{tocdepth}{2}
\newpage
\tableofcontents

\parskip 5pt plus 1pt   \jot = 1.5ex

\section{Introduction\label{sec:intro}}
A conformal field theory is characterized by its spectrum and structure constants. This however does not mean that higher-point functions are inconsequential. By taking various limits of higher-point functions, one can study interesting physical phenomena\footnote{Examples of such interesting physics discussed recently are the Regge limit \cite{ConformalRegge}, the emergence of the bulk locality \cite{MSZ} and chaos \cite{MSS}.} which cannot be explored just by looking at individual two- and three-point functions.
\begin{figure}[t]
\begin{center}
\includegraphics[clip,height=5.7cm]{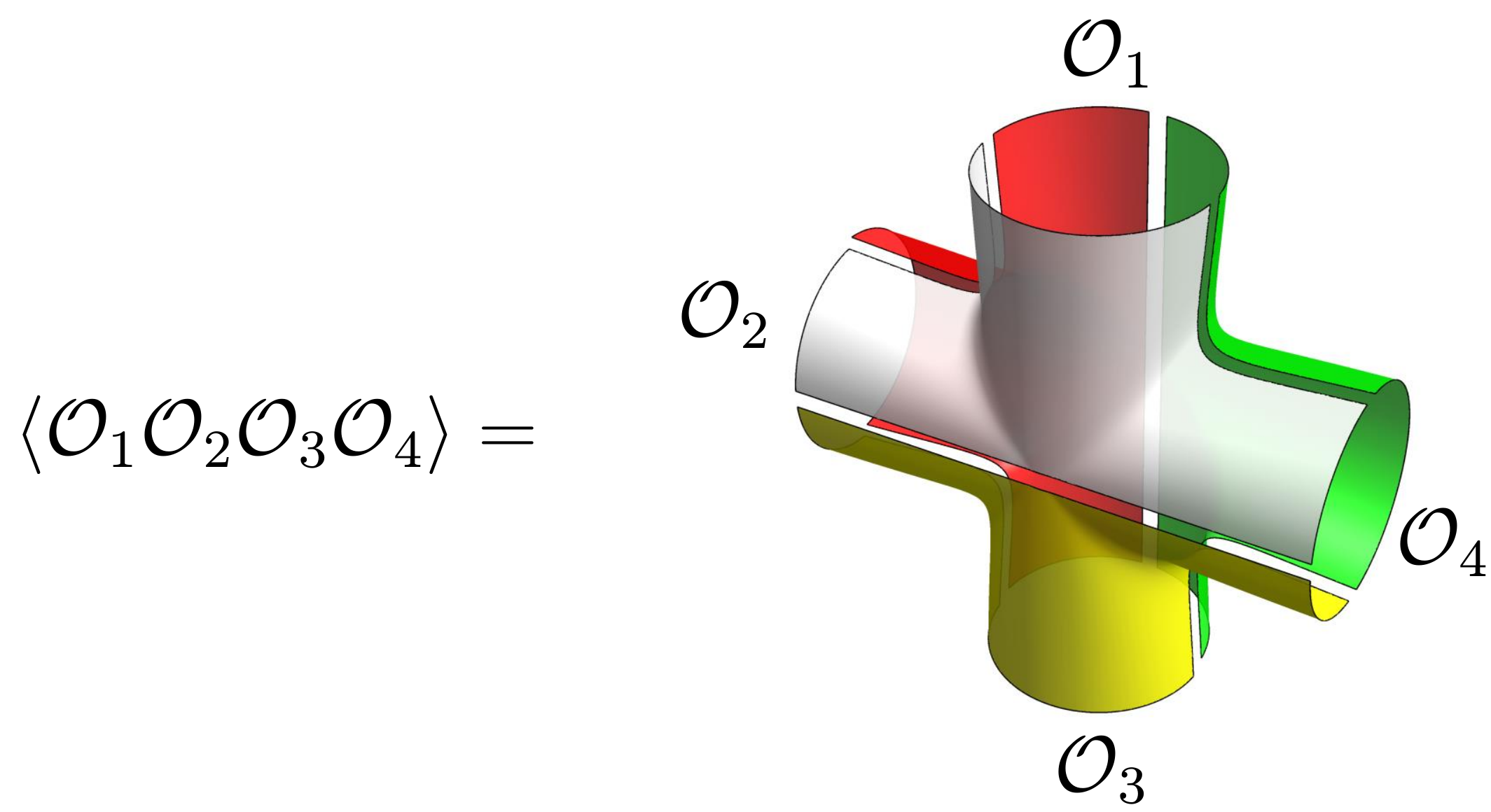} 
\end{center}
\vspace{-0.7cm}
\caption{Hexagonalization of a four-point function: A planar four-point function can be represented as a surface with four holes. The idea of hexagonalization is to cut it into four hexagonal patches as depicted above. The contribution from each patch is given by a hexagon form factor. It is conceptually different from the usual operator product expansion. {{\footnotesize (The colors of the figure represent the two places where this work was done.)}}} \label{Cut4pt}
\vspace{-0.3cm}
\end{figure}

The situation is more interesting, and at the same time, more intricate in large $N$ conformal field theories such as planar $\mathcal{N}=4$ supersymmetric Yang-Mills theory ($\mathcal{N}=4$ SYM). This is because the operator product expansion (OPE) and the large $N$ limit are not quite ``compatible'': Basic observables in large $N$ CFT's are correlation functions of single-trace operators. Even at large $N$, the OPE series of these correlators contains not only single-trace operators but also multi-trace operators. Therefore one cannot compute higher-point functions just by knowing two- and three-point functions of single-trace operators\footnote{There are certain limits where contributions from multi-trace operators are suppressed. In such limits, one can construct (approximate) higher-point functions from two- and three-point functions of single trace operators. See \cite{Asymptotic4pt} for more detailed discussions.}.

This appears to be an inconvenient truth for integrability practitioners: Owing to the remarkable progress in the last ten years, we now have powerful nonperturbative methods to study the spectrum \cite{Review} (see \cite{QSC} for the current state of the art), and the structure constants \cite{BKV} of planar $\mathcal{N}=4$ SYM. However these approaches are so far limited to single-trace operators. The aforementioned fact seems to indicate that we must extend these methods to multi-trace operators before studying higher-point functions. 

This however is {\it not} the case: In this paper, we propose an alternative route to higher-point functions, which does not necessitate explicit information on multi-trace operators. The key idea is to decompose the correlation functions not to two- and three-point functions, but to more fundamental building blocks called the {\it hexagon form factors}. The hexagon form factors were introduced in \cite{BKV} as the building blocks for the three-point function of single-trace operators. They compute a ``square-root'' of the structure constant, which is associated with a hexagonal patch of the string worldsheet. The purpose of this work is to show that these hexagons can compute higher-point functions as well (see figure \ref{Cut4pt}): By gluing $2 (n-2)$ hexagons together with appropriate weight factors, we can determine $n$-point functions of single-trace operators including the dependence on the conformal and the R-symmetry cross ratios. The decomposition bears resemblance to a triangulation of a Riemann surface, and we thus call it hexagonalization. It also shares conceptual similarities with the operator product expansion of the null polygonal Wilson loop \cite{AGMSV,Pentagon}.

The structure of the paper is as follows: After briefly reviewing the hexagon formalism in section \ref{sec:review}, we begin by computing a simple correlator at tree level in section \ref{sec:simple}. The main purpose is to demonstrate that cross ratios can appear in weight factors. Motivated by this observation, we present our proposal, hexagonalization, in section \ref{sec:hexagonalization}. We first determine weight factors for the so-called mirror channel using the superconformal symmetry and then explain how they generalize to the physical channel. We test our proposal against one-loop data in section \ref{sec:4BPS} and \ref{sec:Konishi}, and obtain a complete match. Furthermore, we compute a simple class of contributions at higher loops in section \ref{sec:ladder} and show that they coincide with the so-called ladder integrals. We conclude with discussions of future directions in section \ref{sec:conclusion}. A few appendices are included to explain technical details.

\noindent {\small
{\bf Note added}: After this paper is completed, the paper \cite{ES}, which partially overlaps the result in section \ref{sec:simple}, appeared in the arXiv.}
\vspace{-15pt}
\section{Review of the Hexagon Formalism\label{sec:review}}
\begin{figure}[t]
\begin{center}
\includegraphics[clip,height=4.2cm]{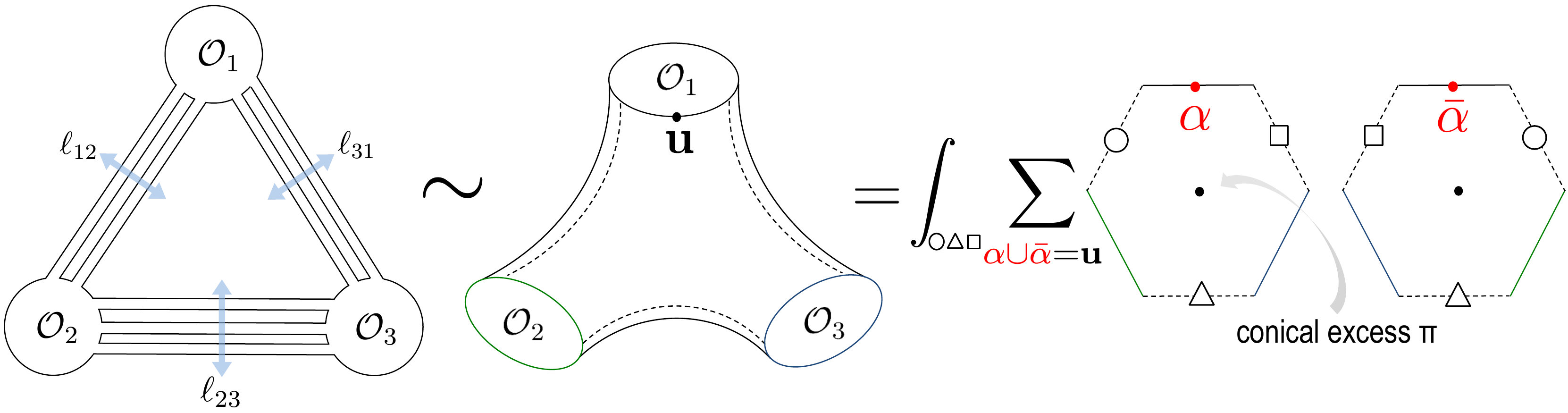}
\end{center}
\vspace{-0.5cm}
\caption{Hexagon formalism for three-point functions: 
The three-point function are represented 
as a pair of pants 
which is cut into two hexagons. These two hexagons are 
separated by 
bridges of lengths $\ell_{ij}\equiv (L_i+L_j-L_k)/2$ ($L_i$ is the 
length of $\mathcal{O}_i$). To compute the three-point function, 
one sums over partitions of physical magnons and the mirror states 
appearing on the dashed edges. Since the physical and the mirror edges intersect by $90$ degrees, the hexagon has a conical excess $\pi$ in its center.\label{fig:3pt}}
\vspace{-0.2cm}
\end{figure}

\vspace{-5pt}
Both in spin chain and in string theory, the structure constant 
of single-trace operators can be represented pictorially by a pair of pants. The key idea in \cite{BKV} is to cut the pair of pants into two hexagonal patches and determine the contribution from each patch, called the {\it hexagon form factor}, using integrability (see figure \ref{fig:3pt}).

When we cut the pair of pants, excitations (magnons) in each 
operator are divided between two hexagons and we need to sum 
over all such possibilities. To bring excitations to the 
second hexagon, we have to move them across 
the {\it bridges}, namely 
propagators connecting two operators. 
This leads to a propagation phase $e^{i p \ell_{ij}}$, with $\ell_{ij}$ 
being the length of the 
bridge between 
$\mathcal{O}_i$ and $\mathcal{O}_j$ and $p$ being 
the 
set of momenta. Upon doing so, we sometimes need to 
reorder excitations. In case it happens, there will be an extra contribution to the phase shift from the S-matrices $S(u,v)$. Altogether, it constitutes the so-called {\it asymptotic part} of the structure constant, which has the following schematic form\footnote{For simplicity, here we consider a three-point function with one non-BPS operator in a rank $1$ sector.} for the configuration depicted in figure \ref{fig:3pt}:
\beq
\left.C_{123}\right|_{\rm asympt} \sim \sum_{\alpha\cup \bar{\alpha}=
\{{\bf u}\}}w_{\alpha,\bar{\alpha}} \times \mathcal{H}_{\alpha}\times\mathcal{H}_{\bar{\alpha}}\period
\eeq
Here $\{{\bf u}\}$ is a set of rapidities of magnons 
and $\mathcal{H}$ denotes the hexagon form factor.
$w_{\alpha,\bar{\alpha}}$ is a partition-dependent prefactor given 
by\footnote{Additional signs 
can appear when the excitations are fermionic, see \cite{fermionic}.}
\beq
w_{\alpha,\bar{\alpha}}= (-1)^{|\bar{\alpha}|} 
\left(\prod_{u_i \in \bar{\alpha}}e^{i p(u_i) \ell_{31}}\right)
\left(\prod_{\substack{i<j\\u_i\in \bar{\alpha},u_j\in \alpha}} S(u_i, u_j)\right)\period
\eeq
This asymptotic part gives the leading contribution when all the operators are long.

To compute the finite-size effects, we need to sum over all possible states appearing on the dashed lines in figure \ref{fig:3pt}, called the {\it mirror edges}. This can be achieved by dressing the mirror edges by magnons and integrating over their momenta. This leads to a series 
\beq\label{mirrorc123}
C_{123}\sim \sum_{\alpha \cup \bar{\alpha}=\{{\bf u}\}} w_{\alpha,\bar{\alpha}}  \left[ \mathcal{H}_{\alpha}\mathcal{H}_{\bar{\alpha}}+\sum_{a}\int \frac{dv}{2\pi}\mu_{a} (v) e^{-\tilde{E}_{a}(v) \ell_{31}}\mathcal{H}_{\alpha;v}\mathcal{H}_{\bar{\alpha};v}  +\cdots\right] \, , 
 \eeq 
 where $\tilde{E}$ is the energy of the mirror magnon and $\mu$ is the measure factor. The subscript $a$ signifies the $a$-th bound state\footnote{Roughly speaking, the bound state index $a$ can be thought of as the Kaluza-Klein mode number arising from the dimensional reduction of $R\times S^3$ to $R$.}, which exists in the spectrum on the mirror edge.
The asymptotic part, discussed above, corresponds to the contribution 
from the vacuum states of the mirror edges. The series \eqref{mirrorc123} can also be regarded as the form factor expansion of a two-point function of {\it hexagon twist operators}, which create an excess angle $\pi$ on the string worldsheet (see figure \ref{fig:3pt}). This is why we refer to $\mathcal{H}$ as the hexagon ``form factor''. 

\begin{figure}[t]
\begin{center}
\begin{minipage}{0.4\hsize}
\begin{flushleft}
\includegraphics[clip,height=4.5cm]{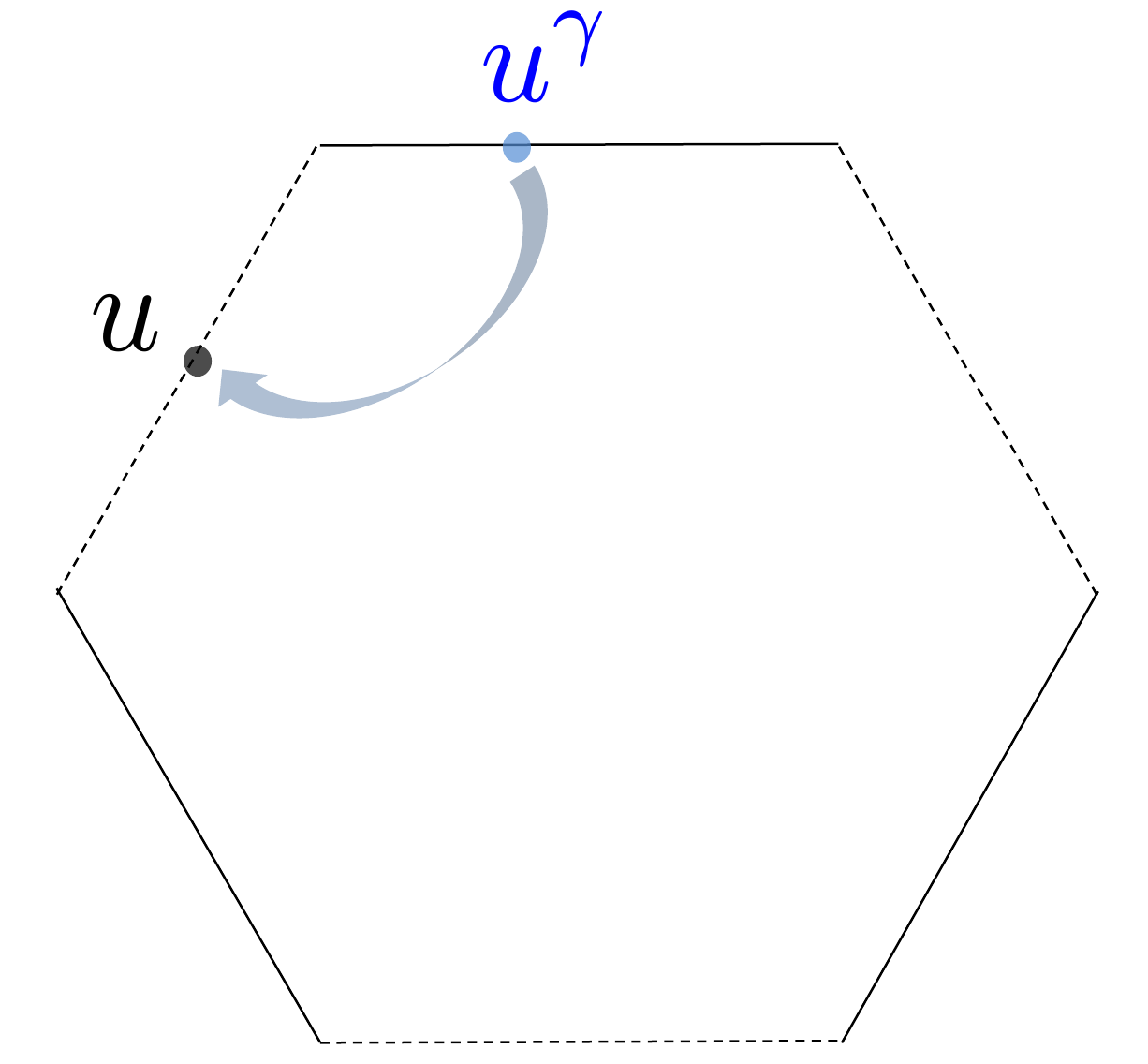}
\end{flushleft}
\end{minipage}
\begin{minipage}{0.4\hsize}
\begin{flushright}
\includegraphics[clip,height=4.5cm]{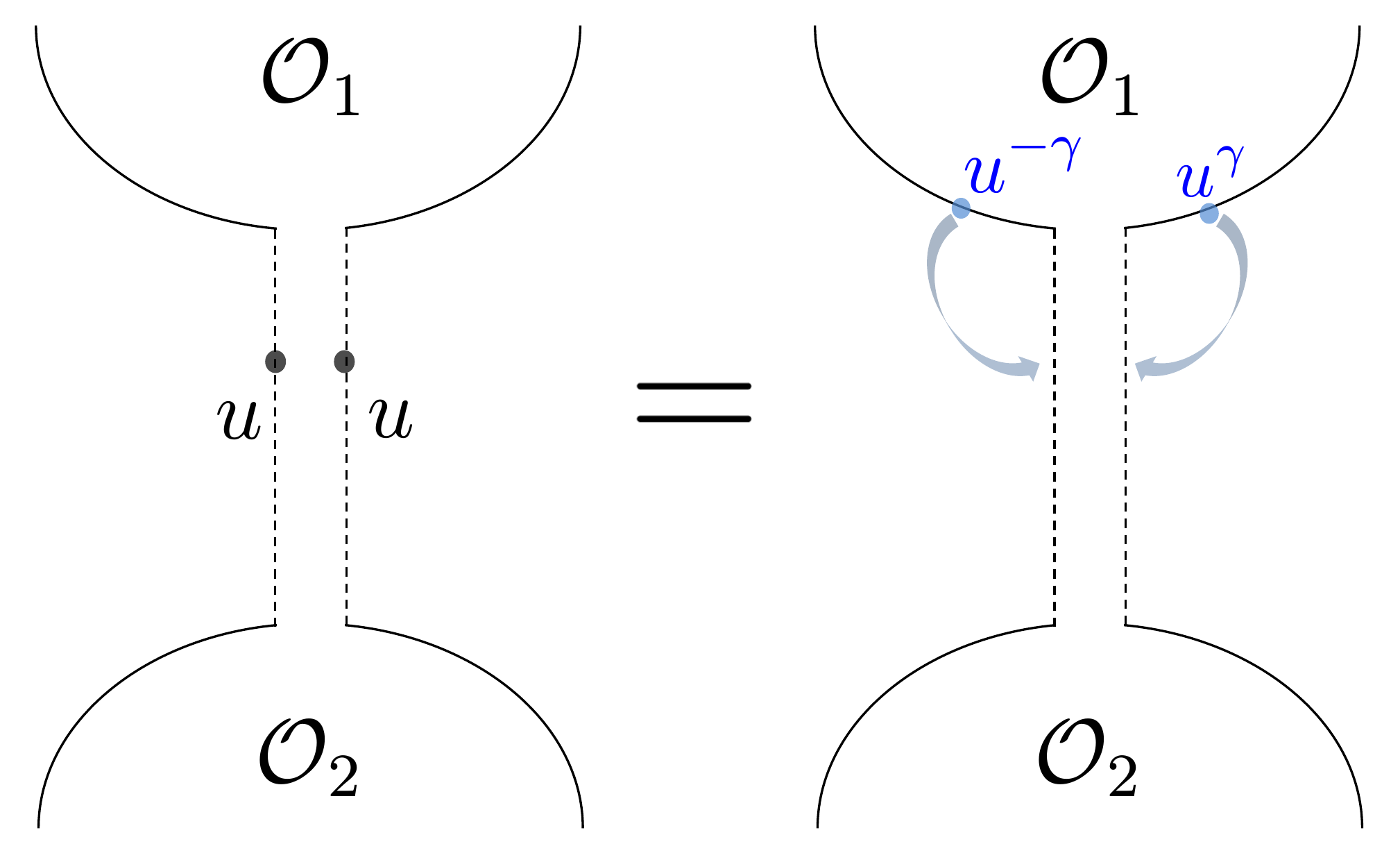}
\end{flushright}
\end{minipage}
\end{center}
\vspace{-0.5cm}
\caption{Mirror transformation. Left: The mirror transformation is an analytic continuation of the rapidity $(u\to u^{\gamma})$, which allows us to move a particle from one edge to another. Right: Alternative viewpoint on the finite size correction. Inserting a complete basis on a mirror edge is equivalent to dressing neighboring physical edges by virtual particle pairs and making them ``entangled''. \label{fig:mirror}}
\end{figure}

It is worth noting that the weight factor for the 
physical particle $(e^{i p \ell})$ and the weight factor for the mirror particle $(e^{-\tilde{E} \ell})$ are related to each other by the analytic continuation called the mirror transformation (see figure \ref{fig:mirror}). We will later see that such a relation exists also for higher-point functions.

The mirror transformation offers another viewpoint on the finite-size correction. As shown in figure \ref{fig:mirror}, inserting a complete basis of states on a mirror edge is equivalent to putting virtual particle pairs on the adjacent physical edges. This stitches two physical edges by making them ``entangled'' with each other. This point of view is often useful in practical computation. 
See for instance Appendix \ref{ap:transfer}.

In the rest of this paper, we will explain how to generalize this formalism to more complicated surfaces which describe higher-point functions.
\section{Simple Exercise at Tree Level\label{sec:simple}}
As a warm up, we compute a tree-level correlation function\footnote{An extensive study of tree-level four-point functions in the SU(2) sector was performed in \cite{Escobedo}.} of three BPS operators and the following single-magnon operator in the SL(2)-sector:
\beq
\mathcal{O}_1 =\sum_{n} e^{i p n}\,\Tr \, \big(\cdots Z \underset{\substack{\uparrow\\n}}{(DZ)} Z\cdots\big)\period
\eeq
Here $D$ is a holomorphic derivative on the $x^2$-$x^3$ plane; $D=(\del_2 -i \del_3)/2$.
Strictly speaking, this operator cannot exist unless $p=0$ since it violates the cyclicity of the trace. However we keep $p$ to be nonzero throughout this section in order to illustrate the main idea in the simplest set-up. We nevertheless impose the Bethe equation $e^{ip L_1}=1$, with $L_1$ being the length of the operator. 
\begin{figure}[t]
\begin{center}
\includegraphics[clip,height=6.5cm]{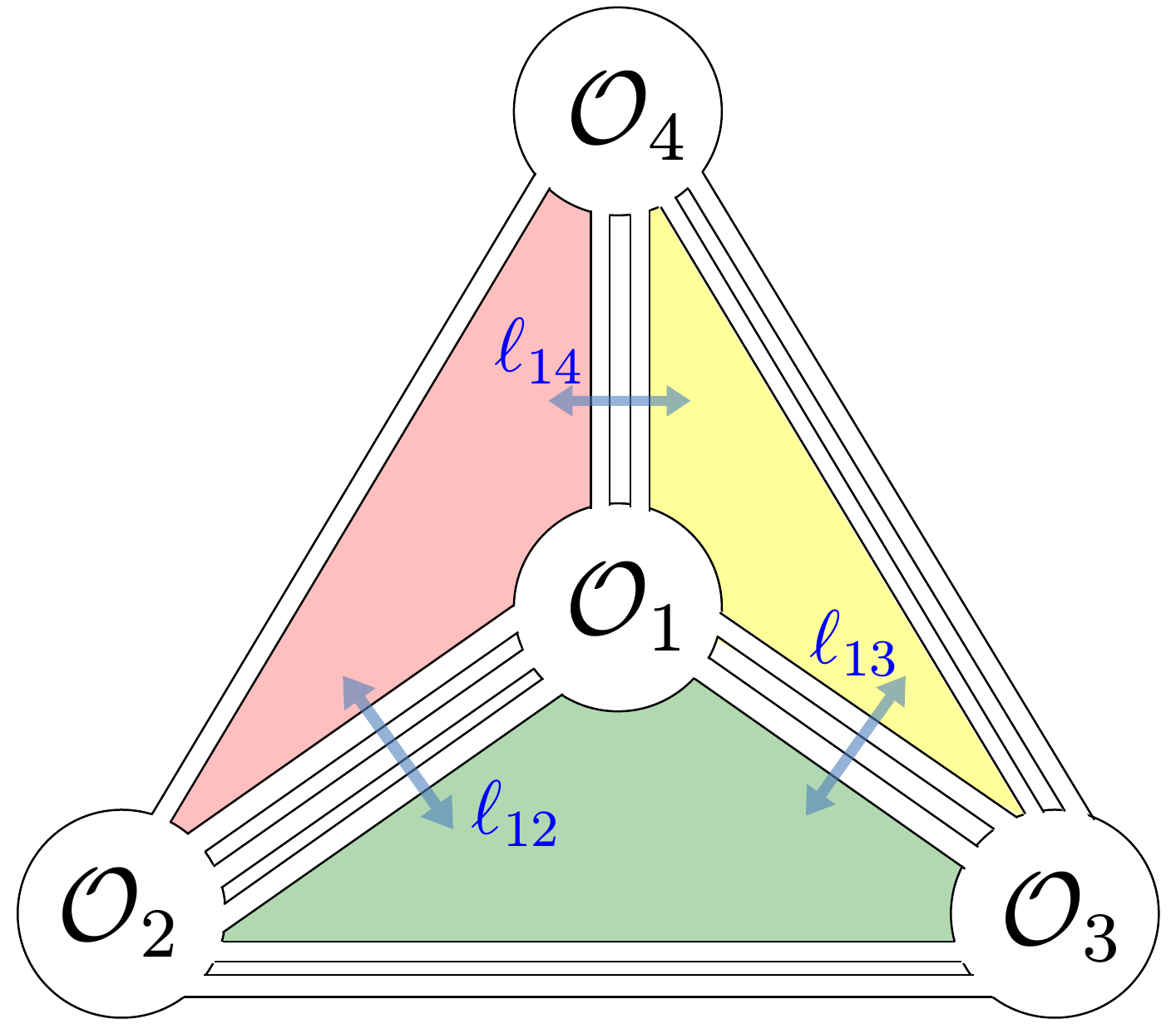}
\end{center}
\vspace{-0.7cm}
\caption{An example of tree-level Wick contraction: A magnon $D$ can live in either one of three bridges ($\ell_{12}$, $\ell_{13}$, or $\ell_{14}$). As shown above, the bridges split the diagram into four hexagonal patches, which are depicted in different colors. Each term in the final result \eqref{treesum2} can be interpreted as a contribution from one of those hexagon patches.\label{fig:simple}}
\end{figure}

The rest of the operators are given by
\beq\label{definitionBPS}
\mathcal{O}_i (x_i,Y_i)=\Tr \left((Y_i\cdot \Phi)^{L_i}(x_i)\right)\comma
\eeq
where $Y_i$'s are six-dimensional null vectors parameterizing the orientation in the R-symmetry space, and the product $(Y_i\cdot \Phi)$ is a standard inner product defined by $\sum_{I=1}^{6}Y_i^{I} \Phi^{I}$. 
As a further simplification, we assume 
that all four operators live on the $x^2$-$x^3$ plane.

At tree level, the four-point function can be computed in two steps: First we list up all possible ways to contract four operators. See figure \ref{fig:simple} for an example. They are specified by the numbers of Wick contractions between operators, which we call the {\it bridge lengths}. Second, for each graph, we sum over the positions of the magnon in $\mathcal{O}_1$.

In the case depicted in figure \ref{fig:simple}, there are three distinct possibilities: When the magnon lives in the bridge $\ell_{14}$, the derivative acts on a propagator between $\mathcal{O}_1$ and $\mathcal{O}_4$ and produces an extra position dependence,
\beq
D\left( \frac{1}{x_{14}^2}\right) = -\frac{1}{x^{+}_{14}}\frac{1}{x_{14}^2}\comma
\eeq
where $x^{\pm}$ denote holomorphic and anti-holomorphic 
coordinates\footnote{In this paper we are using a slightly unusual notation, in order to avoid the conflict of notations.} $x^2 \pm ix^3$, and $x^{\pm}_{ij}$  and $|x_{ij}|$ are given by $x^{\pm}_{ij}\equiv x^{\pm}_i-x^{\pm}_j$ and $x_{ij}\equiv |x_i-x_j|$. Then, the summation over positions of the magnon yields an extra overall factor,
\beq
{\tt first}\equiv-\frac{1}{x_{14}^{+}}\sum_{n=1}^{\ell_{14}} e^{ipn}=
-\frac{\mathcal{N}(p)}{x_{14}^{+}} \left(1-e^{ip\ell_{14}}\right)\comma
  \eeq 
where $\mathcal{N}(p)$ is given by $1/(e^{-ip}-1)$.
  Similarly, when the magnon lives on the bridges $\ell_{12}$ and $\ell_{13}$, we obtain respectively  \begin{align}
{\tt second}&\equiv-\frac{1}{x_{12}^{+}}
\sum_{n=\ell_{14}+1}^{\ell_{14}+\ell_{12}} e^{ipn}=-
\frac{\mathcal{N}(p)}{x_{12}^{+}} \left(e^{ip\ell_{14}}-e^{ip(\ell_{14}+\ell_{12})}\right)\comma\\
{\tt third}&\equiv-\frac{1}{x_{13}^{+}}
\sum_{n=\ell_{14}+\ell_{12}+1}^{L_1} e^{ipn}=-\frac{\mathcal{N}(p)}{x_{13}^{+}} \left(e^{ip(\ell_{14}+\ell_{12})}-1\right)\period\label{thirdbridge}
  \end{align}
In \eqref{thirdbridge}, we used the Bethe equation $e^{ip L_1}=1$.

Adding up all these contributions and reorganizing them, we arrive at
\beq\label{treesum}
{\tt first}+{\tt second}+{\tt third}= \mathcal{N}(p) \left(\textcolor[rgb]{0.7,0.4,0}{\frac{x^{+}_{34}}{x^{+}_{13}x^{+}_{14}}}+\textcolor[rgb]{0.7,0,0}{\frac{x^{+}_{42}}{x^{+}_{12}x^{+}_{14}}e^{ip\ell_{14}}}+\textcolor[rgb]{0,0.5,0}{\frac{x^{+}_{23}}{x^{+}_{12}x^{+}_{13}}e^{ip(\ell_{14}+\ell_{12})}}\right)\period
\eeq
A similar computation  for the three-point function was performed in \cite{BKV}. In that case, different terms corresponded to different ways of distributing magnons among two hexagons, and the exponential prefactors ($e^{ip\ell}$) were interpreted as the phase shift needed to move a magnon from one hexagon to the other. Here as well, we propose to interpret terms in the parenthesis as describing different ways to distribute the magnon 
among several hexagons. For instance, the first term in \eqref{treesum} corresponds to the case where the magnon is in the hexagon formed by $\mathcal{O}_1$, $\mathcal{O}_3$ and $\mathcal{O}_4$ whereas the second term corresponds to the case where the magnon is in the hexagon formed by 
$\mathcal{O}_1$, $\mathcal{O}_4$ and $\mathcal{O}_2$ (see also figure \ref{fig:simple}). 
\begin{figure}[t]
\begin{center}
\includegraphics[clip,height=5.5cm]{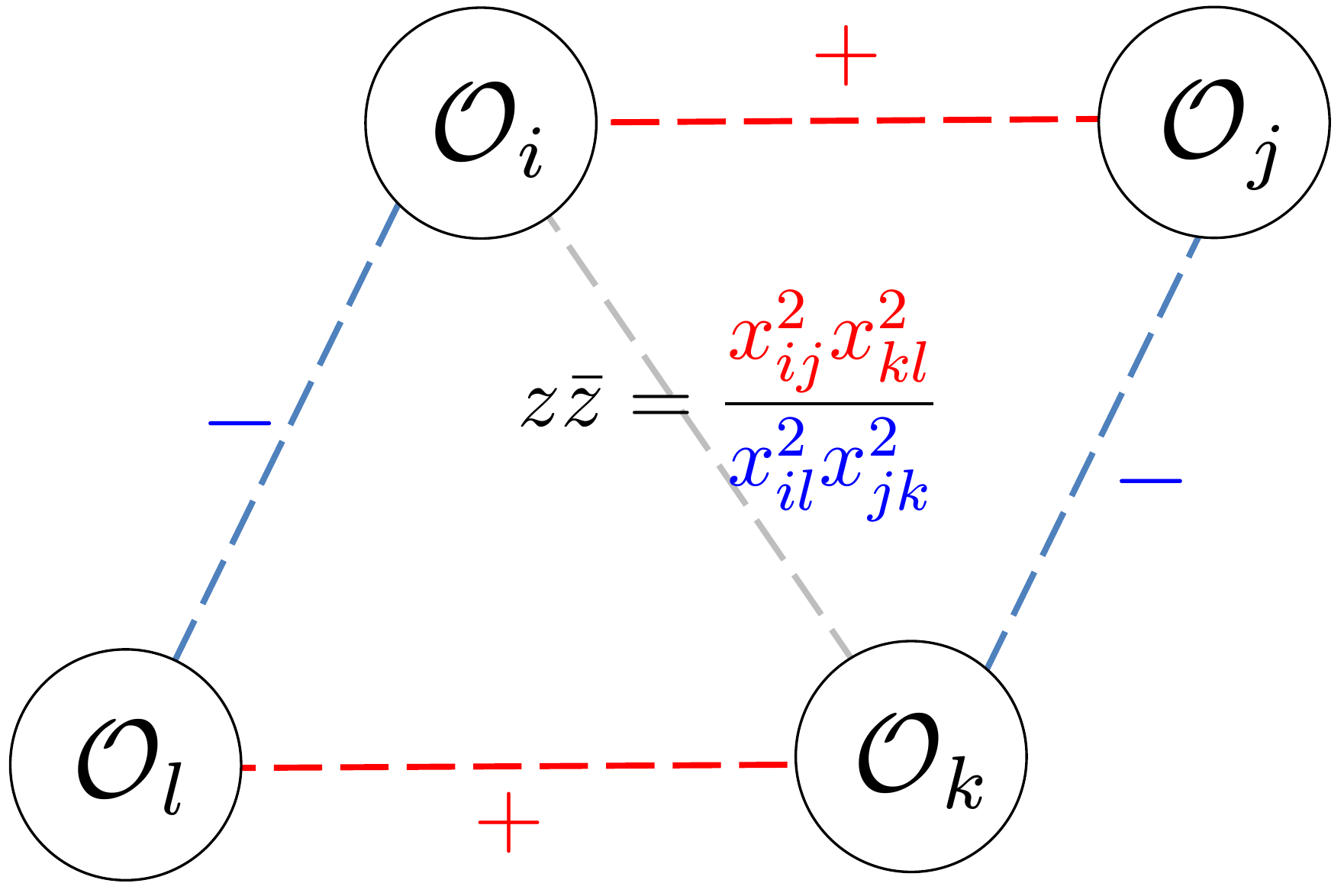}
\end{center}
\vspace{-0.7cm}
\caption{Edge and cross ratio: There is a natural way to associate a cross ratio with each edge. For instance, the cross ratio associated with the black dotted edge is given by the expression above. $\pm$ signs signify whether the corresponding propagators appear in the numerator or in the denominator. In terms of these cross ratios, the weight factor in \eqref{treesum2} reads $\mathcal{W}= 1/z$. This rule of relating an edge and a cross ratio is akin to the definition of the Fock coordinates of the Teichmuller space \cite{Fock} or the Fock-Goncharov coordinates of the moduli space of flat connections \cite{FockGoncharov}. Note also that the Fock-Goncharov coordinates show up in the study of four-point functions at strong coupling \cite{chi-system}.\label{fig:ratiorule}}
\vspace{-0.2cm}
\end{figure}

A crucial difference from the three-point function is that different terms in \eqref{treesum} are dressed by different space-time dependences. To understand its physical implication, it is useful to factor out the first term and rewrite \eqref{treesum} as\footnote{In \cite{BKV}, there are {\it{ad hoc}} additional signs when moving the magnons from one hexagon to the other. Here such signs appear naturally after factoring out $x_{34}^{+}/(x_{13}^{+}x_{14}^{+})$. A similar reasoning can be applied to three-point functions. It will be interesting to check that the signs of \cite{fermionic} for fermionic excitations can also be reproduced in this way.
}
\beq\label{treesum2}
{\tt first}+{\tt second}+{\tt third}=\mathcal{N}(p)\frac{x^{+}_{34}}{x^{+}_{13}x^{+}_{14}} \left(\textcolor[rgb]{0.7,0.4,0}{1}-\textcolor[rgb]{0.7,0,0}{\frac{x_{13}^{+}x_{24}^{+}}{x_{12}^{+}x_{34}^{+}}e^{ip\ell_{14}}}+\textcolor[rgb]{0,0.5,0}{\frac{x_{13}^{+}x_{24}^{+}}{x_{12}^{+}x_{34}^{+}}\frac{x^{+}_{14}x^{+}_{23}}{x^{+}_{13}x^{+}_{24}}e^{ i p( \ell_{14}+\ell_{12})}}\right)\period
\eeq
As can be readily seen, the factors in front of $e^{ip\ell}$'s are precisely the (holomorphic part of) cross ratios. This implies that, besides the phase shift, we should multiply appropriate cross ratios when we move magnons across the bridges. The relation between the bridge and the cross ratio can be understood graphically as shown in figure \ref{fig:ratiorule}. In addition to such factors, there is an overall prefactor $x^{+}_{34}/(x^{+}_{13}x^{+}_{14})$.

As mentioned in section \ref{sec:review}, the weight factors of physical and mirror magnons for the three-point functions are related with each other by the mirror transformation. It is thus tempting to speculate that, in higher-point functions, the cross ratios couple also to mirror magnons. In the next section, we will see that this is indeed the case: We derive the weight factor for mirror magnons based on the symmetry, and show that it incorporates the dependence on the cross ratios.
\section{Hexagonalization\label{sec:hexagonalization}}
\subsection{Main Proposal\label{subsec:main}}
\begin{figure}[t]
\begin{center}
\includegraphics[clip,height=7cm]{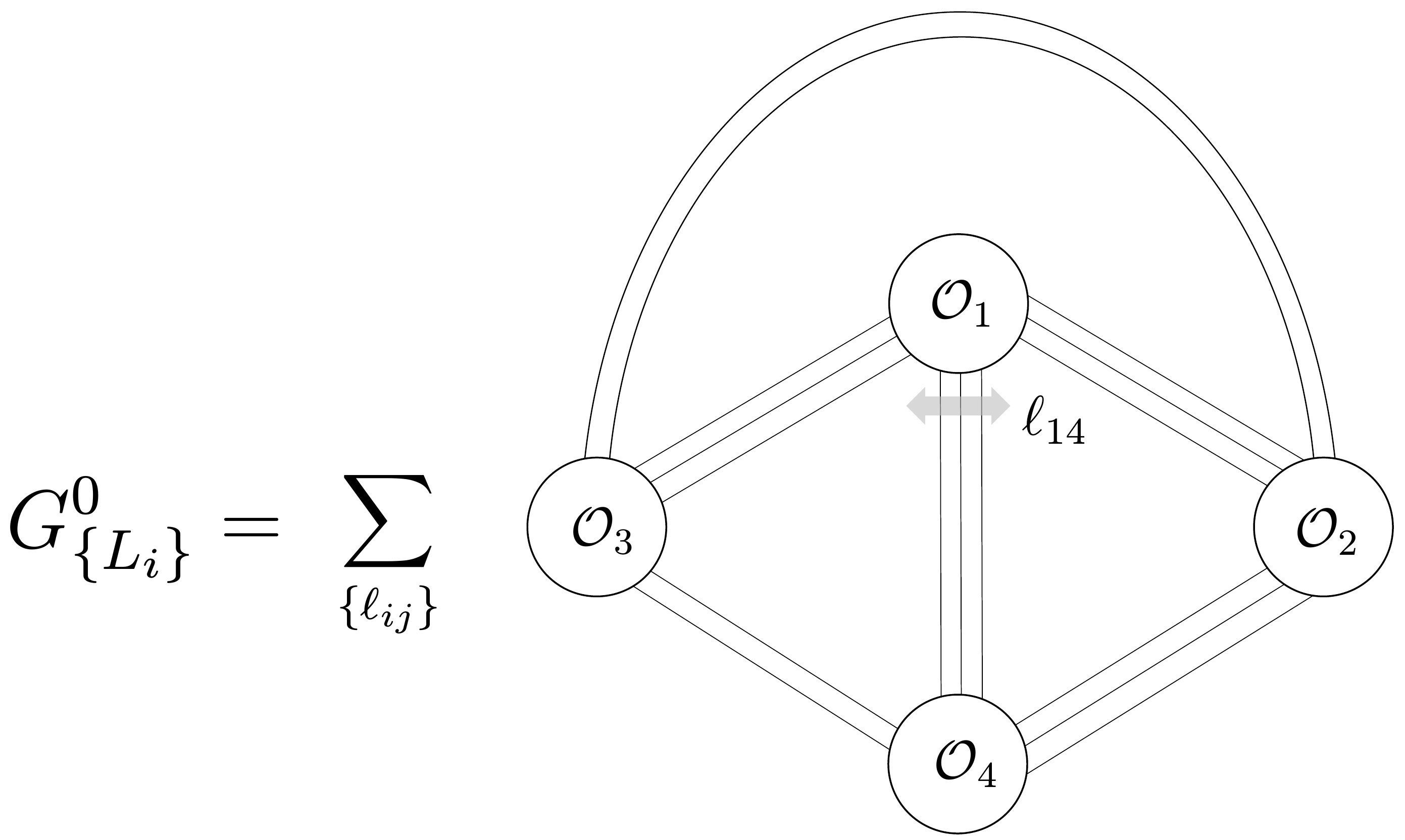}
\end{center}
\vspace{-0.5cm}
\caption{Tree-level correlation function: At tree level, the correlation function is given by a sum over all possible planar graphs. Each graph is characterized by a set of bridge lengths.\label{fig:weakhex}}
\end{figure}
We consider a correlation function of BPS operators of the form \eqref{definitionBPS}.
  Here and below, we normalize the operators as
\beq\label{normalization}
\langle \mathcal{O} (x_1,Y_1)\mathcal{O}(x_2,Y_2)\rangle = (d_{12})^{L}\comma
\eeq
where $L$ is the length of the operator and $d_{ij}$ is a Wick contraction of two scalar fields,
\beq
d_{ij}\equiv\frac{y_{ij}^{2}}{x_{ij}^2}\comma
\eeq
with $y_{ij}^2 = Y_i\cdot Y_j$.

In the large $N_c$ limit, the correlator consists of two parts: One is the disconnected part, which is given by a product of lower-point functions and has a lower power of $1/N_c$. The other is the connected part, which scales as $1/N_c^{n-2}$ and corresponds to a true interaction of $n$ operators:
\beq\label{structuredisc}
\langle \mathcal{O}_1 (x_1,Y_1)\mathcal{O}_2 (x_2,Y_2) \cdots \mathcal{O}_{n}(x_n,Y_n)\rangle={\tt (disconnected)}+\frac{\prod_{i=1}^{n}\sqrt{L_i}}{N_c^{n-2}}\,\,G_{\{L_i\}}\period
\eeq
Here we stripped off the factor $\prod_i\sqrt{L_i}$ from the connected part $G_{\{L_i\}}$ as it always appears\footnote{For a more detailed explanation of  this factor, see introduction of \cite{Tailoring}.} owing to the normalization \eqref{normalization}. The main subject of this paper is 
$G_{\{L_i\}}$.

\begin{figure}[t]
\begin{center}
\includegraphics[clip,height=7cm]{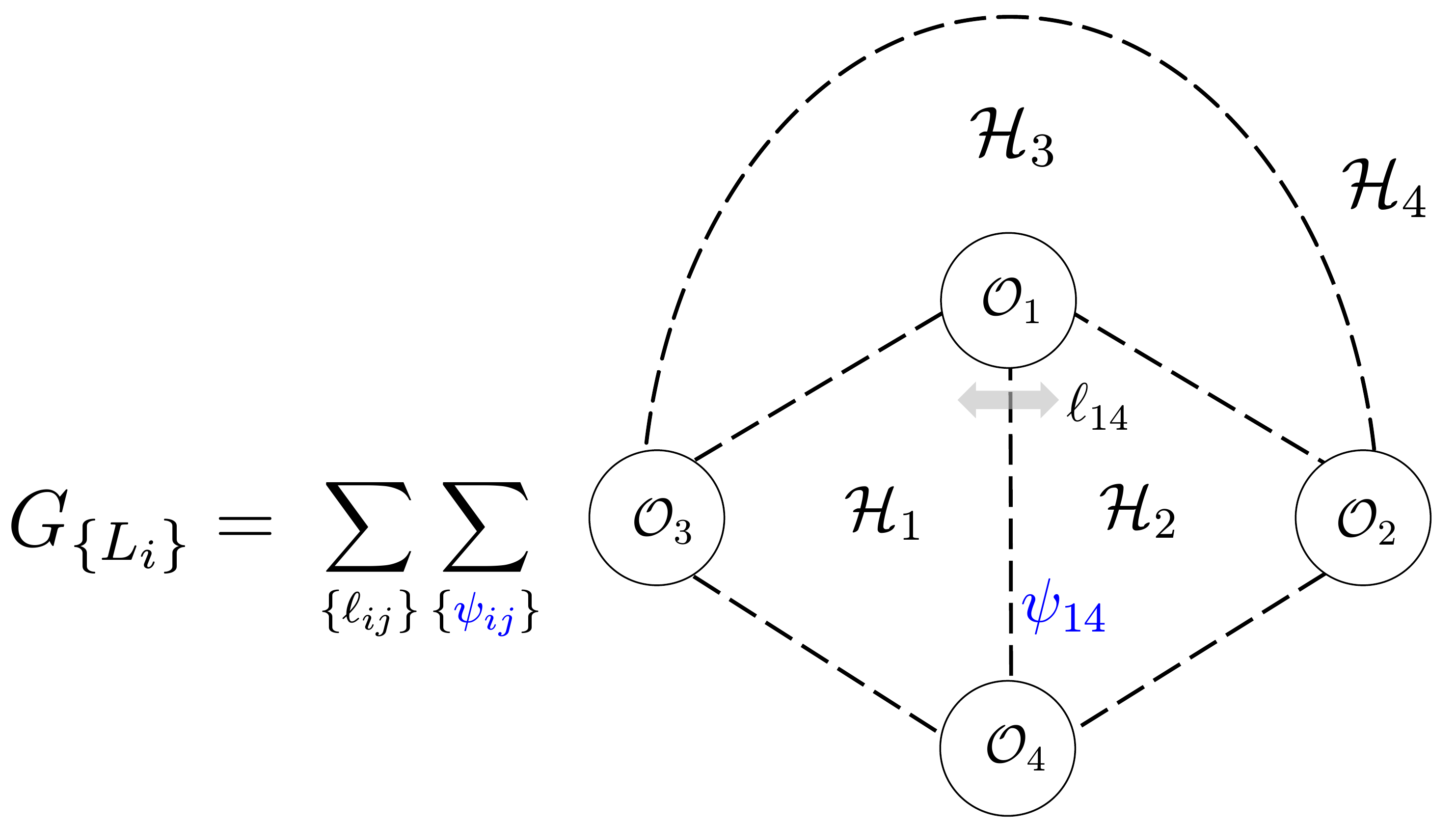}
\end{center}
\vspace{-0.5cm}
\caption{Main proposal \eqref{prop}: BPS correlators at finite coupling are given by a sum over graphs and a sum over all possible mirror states living on the dashed edges. (For simplicity, here we only depicted the state living on the edge $14$.) Each state comes with an appropriate weight factor $\mathcal{W}$, which will be determined in section \ref{subsec:gluing}.
\label{fig:finitehex}}
\end{figure}

At tree level, $G_{\{L_i\}}$ is a sum of all possible planar connected graphs, each of which is specified by a set of bridge lengths $\ell_{ij}$:
\beq\label{treehex}
G^{0}_{\{L_i\}}=\sum_{{\rm graphs}}\,\,\prod_{(i,j)} (d_{ij})^{\ell_{ij}}\period
\eeq
A typical graph divides the planar surface into $2(n-2)$ hexagonal patches as shown in figure \ref{fig:weakhex} and figure \ref{fig:finitehex}. We call this decomposition of the surface {\it hexagonalization} as it bears resemblance\footnote{If we shrink each operator to a point, a hexagonalization reduces to a triangulation.} to a triangulation of a $n$-punctured sphere. The product $\prod_{(i,j)}$ in \eqref{treehex} runs over all the edges of a given hexagonalization.

To compute  $G_{\{L_i\}}$ at finite coupling, we replace each hexagonal patch by the hexagon form factor. More precisely, we conjecture that the connected four-point function at finite coupling can be computed by inserting a complete basis of states to each mirror edge of a hexagonalization, dressing them with appropriate weight factors, and evaluating the contributions from each hexagon. This leads to our main formula (see also figure \ref{fig:finitehex}),
\beq\label{prop}
\begin{aligned}
G_{\{L_i\}}=\sum_{{\rm graphs}}\left(\prod_{(i,j)} (d_{ij})^{\ell_{ij}}\right)\left[ \sum_{\{\psi_{ij}\}}\prod_{(i,j)}\mu_{\psi_{ij}}e^{-\tilde{E}_{\psi_{ij}} \ell_{ij}}\mathcal{W}_{\psi_{ij}} \prod_{(i,j,k)} \mathcal{H}_{\psi_{ij},\psi_{jk},\psi_{ki}}\right]\period
 \end{aligned}
 \eeq
 
Giving a precise meaning to this formula, which is cryptic as it is, is the main goal of the rest of this section. As a preparation, let us clarify what each symbol stands for: $\psi_{ij}$ denotes a state inserted on the edge $i$-$j$, $e^{-\tilde{E}_{\psi_{ij}}\ell_{ij}}$ is a propagation factor of the mirror state, and $\mu_{\psi_{ij}}$ is a measure factor. $\mathcal{W}_{\psi_{ij}}$ is a weight factor encoding kinematics whose explicit form will be derived in section \ref{subsec:gluing}. The symbol $\prod_{(i,j,k)}$ denotes a product over faces of a hexagonalization.  Finally $\mathcal{H}$ is, as before, the hexagon form factor.
\subsection{Symmetry and Gluing Rules\label{subsec:gluing}}
We now determine the weight factor $\mathcal{W}$ from the symmetry-based argument. In the case of three-point functions, the underlying symmetry is most transparent in the {\it canonical configuration}\footnote{In \cite{BKV}, we considered more general configurations using the twisted translation. The canonical configuration is a specialization of it.}, in which the operators take the following form:
\beq\label{twistedframe}
\mathcal{O}_1: \Tr \big(Z^{L_1}\big)|_{x^{\mu}=(0,0,0,0)}\,,\quad \mathcal{O}_2: \Tr \big(\bar{Z}^{L_2}\big)|_{x^{\mu}=\infty}\,,\quad \mathcal{O}_3:\Tr \big(\tilde{Z}^{L_3}\big)|_{x^{\mu}=(0,1,0,0)}\,,
\eeq
Here $\tilde{Z}$ is given by $\tilde{Z}=(Z+\bar{Z}+Y-\bar{Y})/2$. In terms of the polarization vectors, it corresponds to $Y_1 =(1,i,0,0,0,0)$, $Y_2=(1,-i,0,0,0,0)$ and $Y_3=(1,0,0,i,0,0)$ in our convention.
We refer to the hexagon twist operator defined in this configuration as the {\it canonical hexagon}, and denote it by $\hat{\mathcal{H}}$. 

Let us now consider a configuration depicted in figure \ref{fig:finitehex} and try to glue two hexagons $\mathcal{H}_1$ and $\mathcal{H}_2$ through the edge $14$. In this configuration, $\mathcal{H}_{1,2}$ are not canonical since the operators forming these hexagons 
($\mathcal{O}_{1,4,3}$ for $\mathcal{H}_1$ and $\mathcal{O}_{1,2,4}$ for for $\mathcal{H}_2$) are at generic points and have arbitrary R-symmetry polarizations. However, using the conformal and the R-symmetry transformations, we can always relate them to the canonical configuration. Namely, we can express the hexagon operators $\hat{\mathcal{H}}_{1,2}$ in terms of the canonical hexagon as
\beq\label{H and H12}
\hat{\mathcal{H}}_1 = g_1^{-1}\hat{\mathcal{H}}g_1\comma \quad \hat{\mathcal{H}}_2 =g_2^{-1} \hat{\mathcal{H}}g_2\comma
\eeq
where $g_{1,2} \in {\rm PSU}(2,2|4)$ are transformations needed to bring three operators forming each hexagon to the canonical configuration. From this operatorial point of view, gluing two hexagons correspond to considering a sequence of hexagon operators,
\beq
\cdots \hat{\mathcal{H}}_2 e^{-\tilde{E}\ell_{14}}\hat{\mathcal{H}}_1 \cdots
\eeq
and inserting a complete basis of states in between. This leads to an expression\footnote{Here we are using the basis which diagonalizes $g_2g_1^{-1}$.}
\beq
\sum_{\psi}\mu_{\psi} e^{-\tilde{E}_{\psi}\ell_{14}}\left(\cdots \hat{\mathcal{H}}|\psi\rangle  \langle \psi|g|\psi\rangle\langle\psi| \hat{\mathcal{H}} \cdots\right)\comma
\eeq
with $g\equiv g_2 g_1^{-1}$.
The factor in the middle, $\langle \psi|g|\psi\rangle$, gives rise to the weight factor $\mathcal{W}_{\psi}$. 

\begin{figure}[t]
\begin{center}
\includegraphics[clip,height=4cm]{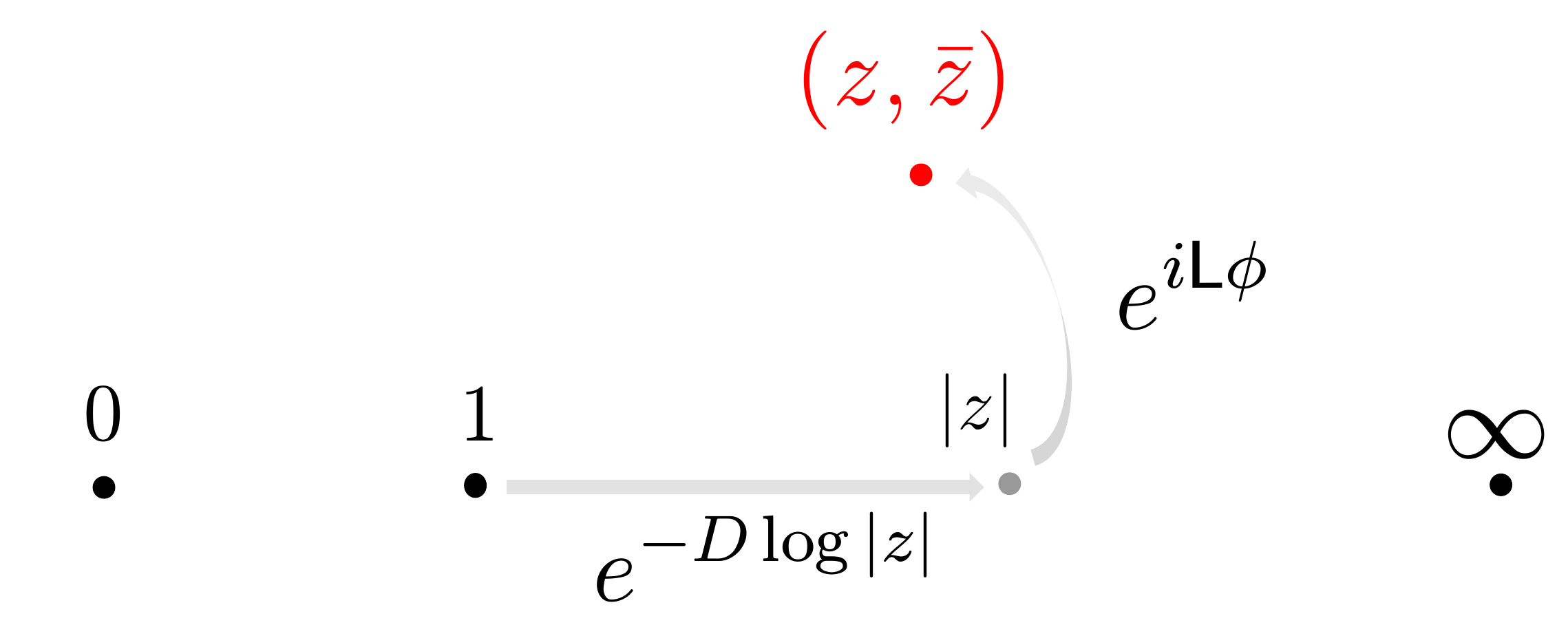}
\end{center}
\vspace{-0.5cm}
\caption{Weight factor and symmetry. In the second hexagon $\mathcal{H}_2$, the operators are positioned at $0$, $(z,\bar{z})$ and $\infty$. To obtain this configuration starting from the ``canonical one'', one needs to perform the transformations, $e^{-D\log|z|}$ and $e^{i{\sf L}\phi}$ as depicted in the figure. Note that these transformations leave the points $0$ and $\infty$ invariant. 
\label{fig:weightsym}}
\end{figure}

Note that $\mathcal{W}$ is invariant under the transformation  $g_{1,2}\to  g_{1,2}h$ with $h\in {\rm PSU}(2,2|4)$. Making use of such transformations, we can bring $\hat{\mathcal{H}}_1$ to be canonical without changing $\mathcal{W}$:
\beq
\begin{aligned}
\mathcal{O}_1:\quad &x_1 = (0,0,0,0)\,,  &&Y_1=(1,\hspace{9pt}i,0,0,0,0)\,,\\
\mathcal{O}_3: \quad&x_3 = (0,1,0,0)\,, &&Y_3=(1,\hspace{9pt}0,0,i,0,0)\,,\\
\mathcal{O}_4:\quad &x_4 = \infty\,,  &&Y_4=(1,-i,0,0,0,0)\,.
\end{aligned}
 \eeq 
On the other hand, $\hat{\mathcal{H}}_2$ is not canonical since the position and the polarization of $\mathcal{O}_2$ in this frame are given in terms of the conformal and the R-symmetry cross ratios as 
\beq
\begin{aligned}
\mathcal{O}_2:\quad 
&x_2 = (0,\text{\small Re} (z),\text{\small Im} (z),0)\,, \\
&Y_2=(2/|\alpha|)((1+\alpha\bar{\alpha})/2,i(1-\alpha\bar{\alpha})/2,i\text{\small Im}(\alpha),i\text{\small Re}(\alpha),0,0)\,,
\end{aligned}
\eeq
where $z$ and $\alpha$ are defined in a standard way as follows:
\beq
\begin{aligned}
z\bar{z}=\frac{x_{12}^2x_{34}^2}{x_{13}^2x_{24}^2}\comma \,\,\,\, (1-z)(1-\bar{z})=\frac{x_{14}^2 x_{23}^2}{x_{13}^2 x_{24}^2}\comma \,\,\,\, \alpha\bar{\alpha}=\frac{y_{12}^2y_{34}^2}{y_{13}^2y_{24}^2}\comma \,\,\,\, (1-\alpha)(1-\bar{\alpha})=\frac{y_{14}^2 y_{23}^2}{y_{13}^2 y_{24}^2}\period
\end{aligned}
\eeq
To obtain the configuration for $\hat{\mathcal{H}}_2$ starting from the canonical configuration, we need to perform the dilatation and the rotation (see figure \ref{fig:weightsym}),
\beq
e^{-D \log |z|} e^{i {\sf L} \phi} \comma
\eeq
where ${\rm L}$ and $\phi$ are given by\footnote{Here $L^{\alpha}{}_{\beta}$ and $\dot{L}^{\dot{\alpha}}{}_{\dot{\beta}}$ are Lorentz generators contained in $\mathfrak{psu}(2|2)^2$.}
\beq
\begin{aligned}
{\sf L}=\frac{1}{2}(L^{1}{}_{1}-L^{2}{}_{2}-L^{\dot{1}}{}_{\dot{1}}+L^{\dot{2}}{}_{\dot{2}})\comma\qquad e^{i\phi}=\sqrt{\frac{z}{\bar{z}}}\period
\end{aligned}
\eeq
The same argument applies also to the R-symmetry part and the full transformation which brings $\hat{\mathcal{H}}$ to $\hat{\mathcal{H}}_2$ is
\beq\label{expression for g}
g=e^{-D \log |z| }e^{i {\sf L}\phi}e^{J \log |\alpha|}e^{i{\sf R}\theta}= e^{-(D-J)\log|z|}e^{J(\log|\alpha|-\log|z|)}e^{i {\sf L}\phi}e^{i{\sf R}\theta}\comma
\eeq
where $J$ is the R-charge which rotates $Z$ and $\bar{Z}$ and {\sf R} and $\theta$ are the R-symmetry analogue of ${\sf L}$ and $\phi$:
\beq
\begin{aligned}
{\sf R}=\frac{1}{2}(R^{1}{}_{1}-R^{2}{}_{2}-R^{\dot{1}}{}_{\dot{1}}+R^{\dot{2}}{}_{\dot{2}})\comma\qquad e^{i\theta}=\sqrt{\frac{\alpha}{\bar{\alpha}}}\period
\end{aligned}
\eeq
Thus, the weight factor can be determined as
\beq\label{chemical potential}
\mathcal{W}_{\psi}=e^{-2i\tilde{p}_{\psi}\log |z|}e^{J_{\psi}\varphi}e^{i{\sf L}_{\psi}\phi}e^{i{\sf R}_{\psi}\theta}\comma
\eeq
with 
\beq
e^{\varphi}=\left|\alpha/z\right|\period
\eeq
Here we used the fact that $D-J$ is related to the spin-chain energy $E$ and the mirror momentum $\tilde{p}$ as
\beq\label{relationDJp}
\frac{D-J}{2}=E = i \tilde{p}\comma
\eeq 
and $J_{\psi}$, ${\sf L}_{\psi}$ and ${\sf R}_{\psi}$ denote the charges of the state $\psi$. The weight factor $\mathcal{W}_{\psi}$ can be regarded as a sort of the chemical potential from the two-dimensional world-sheet point of view. In section \ref{sec:4BPS}, we will explicitly evaluate $\mathcal{W}$ for one magnon state and show that it coincides with a character of $\mathfrak{psu}(2|2)$.
\subsection{Gluing Multiple Channels\label{subsec:multichannel}}
The argument above carries over as long as we glue just one edge: We go to a frame where the edge runs from the origin to infinity and read off the transformation which relates two hexagons. The resulting weight factor is given by \eqref{chemical potential} with the cross ratios replaced appropriately using the rule given in figure \ref{fig:ratiorule}.
\begin{figure}[t]
\begin{center}
\begin{minipage}{0.45\hsize}
\begin{flushleft}
\includegraphics[clip,height=4.5cm]{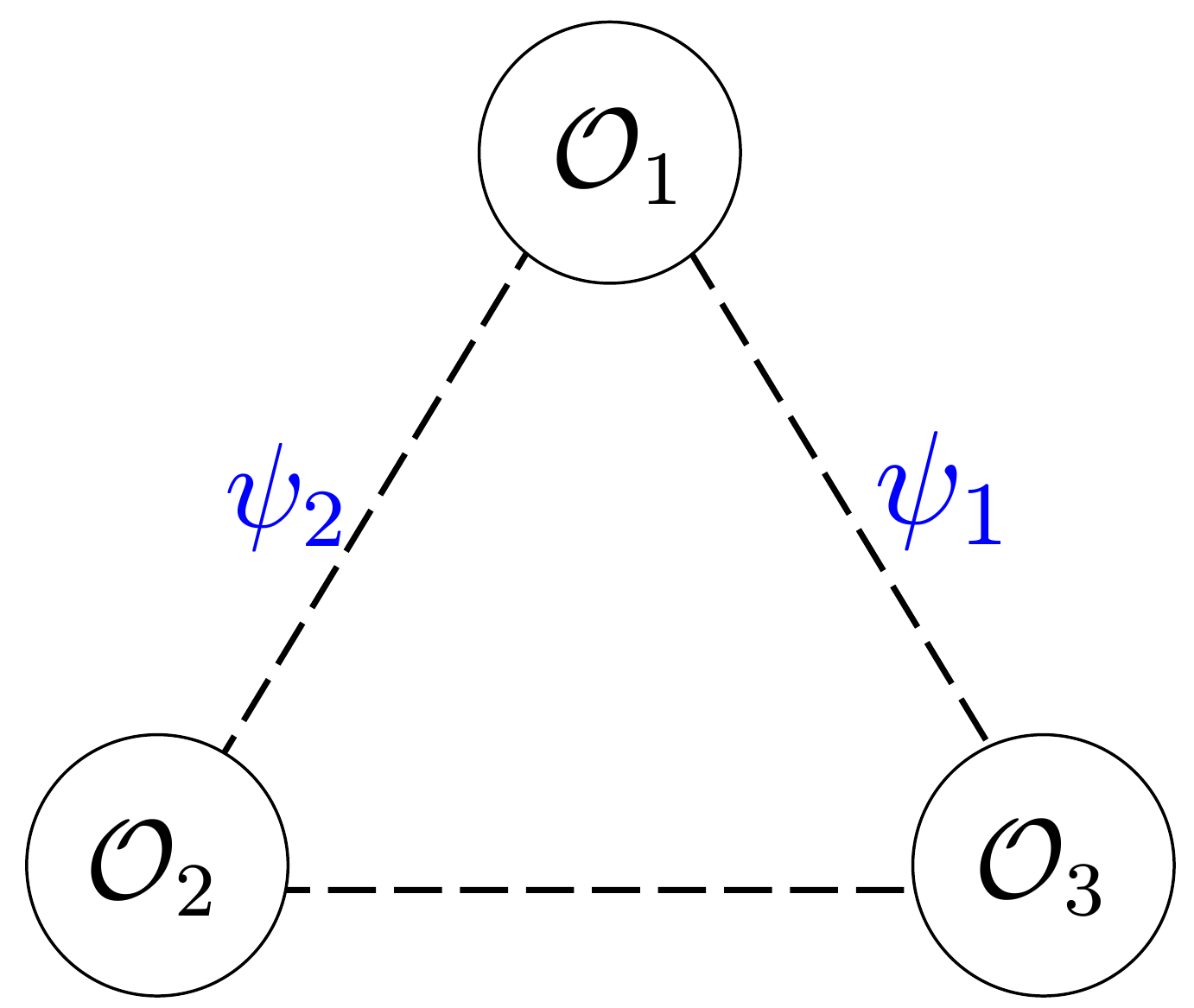}
\end{flushleft}
\end{minipage}
\begin{minipage}{0.45\hsize}
\begin{flushright}
\includegraphics[clip,height=4cm]{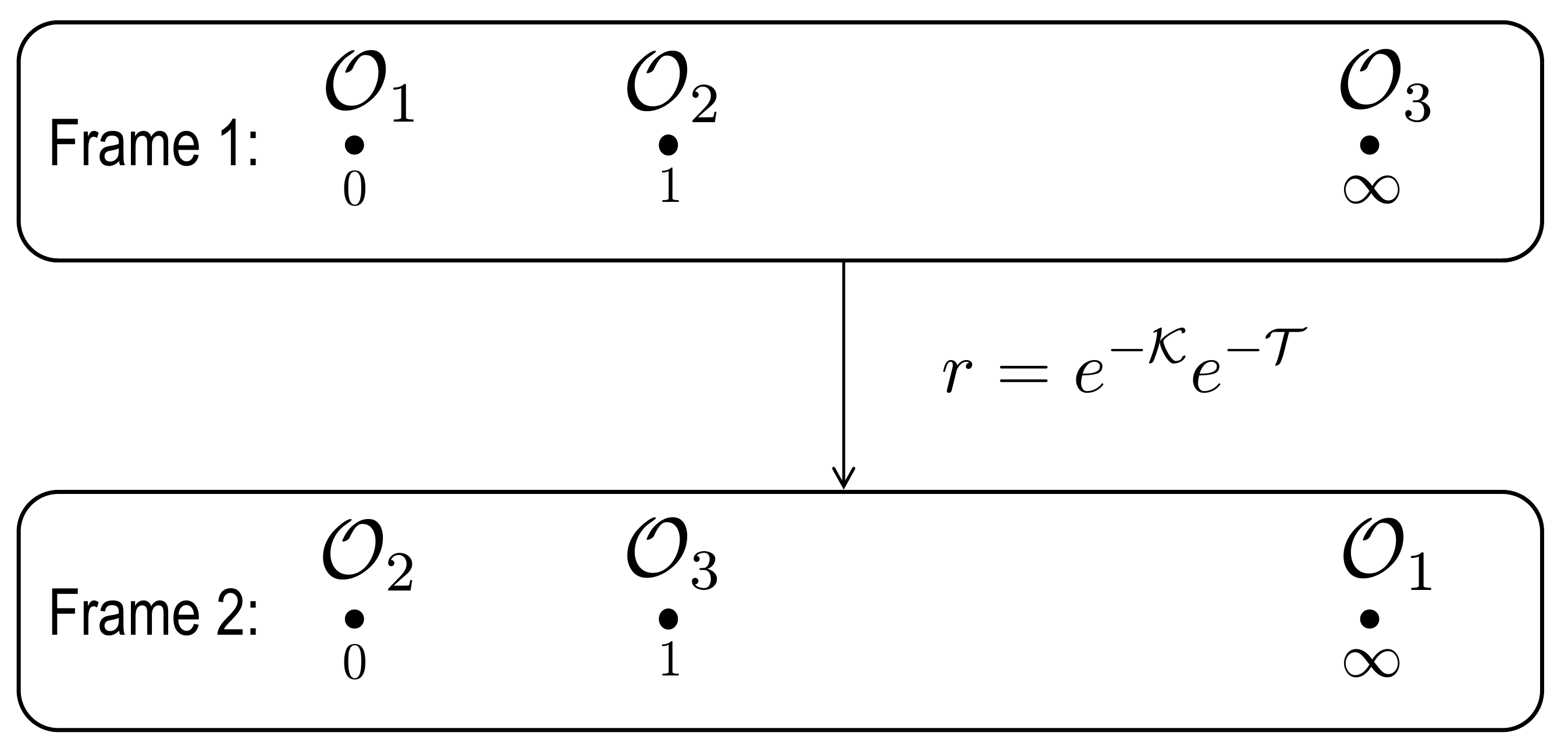}
\end{flushright}
\end{minipage}\\
\begin{minipage}{0.45\hsize}
\begin{center}
\hspace{-65pt}$(a)$
\end{center}
\end{minipage}
\begin{minipage}{0.45\hsize}
\begin{center}
\hspace{40pt}$(b)$
\end{center}
\end{minipage}
\end{center}
\vspace{-0.5cm}
\caption{Gluing multiple edges. We consider two mirror states depicted in $(a)$. The state $\psi_1$ is defined in the frame 1, whereas the state $\psi_2$ is defined in the frame 2. It turns out that the effect of the change of frames is already incorporated into the hexagon formalism as the crossing rule.\label{fig:multi}}
\end{figure}

By contrast, to glue more than one edge, we need to work in different frames at the same time. For instance, if we want to glue two successive channels shown in figure \ref{fig:multi}, we consider the following expansion,
\beq
\sum_{\psi_1,\psi_2}\cdots |\psi_1\rangle \langle\psi_1|g_1 |\psi_1\rangle \langle\psi_1 |\mathcal{H}|\psi_2 \rangle \langle \psi_2 | g_2 |\psi_2\rangle\langle \psi_2 |\cdots\period
\eeq   
In this expression, $|\psi_1\rangle$ is defined in a frame where $\mathcal{O}_1$ is at the origin, $\mathcal{O}_2$ is at $(0,1,0,0)$ and $\mathcal{O}_3$ is at $\infty$, whereas $|\psi_2\rangle$ is in a frame where $\mathcal{O}_2$ is at the origin, $\mathcal{O}_3$ is at $(0,1,0,0)$ and $\mathcal{O}_1$ is at $\infty$. These two frames are related by a nontrivial conformal transformation
\beq
r=e^{-\mathcal{K}}e^{-\mathcal{T}} \comma
\eeq
where $\mathcal{T}$ is the twisted translation \cite{BKV} while $\mathcal{K}$ is the {\it twisted special conformal transformation}:
\beq
\mathcal{T} \equiv i \epsilon_{\alpha\dot{\alpha}}P^{\dot{\alpha}\alpha}+
\epsilon_{\dot{a}a}R^{a\dot{a}}\comma \qquad \mathcal{K} \equiv -i 
\epsilon^{\dot{\alpha}\alpha}K_{\alpha\dot{\alpha}}-\epsilon^{a\dot{a}}R_{\dot{a}a}\period
\eeq
It may seem that such a change of frames substantially complicates the computation of $\langle\psi_1 |\mathcal{H}|\psi_2 \rangle$, which is coupled to both of the states. However, as it turns out, the effect of this transformation is already implemented in the hexagon form factor: 
As shown in Appendix \ref{ap:crossing}, the transformation $r$ essentially swaps two $\mathfrak{psu}(2|2)$'s of the $\mathfrak{psu}(2|2)^2$ symmetry. This turns out to be equivalent to the crossing rule conjectured in \cite{BKV}, which claims that a magnon should swap two indices when we perform a crossing transformation inside a hexagon. This actually provides a physical explanation of the crossing rule in \cite{BKV}. See Appendix \ref{ap:crossing} for details. 

Thus, to summarize, we expect that a change of frame is negligible as long as we use the correct crossing rules. The only thing that matters is that, for $n(\geq 5)$-point functions, we need to include rotations other than ${\sf L}$ in the definition of the weight factor since general $n(\geq 5)$ points cannot be put on a single plane. It will be discussed more in detail in a future publication \cite{FK}.
\subsection{Generalization to Physical Magnons\label{subsec:nonBPS}}
\begin{figure}[t]
\begin{center}
\includegraphics[clip,height=4cm]{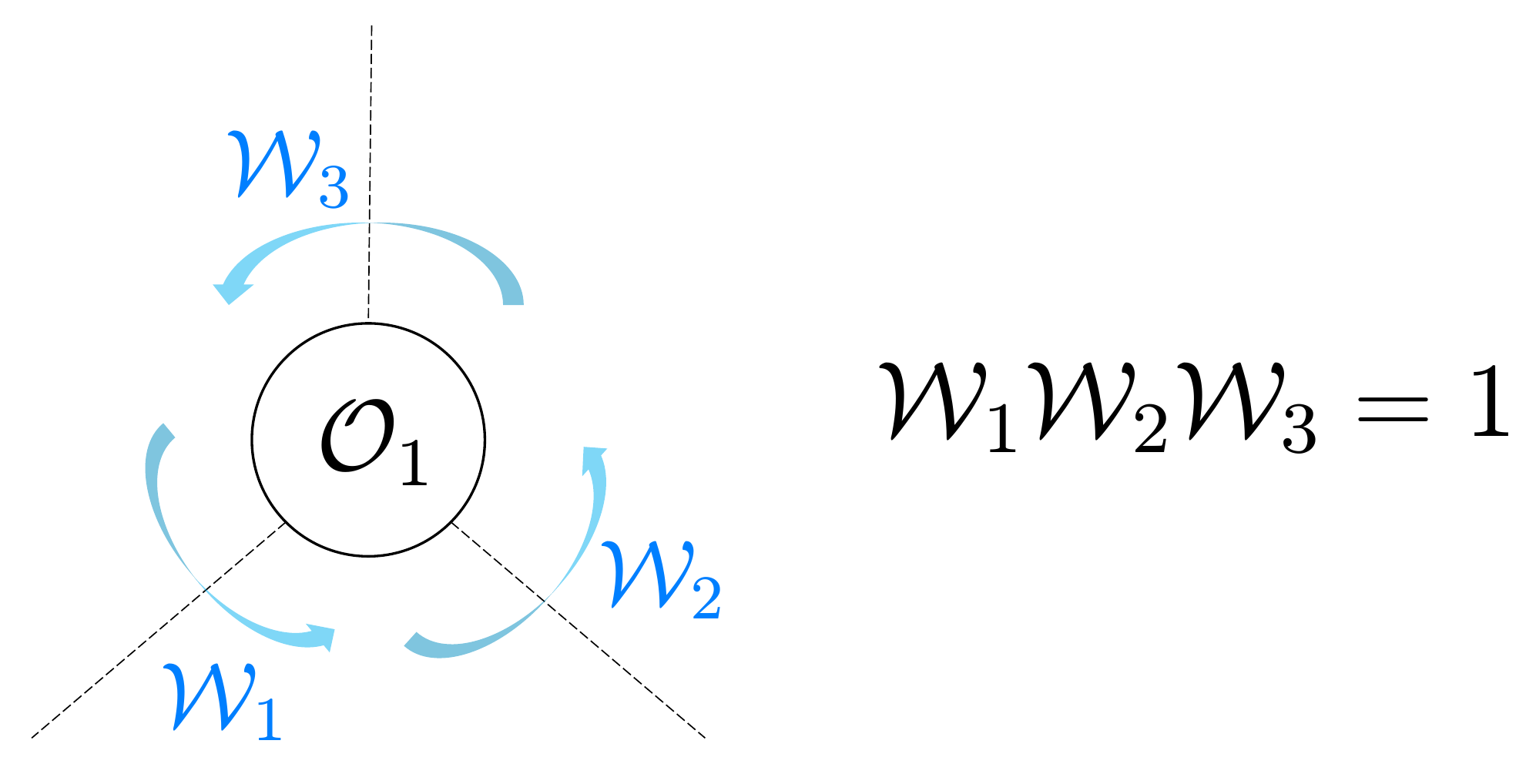}
\end{center}
\vspace{-0.5cm}
\caption{A product of all the weight factors for edges ending at a single operator is always unity. A similar property is observed also in the analysis at strong coupling \cite{chi-system}.
\label{fig:product}}
\end{figure}

We now discuss generalization to the correlators with non-BPS operators. In such cases, in addition to the mirror-particle integrations, we need to perform a sum over partitions of physical magnons. 
We conjecture that the weight factor needed for the physical magnon is related to the weight factor for the mirror particles \eqref{chemical potential} by the mirror transformation. More precisely, the rule is to multiply an extra factor
\beq
\mathcal{W}_{\chi}= e^{-2 E_{\chi}\log |z|} e^{J_{\chi} \varphi}e^{i{\sf L}_{\chi}\phi}e^{i{\sf R}_{\chi}\theta}\comma
 \eeq 
 when we move a magnon $\chi$ in $\mathcal{O}_1$ from $\mathcal{H}_1$ to $\mathcal{H}_2$ in figure \ref{fig:finitehex}. This has a nice property that the product of all the weight factors for edges ending at a single operator is always unity (see figure \ref{fig:product}). Combined with the 
Bethe equations, this property guarantees that the final result is independent of the directions in which we move magnons.
 
 As have been observed in section \ref{sec:simple}, there is also an extra overall space-time (and R-symmetry) dependence when there are physical magnons. This factor depends only on the data of the first hexagon, the hexagon for which we do not multiply a weight factor. As we explain below, it comes from the transformation which relates the first hexagon to the canonical hexagon. For simplicity, we focus on the space-time dependence, but the argument easily carries over to the R-symmetry part. 
 
 Let us first bring the operator $\mathcal{O}_1$, which contains 
 magnons, to $x^{\mu}=0$ 
 by performing a translation. This is of course 
 harmless 
 since the correlation function is invariant under
 translations.
 We then consider a transformation $g^{\prime}$ which keeps $\mathcal{O}_1$ at the origin and maps $\mathcal{O}_3$ and $\mathcal{O}_4$ 
 from the canonical configuration to the configuration we want. In general, $g^{\prime} \in {\rm SO}(2,4)$ can always be expressed as
 \beq
 g^{\prime}=g_{+}g_{0}g_{-} \, , 
 \eeq
 where $g_0$ is generated by the dilatation $D$ and the 
 rotations $L^{\mu}{}_{\nu}$ while $g_{+}$ and $g_{-}$ are generated 
 by the ``upper-triangular'' generators $P_{\mu}$ and 
 the ``lower-triangular'' generators $K_{\mu}$ respectively. Since 
 the upper-triangular generators change the position of $\mathcal{O}_1$, 
 they should not be contained in the transformation $g^{\prime}$ we are 
 studying. In addition, in almost all the cases of interest, the 
 operator $\mathcal{O}_1$ is a conformal primary and 
 the lower-triangular generators act trivially on $\mathcal{O}_1$. 
 Therefore, the only nontrivial effect 
in $\mathcal{O}_1$ 
is brought about by $g_0$. The action of $g_0$ on magnons can be read off easily since $g_0$ belongs to the magnon-symmetry group $\mathfrak{psu}(2|2)^2$. This is the origin of the extra space-time dependence multiplying the sum over partitions.
  
 In the case studied in section \ref{sec:simple}, we find that $g^{\prime}$ is given by
 \beq\label{gprimeg0}
 g^{\prime} = \underbrace{\left(\frac{x_{34}^{+}}{x_{13}^{+}x_{14}^{+}}
 \right)^{(D-{\sf L})/2}
 \left(\frac{x_{34}^{-}}{x_{13}^{-}x_{14}^{-}}
 \right)^{(D+{\sf L})/2}}_{ =\,g_{0}}
 \underbrace{\exp \left[\frac{x_{13}^{+}}{x_{34}^{+}}K_z +\frac{x_{13}^{-}}{x_{34}^{-}}K_{\bar{z}}\right]}_{=\,g_{-}}\comma
 \eeq
  where ${\sf L}$ is the rotation on the $x^2$-$x^3$ plane and $K_z$ ($K_{\bar{z}}$) is the (anti-)holomorphic special conformal transformation on that plane. Applying the aforementioned analysis to this case, we obtain\footnote{Note that the holomorphic derivative has a charge $-1$ under the rotation ${\sf L}$.}
\beq
 f\left(1-\mathcal{W}^{(14)}e^{ip \ell_{14}}+\mathcal{W}^{(14)}\mathcal{W}^{(12)}e^{i p (\ell_{14}+\ell_{12})}\right)\prod_{(ij)}d_{ij}^{\ell_{ij}}\comma
\eeq
where $\mathcal{W}^{(14)}$, $\mathcal{W}^{(12)}$ and $f$ are given by
 \beq
 \begin{aligned}
 \mathcal{W}^{(14)}&=\left(\frac{x_{13}^{+}x_{24}^{+}}{x_{12}^{+}x_{34}^{+}}\right)^{1+\frac{\gamma}{2}}\left(\frac{x_{13}^{-}x_{24}^{-}}{x_{12}^{-}x_{34}^{-}}\right)^{\frac{\gamma}{2}}\comma \quad  \mathcal{W}^{(12)}=\left(\frac{x_{14}^{+}x_{23}^{+}}{x_{13}^{+}x_{24}^{+}}\right)^{1+\frac{\gamma}{2}}\left(\frac{x_{14}^{-}x_{23}^{-}}{x_{13}^{-}x_{24}^{-}}\right)^{\frac{\gamma}{2}}
\comma \\ f&= \left(\frac{x_{34}^{+}}{x_{13}^{+}x_{14}^{+}}\right)^{1+\frac{\gamma}{2}}\left(\frac{x_{34}^{-}}{x_{13}^{-}x_{14}^{-}}\right)^{\frac{\gamma}{2}}\comma
 \end{aligned}
 \eeq
 with $\gamma$ being the anomalous dimension. The prefactor $f$ can be read off by acting the $g_0$ part in \eqref{gprimeg0} on the magnon.
 At tree level, the factors in front of $\prod d_{ij}^{\ell_{ij}}$ coincide with \eqref{treesum2}.
\subsection{A Remark on the Summation over Graphs\label{subsec:graph}}
\begin{figure}[t]
\begin{center}
\begin{minipage}{0.45\hsize}
\begin{center}
\includegraphics[clip,height=4cm]{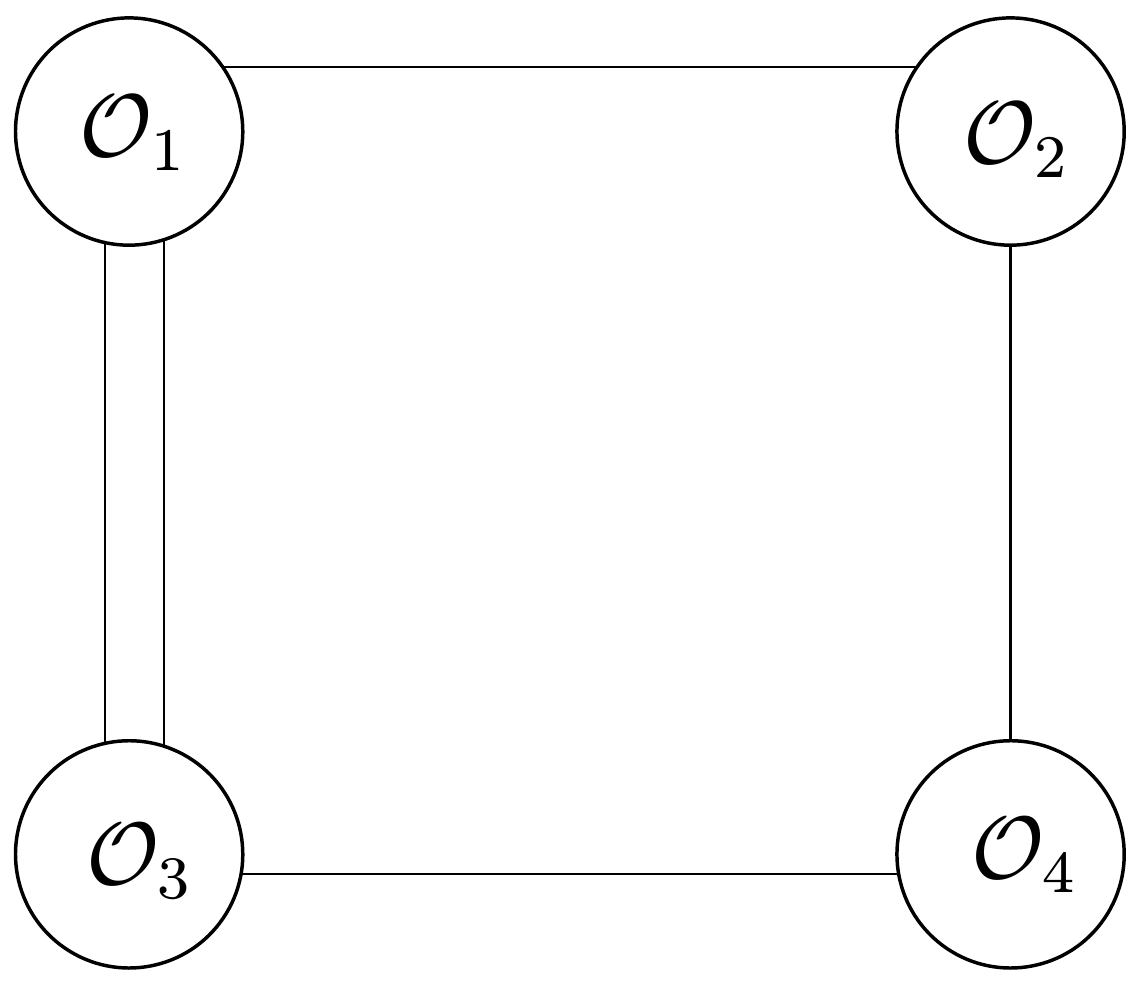}
\end{center}
\end{minipage}
\begin{minipage}{0.45\hsize}
\begin{center}
\includegraphics[clip,height=4cm]{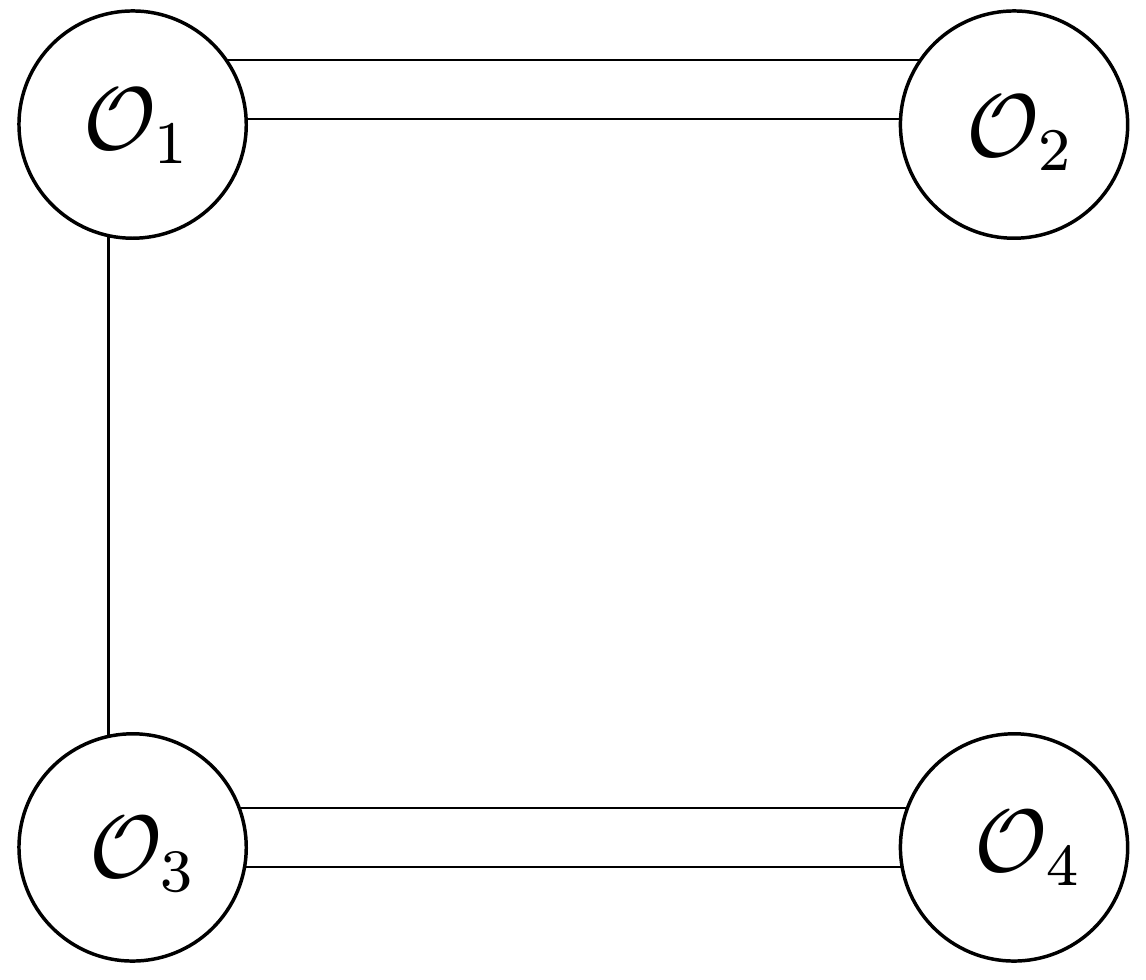}
\end{center}
\end{minipage}\\
\begin{minipage}{0.45\hsize}
\begin{center}
$(a)$
\end{center}
\end{minipage}
\begin{minipage}{0.45\hsize}
\begin{center}
$(b)$
\end{center}
\end{minipage}
\end{center}
\vspace{-0.5cm}
\caption{Examples of a 1EI graph $(a)$ and a non-1EI graph $(b)$ for $G_{2233}$.\label{fig:EI}}
\end{figure}

Let us finally make an important remark about the summation over graphs. For this purpose, it is convenient to introduce a notion of {\it 1-edge irreducible graph} (1EI graph). The 1EI graphs are subsets of connected graphs which are still connected even after we remove all the propagators connecting a pair of points. Examples of 1EI and non-1EI graphs are given in figure \ref{fig:EI}.

When performing a summation over graphs, we should in principle sum over all the connected graphs. However, we found, in all the examples checked so far at one loop, that the correct perturbative result can be reproduced by the following prescription:
\begin{enumerate}
\item For the asymptotic part, sum over all the connected graphs.
\item For the finite-size corrections, sum only over 1EI graphs.
\end{enumerate}
At least in a naive estimate, non-1EI graphs can receive multi-particle mirror corrections already at one loop whereas 1EI graphs only receive one-particle correction at this order. Thus, practically, the restriction to 1EI graphs simplifies our task a lot. We however have not fully understood the origin of such a restriction. It is likely that different mirror-particle contributions cancel out in non-1EI graphs. Such a mechanism, if exists, would be responsible also for the non-renormalization properties of extremal and near-extremal correlators \cite{DFMMR,EHSSW,EHSW} since all the relevant graphs for those correlators are non-1EI. This suggests that our finding may be understood as a consequence of some ``partial'' non-renormalization theorem. Another evidence for our prescription comes from the perturbative analysis in \cite{Topology}, in which they studied several BPS four-point functions at two loops and showed  that the contributions from non-1EI graphs vanish. 

In any case, it would be important to understand the origin of our empirical rule and see if it holds also for more general cases. We postpone the analysis on these points to a future publication \cite{FK}.
  
\section{Four BPS Operators\label{sec:4BPS}}
Here we test our proposal \eqref{prop} against one-loop perturbative data for four-point functions of BPS operators.
\subsection{Flavor-dependent Weight as $\mathfrak{psu}(2|2)$ Character\label{subsec:character}}
At weak coupling, the one-particle measure and the mirror energy scale as
\beq
\mu \sim O(g^2) \comma \quad e^{-\tilde{E}} \sim O(g^2) \comma
\eeq
where $g$ is related to the `t Hooft coupling constant as 
\beq
g^2 =\frac{\lambda}{16\pi^2} \, .
\eeq
Thus the correction at one loop comes only from one-particle states living on an edge with length $0$. To compute such contributions, we just need to evaluate the flavor-dependent part of the weight $\mathcal{W}$, 
\beq
\mathcal{W}_{\rm flavor}=e^{J_{\psi} \varphi}e^{i{\sf L}_{\psi}\phi}e^{i{\sf R}_{\psi}\theta}
\eeq
since all the other factors are known already. Below we first focus on the dependence on ${\sf L}_{\psi}$ and ${\sf R}_{\psi}$ since the determination of $e^{J_{\psi}\varphi}$ is more subtle.

\begin{figure}[t]
\begin{center}
\includegraphics[clip,height=4.5cm]{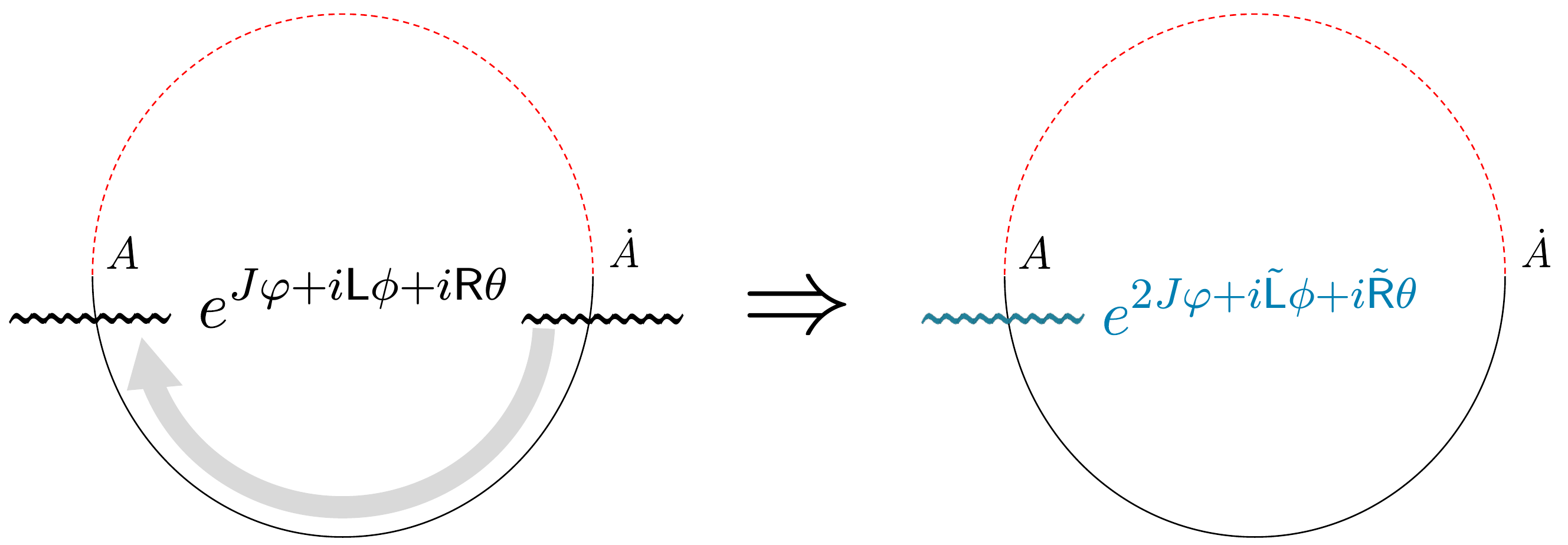}
\end{center}
\vspace{-0.5cm}
\caption{Matrix part and weight factor for the one-particle mirror state. Wavy lines denote the actions of the transformation $e^{J\varphi+i{\sf L}\phi+i{\sf R}\theta}$. It acts both on the left and the right $\mathfrak{psu}(2|2)$ indices. The summation over the flavor indices adds a dashed red curve and makes it into a trace of a single $\mathfrak{psu}(2|2)$. After doing so, we can combine the actions of the transformation into one and rewrite it as $e^{2 J\varphi+i \tilde{\sf L}\phi+i\tilde{\sf R}\theta}$.  
\label{fig:chemical}}
\end{figure}

Let us first consider a fundamental magnon. A fundamental magnon belongs to a bifundamental representation of $\mathfrak{psu}(2|2)^2$, and the charges of the left and the right parts are given in table \ref{tab1}.
\begin{table}[t]
    \begin{minipage}{0.45\hsize}
    \begin{center}
    \begin{tabular}{r||c|c}
    &${\sf L}$&${\sf R}$\\
    $\psi^{1}$&$+1/2$&$0$\\
    $\psi^{2}$&$-1/2$&$0$\\
    $\phi^{1}$&$0$&$+1/2$\\
    $\phi^{2}$&$0$&$-1/2$
     \end{tabular}
    \end{center}
    \end{minipage}
    \begin{minipage}{0.45\hsize}
    \begin{center}
    \begin{tabular}{r||c|c}
    &${\sf L}$&${\sf R}$\\
    $\psi^{\dot{1}}$&$-1/2$&$0$\\
    $\psi^{\dot{2}}$&$+1/2$&$0$\\
    $\phi^{\dot{1}}$&$0$&$-1/2$\\
    $\phi^{\dot{2}}$&$0$&$+1/2$
    \end{tabular}
    \end{center}
    \end{minipage}
    \caption{The charges of a fundamental magnon under the rotations ${\sf L}$ and ${\sf R}$.\label{tab1}}
\end{table}
By straightforward computation, one can confirm that the multiplication of $e^{i{\sf L}_{\psi}\phi}e^{i{\sf R}_{\psi}\theta}$ amounts to modifying the matrix part \cite{BKV} of the mirror particle integrand as
\beq\label{modifymatrix}
\Tr \left[(-1)^{F}\right] \to \Tr \left[(-1)^{F} e^{i \phi \tilde{{\sf L}}+i \theta \tilde{{\sf R}}}\right] \comma
\eeq
where $F$ is a fermion number and $ \tilde{{\sf L}}$ and $ \tilde{{\sf R}}$ are given by
\beq
\tilde{{\sf L}}=L^{1}{}_{1}-L^{2}{}_{2}\comma \quad \tilde{{\sf R}}=R^{1}{}_{1}-R^{2}{}_{2}\period
 \eeq 
This can be evaluated explicitly as
 \beq
 \Tr \left[(-1)^{F}e^{i \phi \tilde{{\sf L}}+i \theta \tilde{{\sf R}}}\right]=-2(\cos \phi- \cos \theta)\period
 \eeq
 It can also be understood graphically as shown in figure \ref{fig:chemical}. In the presence of physical magnons, it will be replaced by a twisted transfer matrix (see section \ref{sec:finite}).

We now perform the same analysis also for the bound states. It is not hard to verify that $e^{i{\sf L}_{\psi}\phi}e^{i{\sf R}_{\psi}\theta}$ for the bound states leads to
\beq
\Tr_{a} \left[(-1)^{F}\right] \to \Tr_{a} \left[(-1)^{F} e^{i \phi \tilde{{\sf L}}+i \theta \tilde{{\sf R}}}\right]\comma
\eeq
where now the trace is taken over the $a$-th anti-symmetric representation. This is nothing but the character of $\mathfrak{psu}(2|2)$ and therefore can be evaluated using the known formula\footnote{See for instance \cite{CMS,Drukker}, where the same character appears in a different context.}. Here however, we take a more pedestrian approach and evaluate the trace using the explicit basis. The basis for the $a$-th anti-symmetric representation is given by
\beq
\begin{aligned}
 &|\psi_{\alpha_1}\cdots \psi_{\alpha_{a}}\rangle +\cdots\comma \quad&&|\phi_{1}\psi_{\alpha_1}\cdots \psi_{\alpha_{a-1}}\rangle+\cdots\comma \\
 &|\phi_{2}\psi_{\alpha_1}\cdots \psi_{\alpha_{a-1}}\rangle+\cdots\comma \quad &&|\phi_{1}\phi_{2}\psi_{\alpha_1}\cdots \psi_{\alpha_{a-2}}\rangle+\cdots\comma
  \end{aligned}
\eeq
with $\alpha_i=1,2$. Computing the trace using this basis, we arrive at
\begin{align}
   \Tr_{a} \left[(-1)^Fe^{i\phi \tilde{\sf L} +i\theta \tilde{\sf R}}\right]&=(-1)^a \left(e^{i a\phi}\sum_{n=0}^{a}e^{-2 in\phi}-2 \cos \theta e^{i (a-1)\phi}\sum_{n=0}^{a-1}e^{-2 in\phi}+e^{i (a-2)\phi}\sum_{n=0}^{a-2}e^{-2 in\phi}\right)\nn\\
  &=2(-1)^a\left( \cos \phi -\cos \theta\right) \frac{\sin a \phi}{\sin \phi}\period
\end{align}
A gratifying feature of this expression is that it vanishes when $\phi=\theta$, as expected from supersymmetry \cite{DP2}.

Let us now turn to the remaining factor $e^{J_{\psi}\varphi}$. In the so-called string frame, we usually assume that the excitations do not carry any $J$-charge since the $J$-charge corresponds to the length of the string, which is fixed once and for all by taking the light-cone gauge. However, it turns out that setting $J_{\psi}=0$ in \eqref{chemical potential} does not lead to a reasonable answer. This is mainly due to supersymmetry: Suppose that we have a state $|\psi_{\alpha}\rangle$ and construct other states in the same multiplet using the supersymmetry transformations. Since the supercharges have $\pm 1/2$ $J$-charges, the states we obtain will have nonzero $J$-charges even if the original state does not. In terms of $Z$-markers introduced by Beisert, it can be expressed also as\footnote{Throughout this paper, we use a ``hybrid'' of the conventional spin-chain frame and the string frame: Although we use $Z$ markers to keep track of non-local effects, the excitations are redefined as in Appendix F of \cite{BKV} so that the S-matrix matches the one for the string frame. This is why the transformations \eqref{transf} are slightly different from the ones given in \cite{Beisert1,Beisert2}.}
\beq\label{transf}
|\psi\rangle \overset{Q,S}{\to} |Z^{\pm 1/2}\phi\rangle \overset{Q,S}{\to} \cdots\period
\eeq
This suggests that the states in the same multiplet can have different $J$-charges. It is however difficult to know what the charges should be since one can repeat acting the supercharges and dress the state with an arbitrary number of $Z$-markers. The one thing we can say for sure is that we should not add too many $Z$ markers: The factor associated with a $Z$-marker, $e^{\varphi}=|\alpha/z|$, can appear as a ratio between different tree-level Wick contractions. It thus implies that dressing with a large number of $Z$-markers will mix the contributions from different graphs and mess up the summation over graphs.

Guided by these considerations, we were led to a ``minimal'' modification of the weight factor, given as follows:
\beq
\begin{aligned}\label{modifiedweight}
   \Tr_{a} \left[(-1)^Fe^{2\varphi J+i\phi \tilde{\sf L} +i\theta \tilde{\sf R}}\right]=&(-1)^a \left(e^{i a\phi}\sum_{n=0}^{a}e^{-2 in\phi}-2 \red{\cosh \varphi} \cos\theta  e^{i (a-1)\phi}\sum_{n=0}^{a-1}e^{-2 in\phi}\right.\\
   &\left.+e^{i (a-2)\phi}\sum_{n=0}^{a-2}e^{-2 in\phi}\right)\\
  =&2(-1)^a\left( \cos \phi -\red{\cosh\varphi} \cos \theta\right) \frac{\sin a \phi}{\sin \phi}\period
\end{aligned}
\eeq
The factor of $2$ in the exponent $e^{2\varphi J}$ comes about when rewriting the weight factor as a trace in a single $\mathfrak{psu}(2|2)$ (see figure \ref{fig:chemical} for an explanation). 
The modification \eqref{modifiedweight} amounts to dressing the states as
\beq
\begin{aligned}\label{modifiedstates}
&|\psi_{\alpha_1}\cdots \psi_{\alpha_{a}}\rangle +\cdots\comma \quad&&|Z^{\pm 1/2}\phi_{1}\psi_{\alpha_1}\cdots \psi_{\alpha_{a-1}}\rangle+\cdots\comma \\
 &|Z^{\pm 1/2}\phi_{2}\psi_{\alpha_1}\cdots \psi_{\alpha_{a-1}}\rangle+\cdots\comma \quad &&|\phi_{1}\phi_{2}\psi_{\alpha_1}\cdots \psi_{\alpha_{a-2}}\rangle+\cdots\comma
\end{aligned}
\eeq
and averaging over the choices of signs. We confirmed a posteriori that this correctly reproduces all the results we checked so far including four-point functions with a Konishi operator (see section \ref{sec:Konishi}). It is however desirable to have a first-principle derivation.

Now, using the weight factor \eqref{modifiedweight}, one can write down a general one-particle integrand\footnote{The factor $(-1)^{a}$ in \eqref{modifiedweight} cancels out with another $(-1)^a$ coming from the hexagon form factor.} for gluing the edge $1$-$4$:
\beq\label{intBPS0}
{\rm int}^{1\text{-}4}_{a}(v)=\frac{2 (\cos \phi -\cosh \varphi \cos \theta)\sin a \phi}{\sin \phi}\mu_{a}(v^{\gamma})e^{-2i \tilde{p}_{a} (v) \log |z|}e^{ -\tilde{E}_{a}(v) \ell}\period
 \eeq 
 By setting $\ell=0$ and going to the weak coupling (see Appendix \ref{ap:weak} for expressions at weak coupling), we get 
 \beq\label{intBPS}
 {\rm int}^{1\text{-}4}_{a}(v)=\frac{2g^2 (\cos \phi -\cosh \varphi \cos \theta)\sin a \phi}{\sin \phi}\frac{a}{(v^2+a^2/4)^{2}}e^{-2i v \log |z|}+O(g^2)\period
 \eeq
 The expressions for other channels can be obtained by replacing the cross ratios with appropriate ones.
 \subsection{Simplest Example: Four ${\bf 20}^{\prime}$}
\begin{figure}[t]
\begin{center}
\begin{minipage}{0.32\hsize}
\begin{center}
\includegraphics[clip,height=4cm]{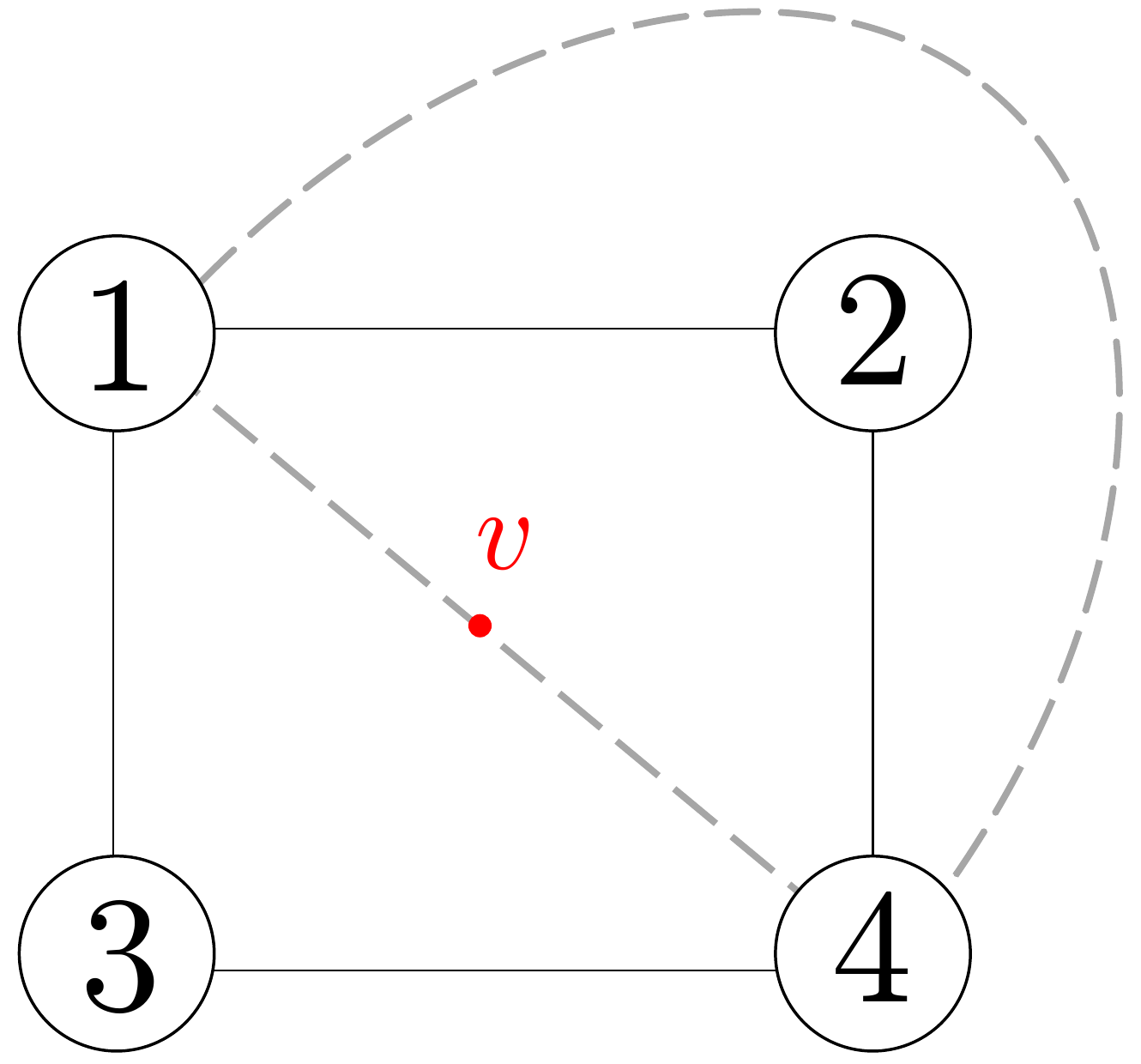}
\end{center}
\end{minipage}
\begin{minipage}{0.32\hsize}
\begin{center}
\includegraphics[clip,height=4cm]{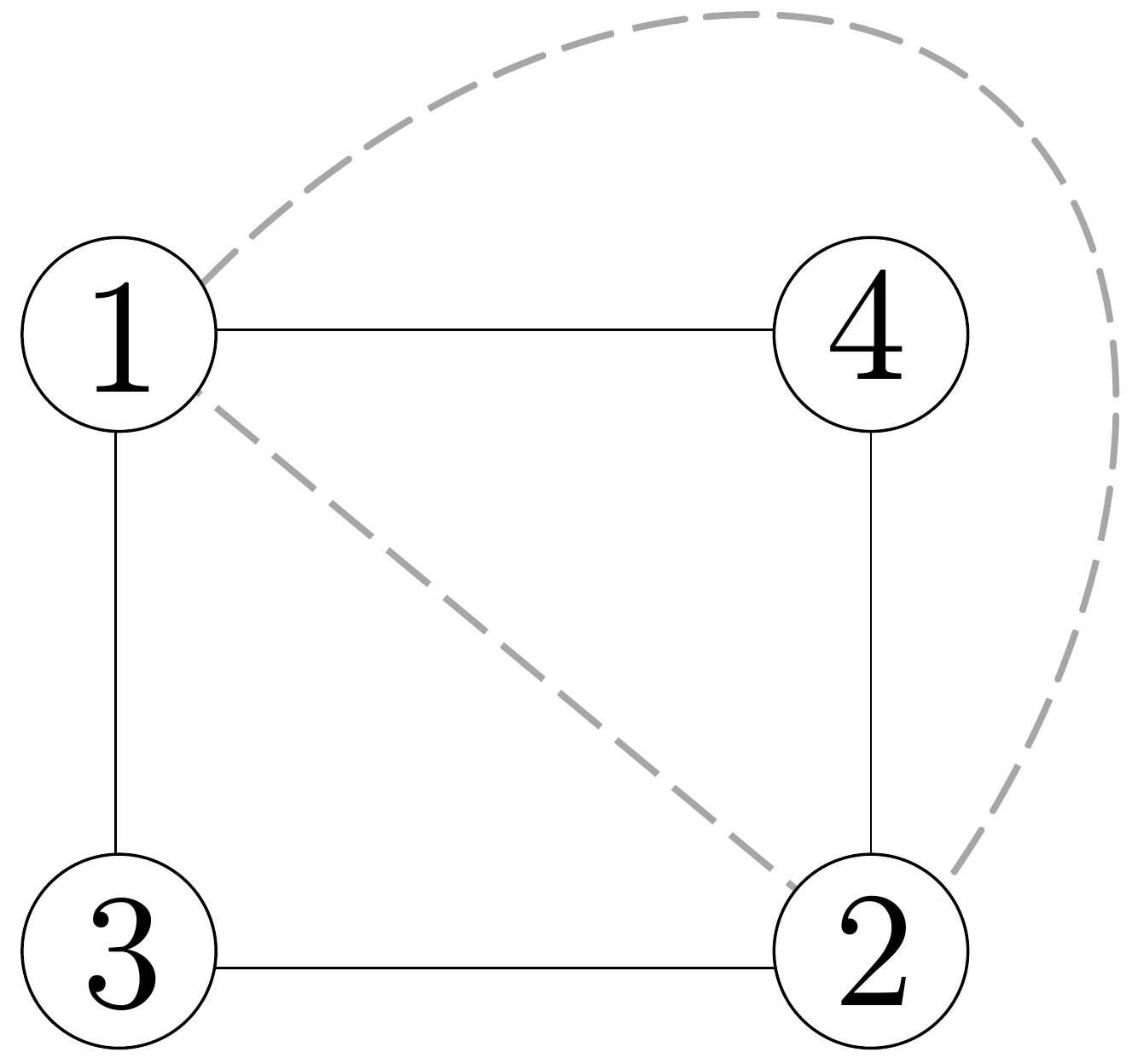}
\end{center}
\end{minipage}
\begin{minipage}{0.32\hsize}
\begin{center}
\includegraphics[clip,height=4cm]{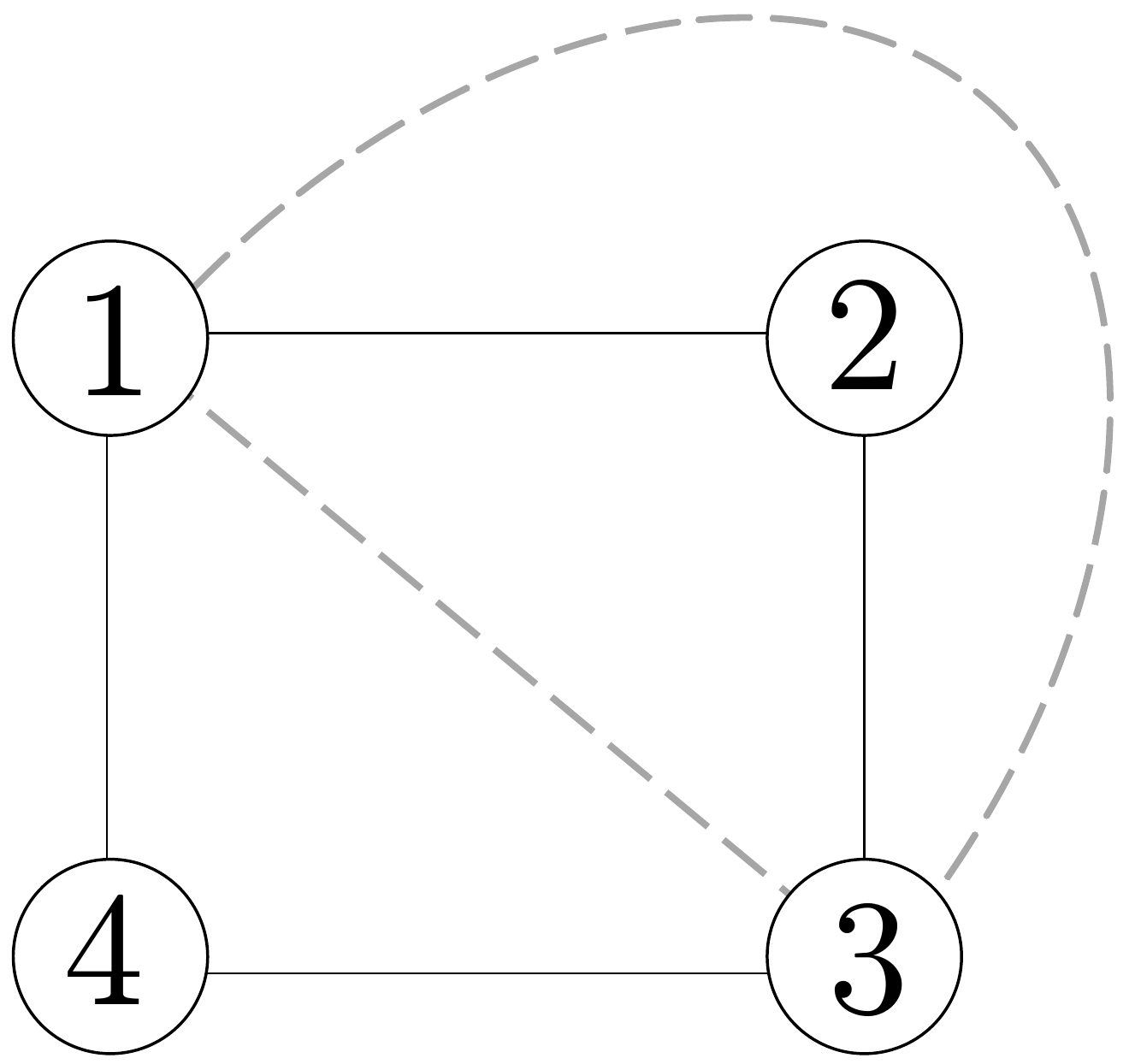}
\end{center}
\end{minipage}\\
\begin{minipage}{0.32\hsize}
\begin{center}
\hspace{-10pt}$(1243)$
\end{center}
\end{minipage}
\begin{minipage}{0.32\hsize}
\begin{center}
\hspace{-10pt}$(1423)$
\end{center}
\end{minipage}
\begin{minipage}{0.32\hsize}
\begin{center}
\hspace{-10pt}$(1234)$
\end{center}
\end{minipage}
\end{center}
\vspace{-0.5cm}
\caption{Three distinct graphs contributing to the four-point function of ${\bf 20}^{\prime}$ operators. Here we just put numbers $i$ to denote the operator $\mathcal{O}_i$. Dashed lines are the zero-length bridges on which we insert a mirror magnon (denoted by a red dot).\label{fig:4bps}}
\end{figure}

 We first compute the simplest four-point function: The four-point function of length 2 BPS operators, also known as ${\bf 20}^{\prime}$ operators. For this correlation function, there are only three distinct planar graphs as depicted in figure \ref{fig:4bps}. To apply the hexagonalization, we split them into four hexagons by adding dashed lines shown in the figure. These lines denote zero-length bridges and the one-loop correction comes from adding a mirror magnon on these lines.

The contribution from each channel can be computed straightforwardly using the integrand \eqref{intBPS}. For instance, two channels (inside and outside the square) in the graph $(1243)$ produce the same contribution and their sum reads  
\beq
\begin{aligned}
2\mathcal{M}_{z,\alpha}&\equiv 2\sum_{a=1}^{\infty}\int \frac{dv}{2\pi} {\rm int}^{1\text{-}4}_{a}(v)\\
&=g^2\left[2(z+\bar{z})-(\alpha^{-1}+\bar{\alpha}^{-1})(z\bar{z}+\alpha\bar{\alpha})\right]F^{(1)}(z,\bar{z})\period
\end{aligned}
\eeq
Here $F^{(1)}$ is the so-called one-loop conformal integral,
\beq
F^{(1)}(z,\bar{z})\equiv \frac{2 {\rm Li}_2 (z)-2 {\rm Li}_2 (\bar{z})+\log z\bar{z} \log \frac{1-z}{1-\bar{z}}}{z-\bar{z}}\quad \left(=\frac{x_{13}^2x_{24}^2}{\pi^2 }\int \frac{d^4 x_5}{x_{15}^2x_{25}^2x_{35}^2 x_{45}^2}\right)\comma
\eeq
which satisfies the following properties:
\beq\label{propF}
\begin{aligned}
F^{(1)}(1-z,1-\bar{z})=F^{(1)}(z,\bar{z})\comma 
\qquad F^{(1)}(1/z,1/\bar{z})=z\bar{z}F^{(1)}(z,\bar{z})\, ,\\
F^{(1)}\big(z/(z-1),\bar{z}/(\bar{z}-1)\big) = (1-z)(1-\bar{z}) 
F^{(1)}(z, \bar{z}) \, .\hspace{30pt} 
\end{aligned}
\eeq
The contribution from other graphs can be obtained by replacing the cross ratios appropriately. Namely, we make the transformation\footnote{Of course we also make the same transformations to other cross ratios $\bar{z}$, $\alpha$ and $\bar{\alpha}$.} $z\to 1-z$ for the graph $(1423)$, and the transformation $z\to z/(z-1)$ for the graph $(1234)$.

To compute the full one-loop four-point function, we dress these mirror contributions by the tree-level correlator for each graph and sum them up. This leads to an expression\footnote{As given in \eqref{structuredisc}, $G_{2222}$ denotes a connected part of the correlation function with trivial combinatorial factors stripped off. In our normalization, the tree-level result reads $G_{2222}^{(0)}=d_{12}d_{24}d_{34}d_{13}+d_{13}d_{23}d_{24}d_{14}
+d_{12}d_{23}d_{34}d_{14}$.}
\beq\label{4length2}
\begin{aligned}
G_{2222}^{(1)}&=2\left(d_{12}d_{24}d_{34}d_{13} \mathcal{M}_{z,\alpha} +d_{13}d_{23}d_{24}d_{14} \mathcal{M}_{1-z,1-\alpha}+d_{12}d_{23}d_{34}d_{14} \mathcal{M}_{\frac{z}{z-1},\frac{\alpha}{\alpha-1}}\right)\\
&=-2g^2\tilde{R}_{1234} F^{(1)}(z,\bar{z})\comma
\end{aligned}
\eeq
where $\tilde{R}_{1234}$ is a universal prefactor\footnote{It is related to the {\it universal rational prefactor} $R_{1234}$ defined in \cite{CDHS} as $\tilde{R}_{1234}=R_{1234}/x_{13}^2x_{24}^2$.}, given by
\beq
\begin{aligned}
\tilde{R}_{1234} &= \frac{(z-\alpha)(z-\bar{\alpha})(\bar{z}-\alpha)(\bar{z}-\bar{\alpha})}{z\bar{z}(1-z)(1-\bar{z})}d_{13}^2 d_{24}^2 \period
\end{aligned}
\eeq
 The result \eqref{4length2} perfectly matches the one computed from perturbation theory \cite{1loop1,1loop2,DP}. It is worth noting that the factor $\tilde{R}_{1234}$, which is manifestation of supersymmetry, comes about only after the summation over different channels. This suggests that supersymmetry is realized in a rather nontrivial manner in the integrability approach. See also  section \ref{subsec:graph}.
  
\begin{figure}[t]
\begin{center}
\includegraphics[clip,height=4cm]{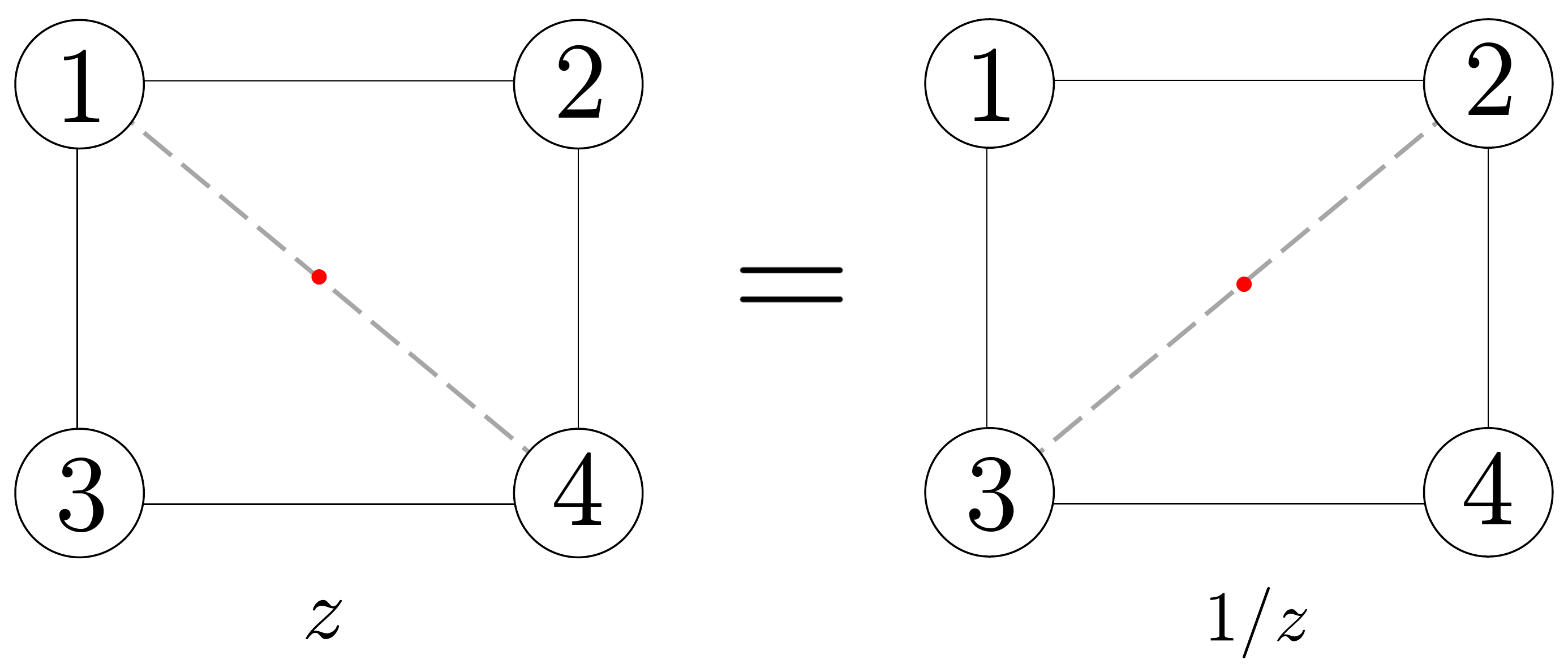}
\end{center}
\vspace{-0.5cm}
\caption{Flip invariance of hexagonalization. Whenever there exists a zero-length bridge, we can cut a correlation function in several different ways. They are related by the {\it flip} transformation, which flips the cross ratio from $z$ to $1/z$. For BPS four-point functions, the flip invariance can be easily verified. 
\label{fig:flipinv}}
\end{figure}

 Before moving to more general correlators, let us make two important comments on the result we got. The first comment is about the {\it flip invariance}: As shown in figure \ref{fig:flipinv}, there are several different ways to cut the four-point function into hexagons. Following the terminology for the Fock coordinates of the Teichmuller space, we refer to the transformation which relates two different cuttings as the {\it flip transformation}. After the flip transformation, the relevant cross ratios change from $z$ to $1/z$. However using \eqref{propF} it is easy to check that the mirror correction is invariant under this change; namely $\mathcal{M}_{z,\alpha}=\mathcal{M}_{z^{-1},\alpha^{-1}}$. This serves as an important consistecy check of our formalism. 

 The second comment is about the relation to the operator product expansion. 
 In the OPE limit $z ,\bar{z}\to 0$, the mirror corrections $\mathcal{M}_{z,\alpha}$ and $\mathcal{M}_{\frac{z}{z-1},\frac{\alpha}{\alpha-1}}$ admit a natural expansion. To see this, we just need to recall that the integrand for the $a$-th bound state (for $\mathcal{M}_{z,\alpha}$) contains a factor
 \beq
 \frac{e^{-iv\log z\bar{z}}}{(v^2+a^2/4)}\comma
 \eeq
 which leads to $|z|^{a}$ upon taking a residue. This shows that, for such mirror corrections, expanding the OPE series corresponds to truncating the sum over the bound states. On the other hand, the remaining contribution $\mathcal{M}_{1-z,1-\alpha}$ does not have a natural expansion in this limit. This is not so problematic as long as we care only about first few terms in the OPE series at weak coupling since the contribution from this graph is non-singular and suppressed in that limit. It would be an interesting future problem to extensively study the connection between the OPE and our approach, especially at finite coupling. 
 
\subsection{General Four BPS Correlators}
We now consider general four-point functions of BPS operators at one loop. A particularly simple expression for such correlators can be found in \cite{CDHS}, which reads in our conventions as follows:
\beq\label{generalfour}
G_{L_1L_2L_3L_4}=-2g^2 \tilde{R}_{1234}F^{(1)}(z,\bar{z})
\sum_{\{b_{ij}\}} \left(\prod_{1\leq i<j\leq4}(d_{ij})^{b_{ij}}\right)\period
\eeq
Here $L_i$ is the length of the $i$-th operator, and the 
nonnegative integers $b_{ij}$ are the set of solutions to the relations
\beq\label{condbij}
b_{ij}=b_{ji}\comma \quad \text{and} \quad \sum_{j\neq i}b_{ij}=L_i-2\period
\eeq
In what follows, we will reproduce the expression \eqref{generalfour} from the integrability side. 

\begin{figure}[t]
\begin{center}
\begin{minipage}{0.32\hsize}
\begin{center}
\includegraphics[clip,height=4cm]{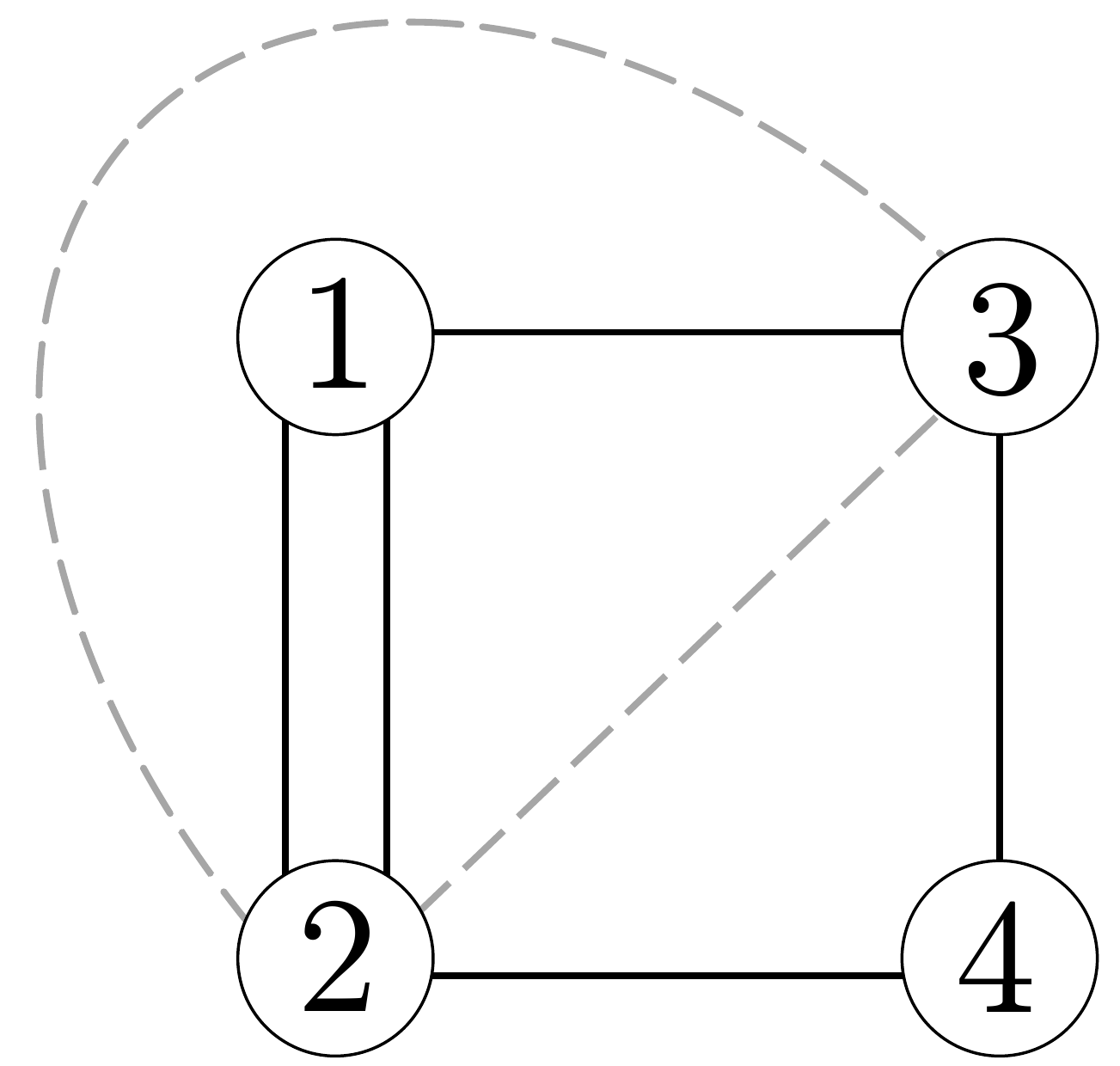}
\end{center}
\end{minipage}
\begin{minipage}{0.32\hsize}
\begin{center}
\includegraphics[clip,height=4cm]{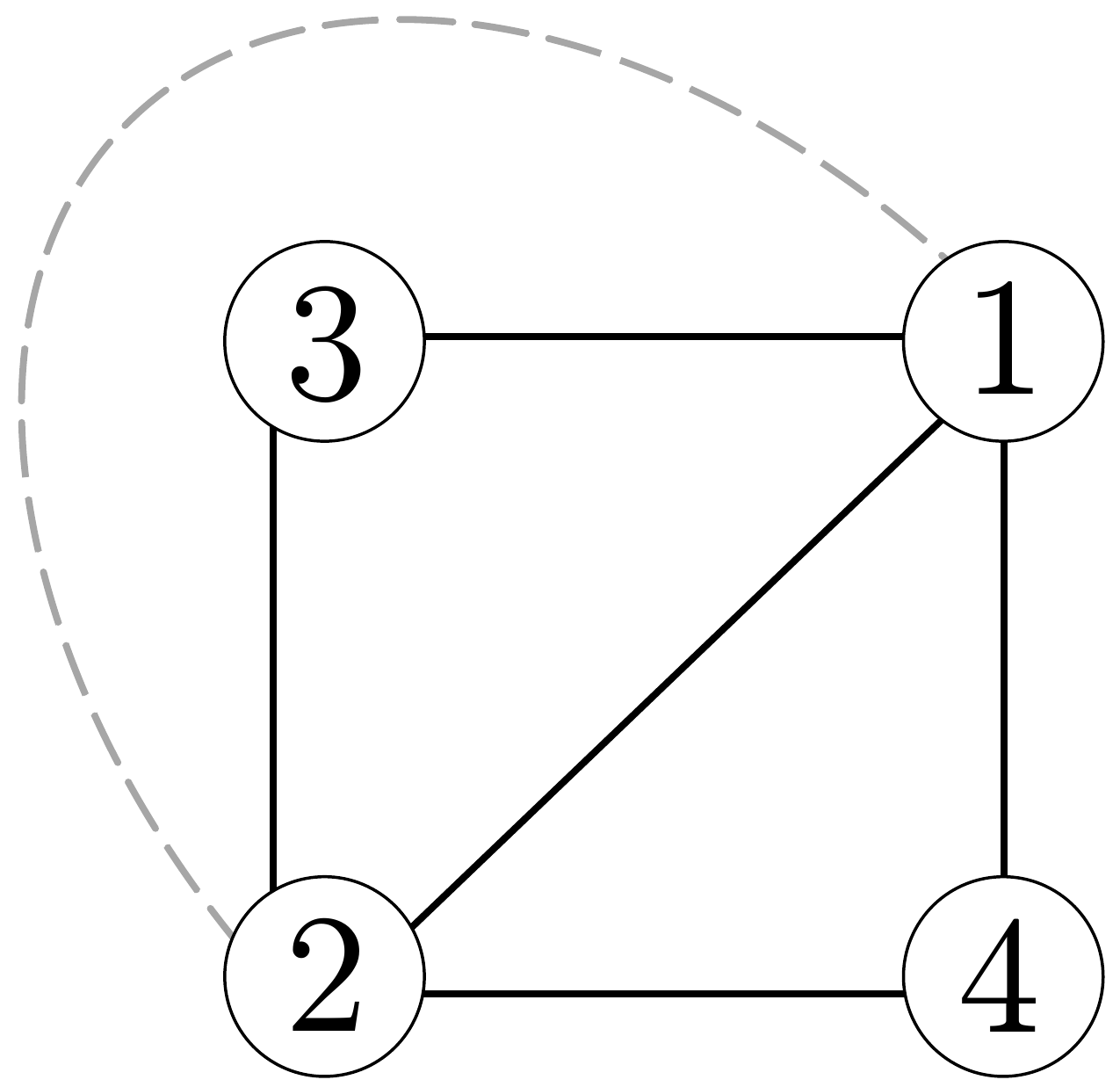}
\end{center}
\end{minipage}
\begin{minipage}{0.32\hsize}
\begin{center}
\includegraphics[clip,height=4cm]{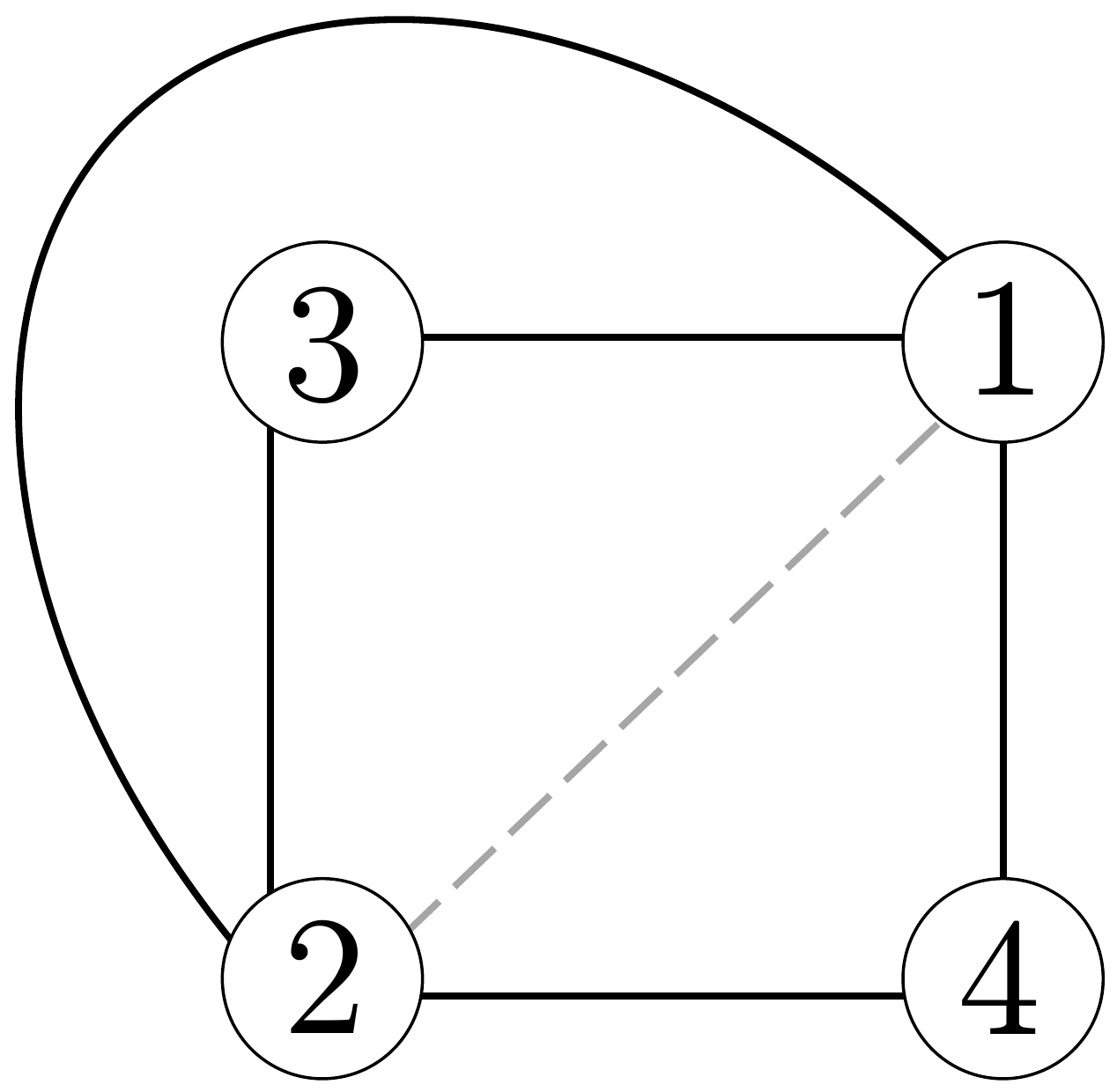}
\end{center}
\end{minipage}\\
\begin{minipage}{0.32\hsize}
\begin{center}
\hspace{20pt}$(a)$
\end{center}
\end{minipage}
\begin{minipage}{0.32\hsize}
\begin{center}
\hspace{20pt}$(b)$
\end{center}
\end{minipage}
\begin{minipage}{0.32\hsize}
\begin{center}
\hspace{20pt}$(c)$
\end{center}
\end{minipage}
\end{center}
\vspace{-0.5cm}
\caption{Explanation for the combinatorial factor in \eqref{t1243}. When there are two zero-length bridges as in $(a)$, a graph has two mirror channels. On the other hand, if there is only one zero-length bridge, we can draw two inequivalent graphs ($(b)$ and $(c)$ in the figure) for a given $b_{ij}$. In both cases, we obtain a factor of $2$ as given in \eqref{t1243}.\label{fig:factor2}}
\end{figure}

As should be clear by now, what we need to do  is to enumerate all planar 1EI graphs with zero-length bridges and dress them by the mirror-particle corrections. To avoid being non-1EI, graphs must contain one of the following three combinations of propagators\footnote{If a graph contains more than one of the three combinations, it clearly has no zero-length bridges.}: 
\beq\label{dcombi}
d_{12}d_{24}d_{34}d_{13} \comma  \qquad 
d_{13}d_{23}d_{24}d_{14} \comma \qquad 
d_{12}d_{23}d_{34}d_{14} \, . 
\eeq
Let us first consider the graphs with $d_{12}d_{24}d_{34}d_{13}$, which 
receive a mirror correction $\mathcal{M}_{z,\alpha}$. Each such a 
graph is characterized by the numbers of the remaining Wick 
contractions, which are nothing but the 
nonnegative integers $b_{ij}$ satisfying 
the condition \eqref{condbij}. However, not all such graphs 
can receive a one-loop correction since, among those, there are 
graphs which do not have zero-length bridges. The ones without 
zero-length bridges are {\it completely connected graphs}, namely the 
graphs in which every pair of points is connected by at least 
one propagator. Completely connected graphs contain a 
propagator factor $d_{12}d_{13}d_{14}d_{23}d_{24}d_{34}$ and are 
characterized by a set of nonnegative integers $c_{ij}$ satisfying
\beq
c_{ij}=c_{ji}\comma \quad {\rm and}\quad \sum_{j\neq i}c_{ij}=L_i-3\period
\eeq
All in all, the contribution from the graphs with $d_{12}d_{24}d_{34}d_{13}$ reads
\beq\label{t1243}
t_{1243}\equiv {\red 2}\mathcal{M}_{z,\alpha}d_{12}d_{24}d_{34}d_{13}\left[\sum_{\{b_{ij}\}}\left(\prod_{1\leq <j\leq 4}(d_{ij})^{b_{ij}}\right)-d_{14}d_{23}\sum_{\{c_{ij}\}}\left(\prod_{1\leq <j\leq 4}(d_{ij})^{c_{ij}}\right)\right]\period
\eeq
The origin of the factor of 2, highlighted in red, is explained in figure \ref{fig:factor2}.

To obtain a full correlator, we should also include graphs which contain the other two combinations in \eqref{dcombi}. The contributions from these graphs can be computed from the previous one by a simple relabelling of the indices. For instance, the contribution from graphs with $d_{13}d_{23}d_{24}d_{14}$ is given by
\beq
t_{1324}\equiv 2\mathcal{M}_{1-z,1-\alpha}d_{13}d_{23}d_{24}d_{14}\left[\sum_{\{b_{ij}\}}\left(\prod_{1\leq <j\leq 4}(d_{ij})^{b_{ij}}\right)-d_{12}d_{34}\sum_{\{c_{ij}\}}\left(\prod_{1\leq <j\leq 4}(d_{ij})^{c_{ij}}\right)\right]\period
 \eeq 
To sum up three contributions, we use the following identity, which can be verified by the straightforward computation:
\beq
\mathcal{M}_{z,\alpha}+\mathcal{M}_{1-z,1-\alpha}+\mathcal{M}_{\frac{z}{z-1},\frac{\alpha}{\alpha-1}}=0\period
\eeq
This identity allows us to get rid of the terms with $\sum_{\{c_{ij}\}}$, and the final result reads
\begin{align}
t_{1243}+t_{1324}+t_{1234}=&2\left(d_{12}d_{24}d_{34}d_{13} \mathcal{M}_{z,\alpha} +d_{13}d_{23}d_{24}d_{14} \mathcal{M}_{1-z,1-\alpha}+d_{12}d_{23}d_{34}d_{14} \mathcal{M}_{\frac{z}{z-1},\frac{\alpha}{\alpha-1}}\right)\nn\\
&\times \sum_{\{b_{ij}\}} \left(\prod_{1\leq i<j\leq4}(d_{ij})^{b_{ij}}\right)\period\label{generalinteg}
\end{align}
Using the equality \eqref{4length2}, we can confirm that \eqref{generalinteg} matches precisely the result \eqref{generalfour}. 

\subsection{Mellin-like Representation\label{subsec:Mellin}}
The integrability result is expressed in terms of mirror states, which have {\it imaginary} eigenvalues of the dilatation operator (see \eqref{relationDJp}). This feature is reminiscent of the {\it Mellin representation} for conformal correlators: In the Mellin representation, the correlation function is expressed as integrals along the imaginary axis of Mellin variables, which can be interpreted as analytically-continued conformal dimensions. Here, with the hope of shedding light on this analogy, we will rewrite the result from integrability into an integral which is akin to (but different from) the Mellin representation.
\begin{figure}[t]
\begin{center}
\includegraphics[clip,height=5cm]{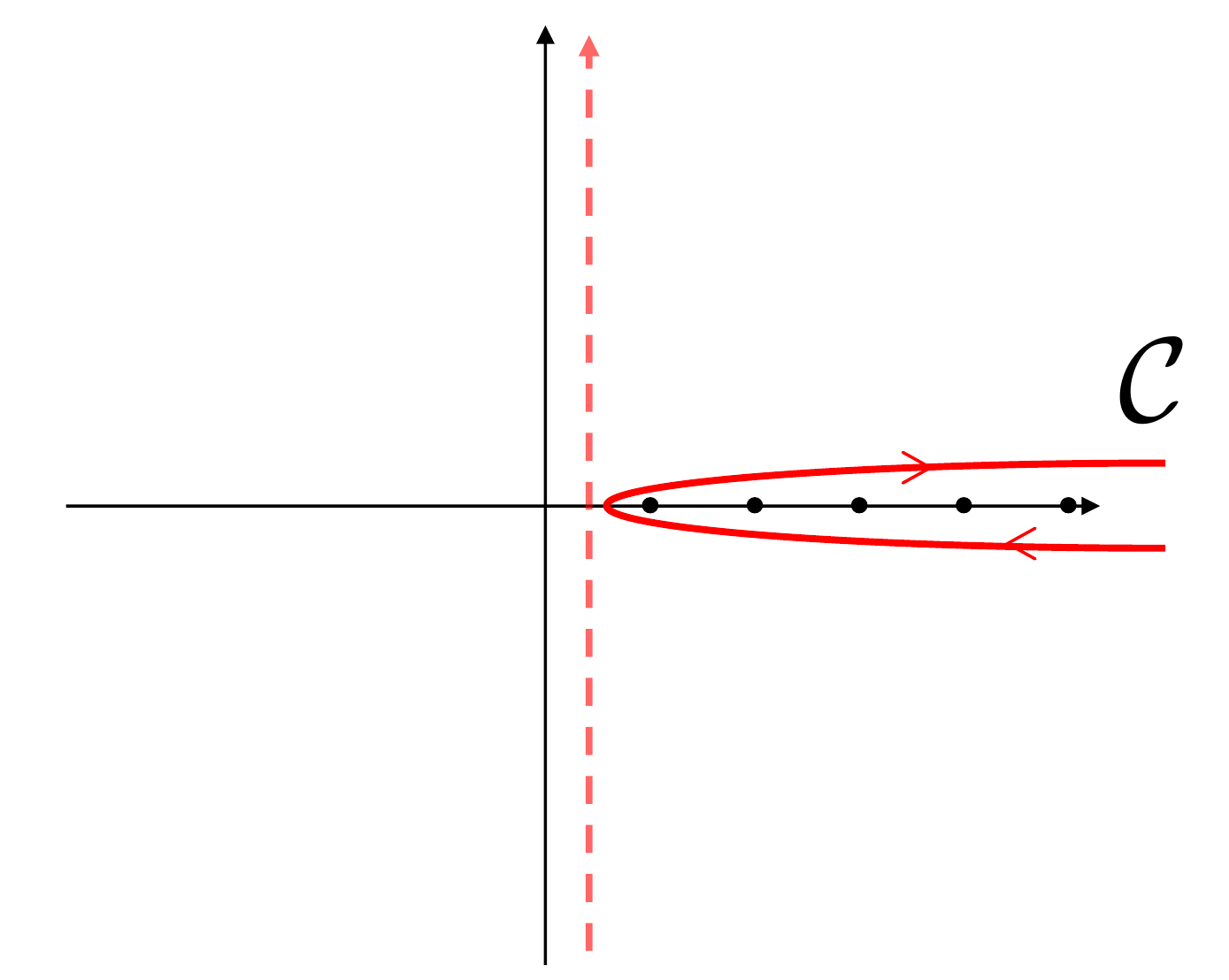}
\end{center}
\vspace{-0.5cm}
\caption{The contour for the Mellin-like representation. The original contour $\mathcal{C}$ encircles all the positive integers. By deforming the contour, one can recast it as a contour along the imaginary axis.   
\label{fig:mellin}}
\end{figure}

Let us consider the one-loop mirror integral,
\beq
\mathcal{M}_{z,\alpha}=
\sum_{a=1}^{\infty}\int \frac{dv}{2\pi} \frac{2g^2 (\cos \phi-\cosh \varphi \cos\theta)\sin a \phi}{\sin\phi} \frac{a}{(v^2+a^2/4)^2}e^{-i v\log z\bar{z}}\period
\eeq
To make a connection with the Mellin representation, we convert the sum over $a$ into an integral. For this purpose, we split the factor $\sin a\phi$ into two parts, $(e^{ia\phi}-e^{-ia\phi})/2i$ and rewrite them as
\beq
\sum_{a}e^{ia\phi}\bullet \,\,\, \to \,\,\, \oint_{\mathcal{C}}du \frac{e^{i\phi u}}{1-e^{2\pi i u}}\bullet\comma \qquad \sum_{a}e^{-ia\phi}\bullet \,\,\, \to \,\,\, \oint_{\mathcal{C}}du \frac{e^{-i\phi u}}{e^{-2\pi i u}-1}\bullet\period
\eeq
where the contour $\mathcal{C}$ is defined in figure \ref{fig:mellin}. Now by deforming the contour, we can express it as an integral along the imaginary axis of $u$. The result can be combined into a single integral,
\beq
\mathcal{M}_{z,\alpha}=\frac{g^2 (\cos \phi-\cosh\varphi \cos\theta)}{\pi i\sin \phi}\int^{i\infty}_{-i\infty} du \int^{\infty}_{-\infty} dv \frac{e^{i\phi u}}{1-e^{2\pi i u}}\frac{u}{(v^2+u^2/4)^2}e^{-iv \log z\bar{z}}\period
\eeq
Changing the integration variable as $u\to 2u$ and $v\to -iv$, we can rewrite it as
\begin{align}
\mathcal{M}_{z,\alpha}&=\frac{4g^2 (\cos \phi-\cosh\varphi \cos\theta)}{\pi \sin \phi}\int^{i\infty}_{-i\infty} du \int^{i\infty}_{-i\infty} dv \frac{e^{2i\phi u}}{e^{4\pi i u}-1}\frac{u}{(u^2-v^2)^2}e^{-v \log z\bar{z}}\\
&\hspace{-10pt}=\frac{g^2 (\cos \phi-\cosh\varphi \cos\theta)}{\pi i\sin \phi}\int^{i\infty}_{-i\infty} du \int^{i\infty}_{-i\infty} dv \frac{z^{u-v}\bar{z}^{-(u+v)}}{(u-v)^2(u+v)^2}\left[\frac{\Gamma[1+2u]\Gamma[1-2u]}{\Gamma[\frac{1}{2}+2u]\Gamma[\frac{1}{2}-2u]}-2 i u\right]\comma\nn
\end{align}
where we used the relation
\beq
\frac{u}{e^{4\pi i u}-1}=\frac{1}{4i}\left[\frac{\Gamma[1+2u]\Gamma[1-2u]}{\Gamma[\frac{1}{2}+2u]\Gamma[\frac{1}{2}-2u]}-2 i u\right]\period
 \eeq 
 Then, by redefining the integration variables from $(u,v)$ to $(s,t)=(u-v,-(u+v))$, we arrive at
 \beq\label{finalMellin}
 \mathcal{M}_{z,\alpha}=c(z,\alpha)\int^{i\infty}_{-i\infty} ds \int^{i\infty}_{-i\infty} dt \frac{z^{s}\bar{z}^{t}}{(st)^2}\left[\frac{\Gamma[1+(s-t)]\Gamma[1-(s-t)]}{\Gamma[\frac{1}{2}+(s-t)]\Gamma[\frac{1}{2}-(s-t)]}- i (s-t)\right]\comma
 \eeq
 with 
 \beq
 c(z,\alpha)=\frac{g^2 (\cos \phi-\cosh\varphi \cos\theta)}{4\pi i\sin \phi}=g^2\frac{2(z+\bar{z})-(\alpha^{-1}+\bar{\alpha}^{-1})(z\bar{z}+\alpha\bar{\alpha})}{8\pi (z-\bar{z})}\period
 \eeq
 
 The representation \eqref{finalMellin} appears similar to the Mellin representation. It is a double integral along the imaginary axis and the OPE series is generated by taking the residues of the integrand. However there are also important differences: Unlike the Mellin representation, the expression \eqref{finalMellin} is given by the Mellin transform of $z$ and $\bar{z}$. This makes it harder to study the crossing symmetry, namely the transformation property under $z\to 1-z$. To some extent, this is already expected  since \eqref{finalMellin} is the result for just one channel, and to obtain a full correlator, one has to sum different channels. It would be interesting to see if there is a natural way to combine contributions from different channels. If so, it may help us to understand the Mellin representation in more physically terms.
\section{One Konishi and Three ${\bf 20}^{\prime}$s\label{sec:Konishi}}
We now test our proposal for physical magnons by studying a correlation function of one Konishi and three ${\bf 20}^{\prime}$ operators. More precisely, we consider a Konishi operator in the SL$(2)$-sector,
\beq
K\propto \Tr \big(D^2 (Y_1\cdot \Phi)\, \,(Y_1\cdot \Phi) \big) - 2\Tr \big( D(Y_1\cdot \Phi) D(Y_1\cdot \Phi)\big)\quad \qquad  D = (\del_2 -i\del_3)/2\comma
\eeq
and put all the operators on the $x^2$-$x^3$ plane. As explained in Appendix \ref{ap:fivept}, this correlation function can be computed by the OPE decomposition of a five-point function studied in \cite{DP}. The connected part of this correlator reads
\beq
\left.\langle K(x_1) \mathcal{O}_{\bf 20^{\prime}}(x_2)\mathcal{O}_{\bf 20^{\prime}}(x_3)\mathcal{O}_{\bf 20^{\prime}}(x_4)\rangle\right|_{\rm connected}=\frac{(\sqrt{2})^4}{N_c^2}G_{K222}(x_i,Y_i)
\eeq
where $(\sqrt{2})^4$ is the usual length-dependent factor (see \eqref{structuredisc}) and $G_{K222}(x_i,Y_i)$ is given by
\beq
G_{K222}(x_i,Y_i)=d_{12}d_{23}d_{34}d_{41}\left(\frac{x_{34}^{+}}{x_{13}^{+}x_{14}^{+}}\right)^{2+\gamma/2}\left(\frac{x_{34}^{-}}{x_{13}^{-}x_{14}^{-}}\right)^{\gamma/2} f(z,\alpha)\period
\eeq
with $\gamma$ being the anomalous dimension of the Konishi operator. At weak coupling, $f(z,\alpha)$ can be expressed in terms of conformal integrals as
\beq\label{dataKonishi}
\begin{aligned}
f(z,\alpha)=&\frac{1}{\sqrt{6}}\left[\left(\frac{1}{z^2}+\frac{(1-z)^3(1-\bar{z})}{z^2 (1-\alpha)(1-\bar{\alpha})}+\frac{z\bar{z}}{\alpha\bar{\alpha}}\right)\right.
\\&\left.\qquad +g^2 \left(c_0 +c_1 F^{(1)} (z,\bar{z})+c_2 \del_z F^{(1)} (z,\bar{z})+c_3\del^2_{z}F^{(1)} (z,\bar{z})\right) +O(g^4)\right]\comma
\end{aligned}
\eeq
where $c_{0}$-$c_{3}$ are given by
{{\small
\beq
\begin{aligned}
c_0=&-6\left(\frac{1}{z^2}+\frac{(1-z)^3(1-\bar{z})}{z^2 (1-\alpha)(1-\bar{\alpha})}+\frac{z\bar{z}}{\alpha\bar{\alpha}}\right)\comma\\
c_1=&\frac{2 (1-z)^3 \left(1-\bar{z}\right) \left(\bar{z}-3 z-\alpha-\bar{\alpha} +4\right)}{ z^2(1-\alpha )
   \left(1-\bar{\alpha }\right)}-\frac{2 z \bar{z} \left(\bar{z}+3 z-\alpha-\bar{\alpha} -3\right)}{\alpha
    \bar{\alpha }}+\frac{2(2-6z+3z^2)}{z^2}\comma \\
c_2=&\frac{2(1-z)}{z^2} \left[7z-8z^2-3\bar{z}+6z\bar{z}+\frac{z^3 \bar{z} \left(3 z-\bar{z}-2 \alpha-2\bar{\alpha} \right)}{\alpha  \bar{\alpha }}\right.\\
&\left.-\frac{(1-z)^2
   \left(1-\bar{z}\right) \left(3 \bar{z}+z \left(-11+9z-5 \bar{z}+2 
   \alpha +2\bar{\alpha}\right)\right)}{(1-\alpha )
   \left(1-\bar{\alpha }\right)}\right]\comma\\
c_3=&(1-z)^2\left[\frac{6(z-\bar{z})}{z}\left(1+\frac{(1-z)^2(1-\bar{z})}{(1-\alpha)(1-\bar{\alpha})}\right)-\frac{(z-\alpha)(z-\bar{\alpha})(\bar{z}-\alpha)(\bar{z}-\bar{\alpha})}{\alpha\bar{\alpha}(1-\alpha)(1-\bar{\alpha})}\right]\period
\end{aligned}
\eeq
}}
In what follows, we reproduce \eqref{dataKonishi} from integrability.
\subsection{Asymptotic Part\label{sec:asympt}}
\begin{figure}[t]
\begin{center}
\begin{minipage}{0.32\hsize}
\begin{center}
\includegraphics[clip,height=4cm]{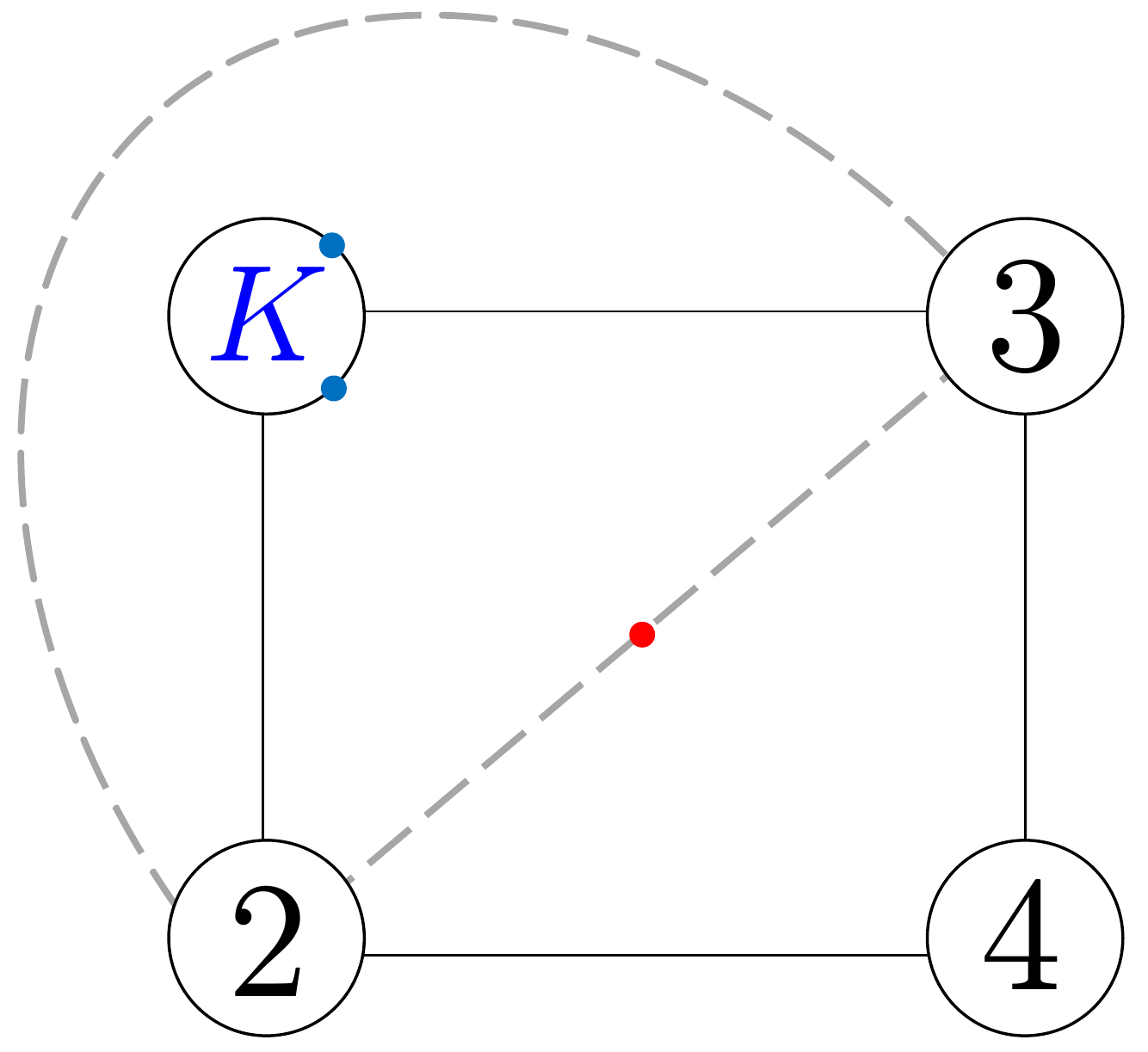}
\end{center}
\end{minipage}
\begin{minipage}{0.32\hsize}
\begin{center}
\includegraphics[clip,height=4cm]{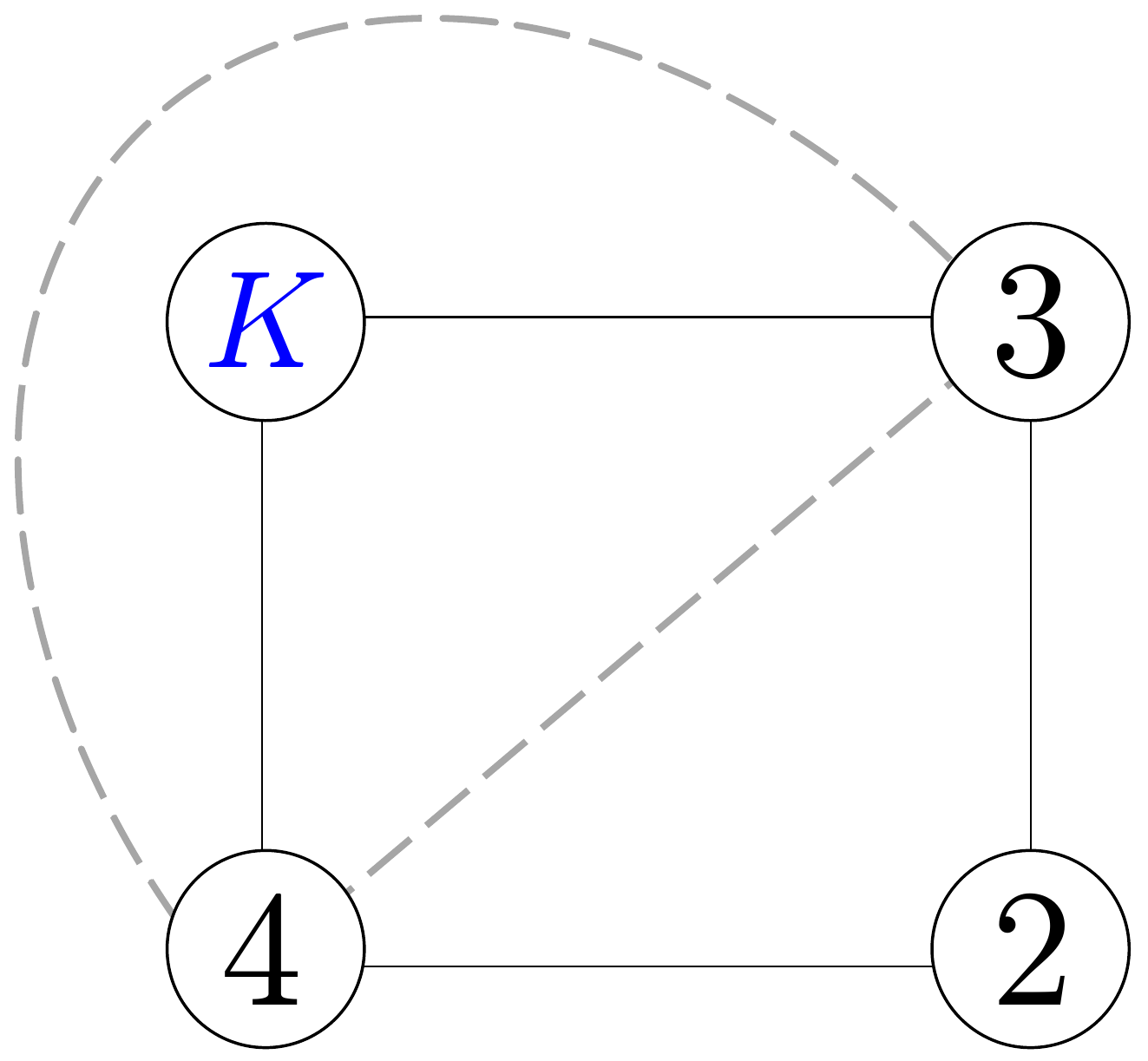}
\end{center}
\end{minipage}
\begin{minipage}{0.32\hsize}
\begin{center}
\includegraphics[clip,height=4cm]{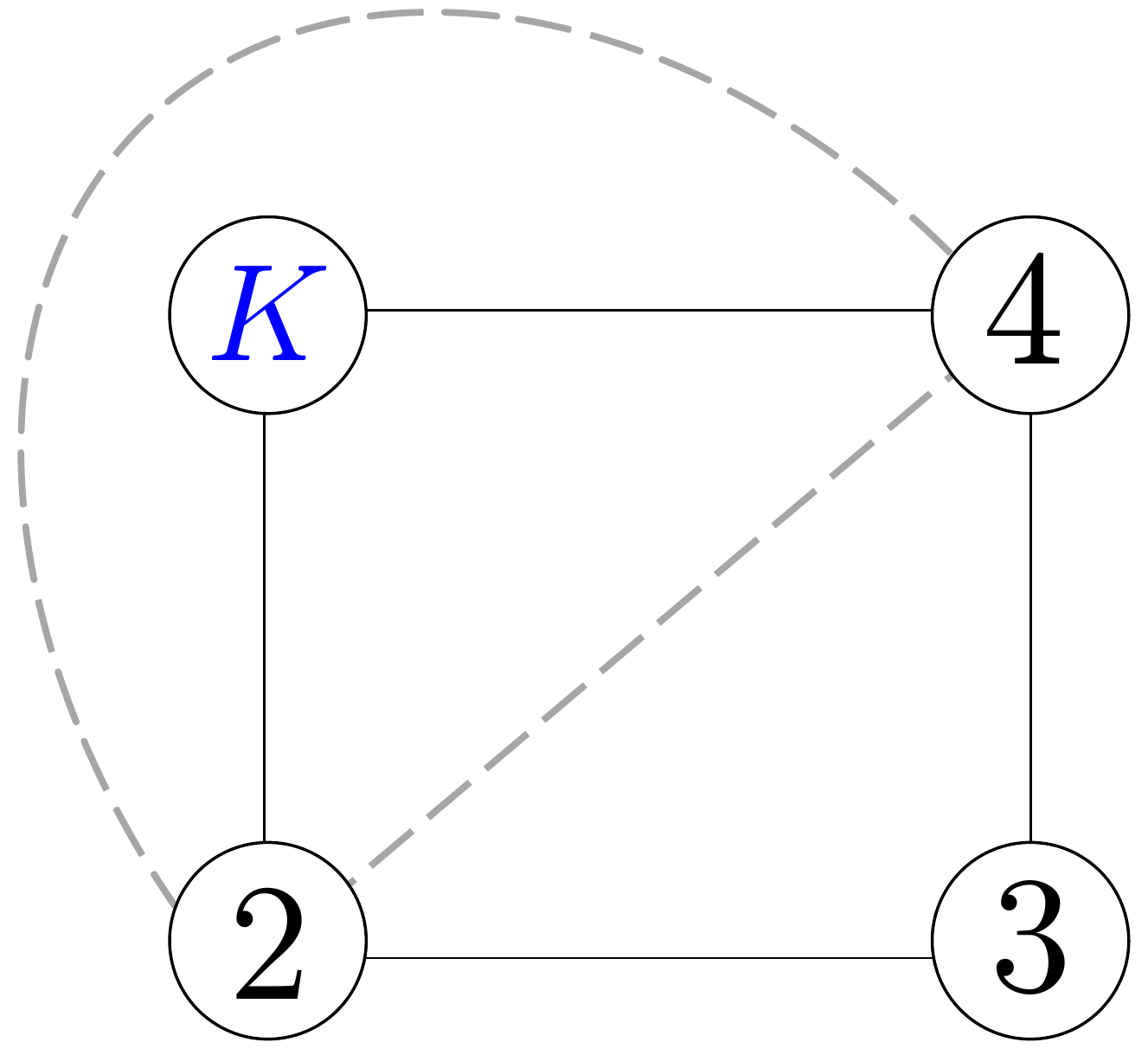}
\end{center}
\end{minipage}\\
\begin{minipage}{0.32\hsize}
\begin{center}
\hspace{10pt}$(1243)$
\end{center}
\end{minipage}
\begin{minipage}{0.32\hsize}
\begin{center}
\hspace{10pt}$(1432)$
\end{center}
\end{minipage}
\begin{minipage}{0.32\hsize}
\begin{center}
\hspace{10pt}$(1234)$
\end{center}
\end{minipage}
\end{center}
\vspace{-0.5cm}
\caption{Graphs for the correlator of a Konishi and three ${\bf 20}^{\prime}$s. Here as well, dashed lines denote zero-length bridges. In the leftmost figure, we depicted the mirror (red) and the physical (blue) particles. As is clear from the figure, the mirror particle only interacts with the physical particles which are in the same hexagon. \label{fig:konishi}}
\end{figure}

The asymptotic contribution can be computed by performing a sum over partitions for each connected graph. In the case at hand, there are three graphs as shown in figure \ref{fig:konishi}. To compute the contribution from each graph, one needs to cut them into hexagons. There are several different ways to achieve this, but the simplest way is the one shown in figure \ref{fig:konishi}, in which the Konishi operator is cut only into two segments. In this way of cutting, the cross ratios appearing in the asymptotic part are all $1$. Thus, there is no extra cross-ratio dependent weight and the sum over partition reduces to the one for the structure constant,
\beq
\mathcal{A}_{\ell=1}=\prod_{i<j}h(u_i,u_j) \sum_{\alpha\cup \bar{\alpha}=\{\bf u\}}(-1)^{|\bar{\alpha}|} \prod_{j\in \bar{\alpha}}e^{i p(u_j)\ell} \prod_{i\in \alpha,j\in\bar{\alpha}}\frac{1}{h(u_i,u_j)}\comma
\eeq
where $\ell=1$ and $h(u,v)$ is the SL$(2)$ hexagon form factor given in Appendix \ref{ap:weak}. For the Konishi state, the rapidities  are given by $\{{\bf u}\}=\{u,-u\}$ with
\beq
u=\frac{1}{2\sqrt{3}}+\frac{4 g^2}{\sqrt{3}}+O(g^4)\period
\eeq
To obtain the full result, we need to multiply the space-time dependences coming from bridges and magnons, and sum over graphs. The result reads
\beq\label{asymptKonishi}
\begin{aligned}
&\left.\langle K\mathcal{O}_{{\bf 20}^{\prime}}\mathcal{O}_{{\bf 20}^{\prime}}\mathcal{O}_{{\bf 20}^{\prime}}\rangle\right|_{\rm asym}=\qquad\\
&\underbrace{\sqrt{\frac{\prod_{i=1}^{2}\mu(u_i)}{\det \del_{u_i} \varphi_j\prod_{i<j}S(u_i,u_j)}}\mathcal{A}_{\ell=1}}_{=C_{123}^{\bullet \circ\circ}/C_{123}^{\circ\circ\circ}}\left[\red{d_{12}d_{23}d_{34}d_{41}}\blue{\left(\frac{x^{+}_{24}}{x^{+}_{12}x^{+}_{14}}\right)^{2+\frac{\gamma}{2}}\left(\frac{x^{-}_{24}}{x^{-}_{12}x^{-}_{14}}\right)^{\frac{\gamma}{2}}}\right.\\
&\left.+\red{d_{12}d_{24}d_{34}d_{13}}\blue{\left(\frac{x^{+}_{23}}{x^{+}_{12}x^{+}_{13}}\right)^{2+\frac{\gamma}{2}}\left(\frac{x^{-}_{23}}{x^{-}_{12}x^{-}_{13}}\right)^{\frac{\gamma}{2}}}+\red{d_{13}d_{23}d_{24}d_{41}}\blue{\left(\frac{x^{+}_{34}}{x^{+}_{13}x^{+}_{14}}\right)^{2+\frac{\gamma}{2}}\left(\frac{x^{-}_{34}}{x^{-}_{13}x^{-}_{14}}\right)^{\frac{\gamma}{2}}}\right]\period
\end{aligned}
\eeq
Here $\gamma$ is the anomalous dimension of the Konishi operator and the factor denoted in red comes from bridges while the factor denoted in blue comes from (physical) magnons. The factor with a square root accounts for the normalization of the Konishi state, where $\varphi_j$ defined by
\beq
e^{i \varphi_j}\equiv e^{i p_j L}\prod_{k\neq j}S(u_j, u_k)\period
\eeq
with $S(u,v)$ being the S-matrix in the SL(2) sector, and $L=2$.

Using the results from integrability \cite{BKV},
\beq
\begin{aligned}
\gamma = 12 g^2 +O(g^4)\comma \qquad 
C_{123}^{\bullet\circ\circ}/C_{123}=\frac{1}{\sqrt{6}}-\sqrt{6}g^2+O(g^4)\comma  
 \end{aligned}
 \eeq 
 we can rewrite \eqref{asymptKonishi} as
 \beq
 \left.\langle K\mathcal{O}_{{\bf 20}^{\prime}}\mathcal{O}_{{\bf 20}^{\prime}}\mathcal{O}_{{\bf 20}^{\prime}}\rangle\right|_{\rm asym}= d_{12}d_{23}d_{34}d_{41}\left(\frac{x_{34}^{+}}{x_{13}^{+}x_{14}^{+}}\right)^{2+\gamma/2}\left(\frac{x_{34}^{-}}{x_{13}^{-}x_{14}^{-}}\right)^{\gamma/2} f_{\rm asym}
 \eeq
 with
 \beq
 f_{\rm asym}=\left(\frac{1}{\sqrt{6}}-\sqrt{6}g^2\right) \left[\frac{1}{z^{2+6g^2}\bar{z}^{6g^2}}+\frac{(1-z)^{3+6g^2}(1-\bar{z})^{1+6g^2}}{z^{2+6g^2}\bar{z}^{6g^2}(1-\alpha)(1-\bar{\alpha})}+\frac{z\bar{z}}{\alpha\bar{\alpha}}\right]+O(g^4)\period
 \eeq
 At tree level, $f_{\rm asym}$ matches precisely the OPE data \eqref{dataKonishi} while at one loop it gives
 \beq
 \begin{aligned}
 &\!\!f_{\rm asym}^{(1)}\!=\!-\sqrt{6}g^2\!\left[\frac{1}{z^2} +\!\frac{(1-z)^3(1-\bar{z})}{z^2(1-\alpha )  \left(1-\bar{\alpha
   }\right)}+\!\frac{\bar{z} z}{\alpha  \bar{\alpha }}+\!\frac{ (1-z)^3(1-\bar{z}) \log \left|\frac{z}{1-z}\right|^2}{z^2(1-\alpha ) 
   \left(1-\bar{\alpha }\right)}+\!\frac{ \log |z|^2 }{ z^2 }\right].
   \end{aligned}
 \eeq
\subsection{Finite-size Correction\label{sec:finite}}
To reproduce the full answer at one loop, we also need to compute the mirror-particle correction.

For each graph in figure \ref{fig:konishi}, there are two mirror channels which contribute at one loop. For definiteness, let us first focus on the channel inside a square in the graph $( 1243)$. As is clear from figure \ref{fig:konishi}, the mirror particle only talks to a part of physical magnons which are in the same hexagon\footnote{By contrast, in three-point functions, all physical particles interact with the mirror particle since there are only two hexagons and physical particles always share one hexagon with the mirror particle. This feature helps to simplify the computation of three-point functions since one can make use of the zero-momentum condition to get rid of phase factors $\prod_j e^{ip (u_j)}$. On the other hand, for four-point functions, one cannot use such simplification and has to keep track of phase factors in order to get the correct results.}. As a result, each term in the sum receives different mirror-particle corrections. By making use of mirror transformations, we can compute the integrand as
\beq
\begin{aligned}
&{\rm int}_a^{2\text{-}3} (v) =\mu_{a} (v^{\gamma}) e^{-iv\log z\bar{z}}\\
&\quad \times \prod_{i<j}h(u_i,u_j)\!\!\!\sum_{\alpha\cup \bar{\alpha}=\{{\bf u}\}}\!\!\!\!\!(-1)^a\tilde{T}_{a}(v^{-3\gamma};\alpha)\prod_{j\in \alpha}h_{a}(u_j,v^{-3\gamma})\prod_{j\in \bar{\alpha}}(-e^{i p (u_j)\ell})\prod_{i\in \alpha,j\in\bar{\alpha}}\frac{1}{h(u_i,u_j)}\,.
\end{aligned}
\eeq
Here $\ell=1$ and $h_{a}(u,v^{-3\gamma})$  and $\tilde{T}_{a}(v^{-3\gamma};\alpha)$ are the dynamical part and the matrix part of the interaction between physical and mirror magnons\footnote{The factor $(-1)^{a}$ comes from the hexagon form factor. More precisely, it arises when we perform the crossing transformations to the mirror particle living in the second hexagon.}.

$\tilde{T}_{a}(v^{-3\gamma};\alpha)$ is essentially a twisted transfer matrix with twists given by cross ratios:
\beq
\tilde{T}_{a}(v^{-3\gamma};{\bf u})\sim \Tr_{a}\left[(-1)^{F}e^{
2 \varphi J +i\phi \tilde{\sf L} +i\theta \tilde{\sf R}}\prod_{u_i\in {\bf u}} \mathcal{S}_{1a}(u_i,v^{-3\gamma})\right]\comma
\eeq
Here $\mathcal{S}_{1a}(u,v)$ is the $\mathfrak{psu}(2|2)$ S-matrix (without the dynamical phase). To be precise, however, one needs to dress the states with $Z$ markers as explained in section \ref{subsec:character}. The effects of $Z$ markers are of two folds: First they produce the dependence on the cross ratio $|\alpha/z|$ as discussed in section \ref{subsec:character}. Second, in the presence of other magnons, the markers bring about an extra phase factor as shown in \cite{BKV}. By taking into account these effects, we can write down $\tilde{T}_{a}(v^{-3\gamma};{\bf u})$ at weak coupling as (see Appendix \ref{ap:transfer} for details),

{{\small
\beq\label{trasferweak}
\begin{aligned}
\tilde{T}_{a}(v^{-3\gamma};{\bf u}) =&\frac{(-1)^{a}}{Q^{[-1-a]}}\left[e^{i(a\phi +P)}Q^{[a+1]}+\sum_{n=1-\frac{a}{2}}^{\frac{a}{2}}e^{-2i n\phi }Q^{[-2n-1]}\right.\\
&\left.\hspace{-20pt}-\cos \theta \left(e^{\varphi+iP}+e^{-\varphi}\right)\sum_{n=\frac{1-a}{2}}^{\frac{a-1}{2}}e^{-2in\phi }Q^{[-2n]}+e^{iP}\sum_{n=1-\frac{a}{2}}^{\frac{a}{2}-1}e^{-2in \phi }Q^{[-2n+1]}\right]\comma
\end{aligned}
\eeq
}}
with $Q(v)=\prod_{u_i\in {\bf u}}(v-u_i)$ and $f^{[n]}=f(v+in/2)$.

It turns out that the other channel gives exactly the same contribution. Adding up two contributions and substituting the rapidities of the Konishi state, we obtain the following mirror correction (divided by $\mathcal{A}_{\ell=1}$) for the graph $(1243)$:
\beq
\begin{aligned}
2\mathcal{M}^{\rm K}_{z,\alpha}&\!\equiv\!  \frac{2}{\mathcal{A}_{\ell=1}} \sum_{a=1}^{\infty} \int_{-\infty}^{\infty} \frac{dv}{2\pi}\,\, {\rm int}_{a}^{2\text{-}3}(v)\!=\!g^2\frac{m_1\log |z|^2+m_2\log |1-z|^2+m_3F^{(1)}(z,\bar{z})}{(z-\bar{z})^2}\comma
\end{aligned}
\eeq
with
{\footnotesize
\beq
\begin{aligned}
m_1&=-\frac{z^2m_2+z\left[2\bar{z}(\bar{z}+\alpha+\bar{\alpha})-z(10\bar{z}+\alpha+\bar{\alpha})+z^2 \left(4+\bar{z}(\alpha^{-1}+\bar{\alpha}^{-1})\right)\right]}{(1-z)^2}\comma\\
m_2&=(\alpha^{-1}+\bar{\alpha}^{-1})\left[\frac{z}{2}(\bar{z}^2+\alpha\bar{\alpha})-\frac{3\bar{z}}{2}(z^2+\alpha\bar{\alpha})\right]-\left(3z^{2}-10z\bar{z}+3\bar{z}^2\right)\comma\\
m_3&=2\left(z^3+\bar{z}^3\right)-\bar{z}\left(\alpha^{-1}+\bar{\alpha}^{-1}\right)
\left(z^{3}+\bar{z}\alpha\bar{\alpha}\right)\period
\end{aligned}
\eeq
}
Summing up all the graphs, we finally obtain the finite-size correction at one loop,
\beq
\begin{aligned}
\left.\langle K\mathcal{O}_{{\bf 20}^{\prime}}\mathcal{O}_{{\bf 20}^{\prime}}\mathcal{O}_{{\bf 20}^{\prime}}\rangle\right|_{\rm mirror}\!\!=\!
2\frac{C_{123}^{\bullet\circ\circ}}{C_{123}^{\circ\circ\circ}}&\left[\mathcal{M}^{\rm K}_{\frac{z}{z-1},\frac{\alpha}{\alpha-1}}\,d_{12}d_{23}d_{34}d_{41}\left(\frac{x^{+}_{24}}{x^{+}_{12}x^{+}_{14}}\right)^{2+\frac{\gamma}{2}}\left(\frac{x^{-}_{24}}{x^{-}_{12}x^{-}_{14}}\right)^{\frac{\gamma}{2}}\right.\\
&+\mathcal{M}^{\rm K}_{z,\alpha}\,d_{12}d_{24}d_{34}d_{13}\left(\frac{x^{+}_{23}}{x^{+}_{12}x^{+}_{13}}\right)^{2+\frac{\gamma}{2}}\left(\frac{x^{-}_{23}}{x^{-}_{12}x^{-}_{13}}\right)^{\frac{\gamma}{2}}\\
&\left.+\mathcal{M}^{\rm K}_{1-z, 1-\alpha}\,d_{13}d_{23}d_{24}d_{41}\left(\frac{x^{+}_{34}}{x^{+}_{13}x^{+}_{14}}\right)^{2+\frac{\gamma}{2}}\left(\frac{x^{-}_{34}}{x^{-}_{13}x^{-}_{14}}\right)^{\frac{\gamma}{2}}\right]\comma
\end{aligned}
\eeq
which leads to
\beq
f_{\rm mirror}^{(1)}=\frac{2}{\sqrt{6}}\left[\frac{1}{z^2} \mathcal{M}^{\rm K}_{\frac{z}{z-1},\frac{\alpha}{\alpha-1}} +\frac{(1-z)^3(1-\bar{z})}{z^2(1-\alpha )  \left(1-\bar{\alpha
   }\right)}\mathcal{M}^{\rm K}_{z,\alpha}+
   \frac{\bar{z} z}{\alpha  \bar{\alpha }}
   \mathcal{M}^{\rm K}_{1-z,1-\alpha}\right]\period
\eeq
Remarkably, a sum of $f_{\rm asym}^{(1)}$ and $f_{\rm mirror}^{(1)}$ precisely matches the OPE result \eqref{dataKonishi}! This is another strong support for our proposal.

Let us make two remarks before closing this section: In \cite{DP}, several other five-point functions, which involve longer BPS operators, were computed. By the OPE expansion of those results, we can compute correlators of one Konishi and three longer BPS operators. We confirmed that they also match the integrability predictions. See Appendix \ref{ap:longer} for details.
For BPS correlators, we showed that the integrability result is ``flip-invariant''; namely it is independent of how we cut a four-point function into hexagons. In Appendix \ref{ap:flip}, we show that the flip invariance holds also in the presence of physical magnons. It is an important consistency check of our construction. 

\section{Ladder Integrals from Integrability\label{sec:ladder}}
As we have seen in section \ref{sec:4BPS}, the one-loop conformal integral can be reconstructed from the integration of the mirror momentum and the summation over the bound-state index. It provides an alternative representation of the conformal integral, which can be recast into a Mellin-like representation (see section \ref{subsec:Mellin}).

Such nice properties seem to persist at higher loops. To get a glimpse of it, let us consider the $L$-loop contribution from a one-particle mirror correction with the bridge length $L-1$. Since the length of the bridge is the maximum possible for a given loop order, we can substitute quantities in the integrand \eqref{intBPS0} with their leading order expressions:
 \beq\label{leadingintL}
 {\rm int}^{(L)}_{a}(v)=\frac{2g^{2L} (\cos \phi -\cosh \varphi \cos \theta)\sin a \phi}{\sin \phi}\frac{a}{(v^2+a^2/4)^{1+L}}e^{-2i v \log |z|}+O(g^{2(L+1)})\period
 \eeq
 By performing the integral, we obtain
 \beq\label{L-loop1-p}
 \int_{-\infty}^{\infty} \frac{dv}{2\pi} {\rm int}^{(L)}_{a}(v)=
 \frac{g^{2L} (\cos \phi -\cosh \varphi \cos \theta)}{i \sin \phi}\left[\sum_{k=0}^{L}\frac{(-1)^k(2L-k)!}{L!(L-k)!k!}\frac{\log^k (z\bar{z})}{a^{2L-k}}(z^a-\bar{z}^{a})\right]\period
 \eeq
 We then perform a sum over $a$ to get
 \beq
 \begin{aligned}
 &\sum_{a=1}^{\infty}\int_{-\infty}^{\infty} 
 \frac{dv}{2\pi} {\rm int}^{(L)}_{a}(v)=g^{2L}
 \frac{2(z+\bar{z})
 \alpha\bar{\alpha}-(\alpha+\bar{\alpha})(z\bar{z}+\alpha\bar{\alpha})}
 {2 \alpha\bar{\alpha}}F^{(L)}(z,\bar{z})\comma
 \end{aligned}
 \eeq
where $F^{(L)}$ is given by 
 \beq
 F^{(L)}(z,\bar{z})=\frac{1}{z-\bar{z}}\left[\sum_{k=0}^{L}\frac{(-1)^k(2L-k)!}{L!(L-k)!k!}\log^k (z\bar{z})({\rm Li}_{2L-k}(z)-{\rm Li}_{2L-k}(\bar{z}))\right]\period
 \eeq
\begin{figure}[t]
\begin{center}
\includegraphics[clip,height=4cm]{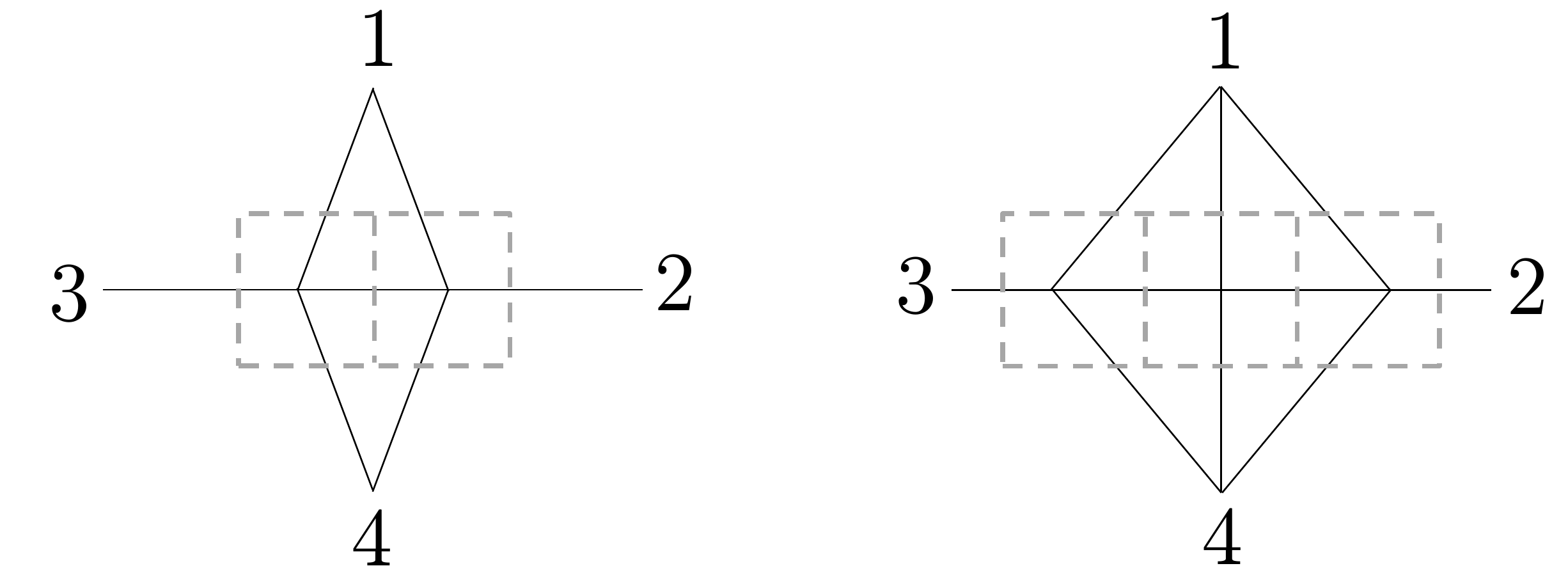}
\end{center}
\vspace{-0.5cm}
\caption{Examples of ladder integrals: $L=2$ (left) and $L=3$ (right). The propagators are denoted by the solid lines and the dual graph is denoted by the dashed lines. The dual graph has a shape of a ladder diagram. (In the context of scattering amplitudes, they correspond to ``four-mass'' ladder diagrams.)
\label{fig:ladder}}
\end{figure}
 What is interesting is that the function $F^{(L)}$ coincides with the so-called $L$-loop
 conformal integral, which is obtained in \cite{UD} by computing a diagram given in figure \ref{fig:ladder}:
 \beq
 (\text{Figure \ref{fig:ladder}})=\frac{F^{(L)}(z,\bar{z})}{\pi^{2L}x^2_{13}x_{24}^2x_{14}^{2(L-1)}}\period
 \eeq
Following the argument in section \ref{subsec:Mellin}, we can also recast \eqref{L-loop1-p} into a Mellin-like representation:
\beq
F^{(L)}=\frac{-1}{2(z-\bar{z})}\int^{i\infty}_{-i\infty} ds \int^{i\infty}_{-i\infty} dt \frac{z^{s}\bar{z}^{t}}{(st)^{L+1}}\left[\frac{\Gamma[1+(s-t)]\Gamma[1-(s-t)]}{\Gamma[\frac{1}{2}+(s-t)]\Gamma[\frac{1}{2}-(s-t)]}- i (s-t)\right]\period
\eeq

This provides an example where integrability makes a connection with perturbation theory. Typically the expressions coming from integrability have less number of integration variables as compared to the ones obtained directly from Feynman diagrams. It would be an interesting future problem to study multi-particle mirror corrections and see if they reproduce more complicated conformal integrals, such as Easy and Hard integrals \cite{DDEHPS}, which appear at three loops \cite{Hidden}. 
It would be even nicer if we can use integrability for computing/predicting integrals at four loops and beyond \cite{EHKS} which have never been evaluated.

Even more amusingly, we found that subleading corrections to \eqref{leadingintL} also produce ladder integrals but this time with higher transcendentality. For instance, $O(g^{2(L+1)})$ correction can be computed by expanding \eqref{intBPS0} as
\beq
g^2{\rm int}^{(L)}_{a}(v)\left(\frac{2 i v \log z\bar{z}}{v^2+a^2/4}+\frac{8v^2-a^2}{(v^2+a^2/4)^2}+(L-1)\frac{2(v^2-a^2/4)}{(v^2+a^2/4)^2}\right)\period
\eeq
By integrating by parts, one can show that this gives a ladder integral with one trascendentality higher, $F^{(L+1)}$. Such a pattern seems to persist at least for the first few subleading corrections. This suggests that the full non-perturbative integral \eqref{intBPS0} resums all the ladder integrals. Presumably such an integral would be an important building block for finite-coupling correlators and studying its analytic property will allow us to extract interesting nonperturbative physics.

Before drawing to an end, let us mention two interesting related works in this context: One is a recent work \cite{Kazakov}, in which they used a double-scaling limit of strongly twisted and weakly coupled $\mathcal{N}=4$ SYM theory to extract, using AdS/CFT integrability tools, the explicit results for particular Feynman graphs, such as wheel graphs. The other is \cite{Isaev}, in which the ladder integral is computed using the conformal quantum mechanics. All these results are pointing towards some deeper relation between integrability and perturbation. It would be worth trying to uncover it. In particular, it would demystify the physical origin of mirror particles.
\section{Conclusion and Prospects\label{sec:conclusion}}
In this paper, we proposed a framework to study correlation functions of single-trace operators in planar $\mathcal{N}=4$ SYM. The method proceeds in two steps: First we decompose the correlation functions into so-called hexagon form factors introduced in \cite{BKV}. We then glue them back together with appropriate weight factors, which are determined by the symmetry. As shown in several examples at one loop, it reproduces the full four-point function including the space-time and the R-symmetry dependence. 

An immediate next step is to study higher points/loops. It is interesting and important to see if our proposal works also in those cases \cite{FK}. Also important is to study four-point functions with more than one non-BPS operators. At least at one loop, we can compare them with the OPE decomposition of six- and higher-point functions \cite{DP}.

Equally important would be to ask how general our formalism is. We expect that a similar decomposition is possible for more general conformal gauge theories at large $N$. In addition, it may also be applicable to 2d CFT's.

As discussed in introduction, the OPE decomposition of 
higher-point functions encodes the information of 
multi-trace operators. In this sense, our result is already hinting 
that 
non-planar quantities 
are amenable to the integrability machinery.  
It would be fascinating if we can directly construct 
non-planar surfaces using the hexagonalization \cite{InProgress}. Once we succeed, we 
will for the first time obtain a 
method to  accurately compute the quantum gravitational effects in AdS.

It is also interesting to explore the relation to the conventional OPE more in detail. In particular, when the lengths of the operators are large, we may be able to relate our method to the approach of \cite{Asymptotic4pt}, in which they used the combination of OPE and integrability to study four-point functions.

Another important direction is to investigate various limits such as the strong coupling, the Regge limit, and the near-BPS limit. In particular, it would be important to see how the locality in AdS emerges at strong coupling \cite{MSZ}. It would also be interesting to reproduce a recent beautiful conjecture on the four-point functions at strong coupling \cite{Rastelli}. Also important is to study the double light-cone limit \cite{AB}. All these may require the resummation of finite-size corrections \cite{JKKS}. In the case of structure constants, such resummation is hindered by the double-pole singularities in the mirror-particle integrand \cite{4loop}, which physically come from the wrapping corrections to the spectrum. By contrast, for higher-point functions of BPS operators there is no physical reason to expect such singularities. This suggests that studying higher-point functions might be easier, although counter-intuitive it may seem.

There are also some loose ends in our story: One is the summation range of graphs and the other is the $J$-charge dependence of the weight factor. They are conjectured through the comparison with the data and there is no rigorous derivation yet. To understand them, it would be helpful to study the roles of supersymmetry since both are somehow related to it. 

\begin{figure}[t]
\begin{center}
\includegraphics[clip,height=4cm]{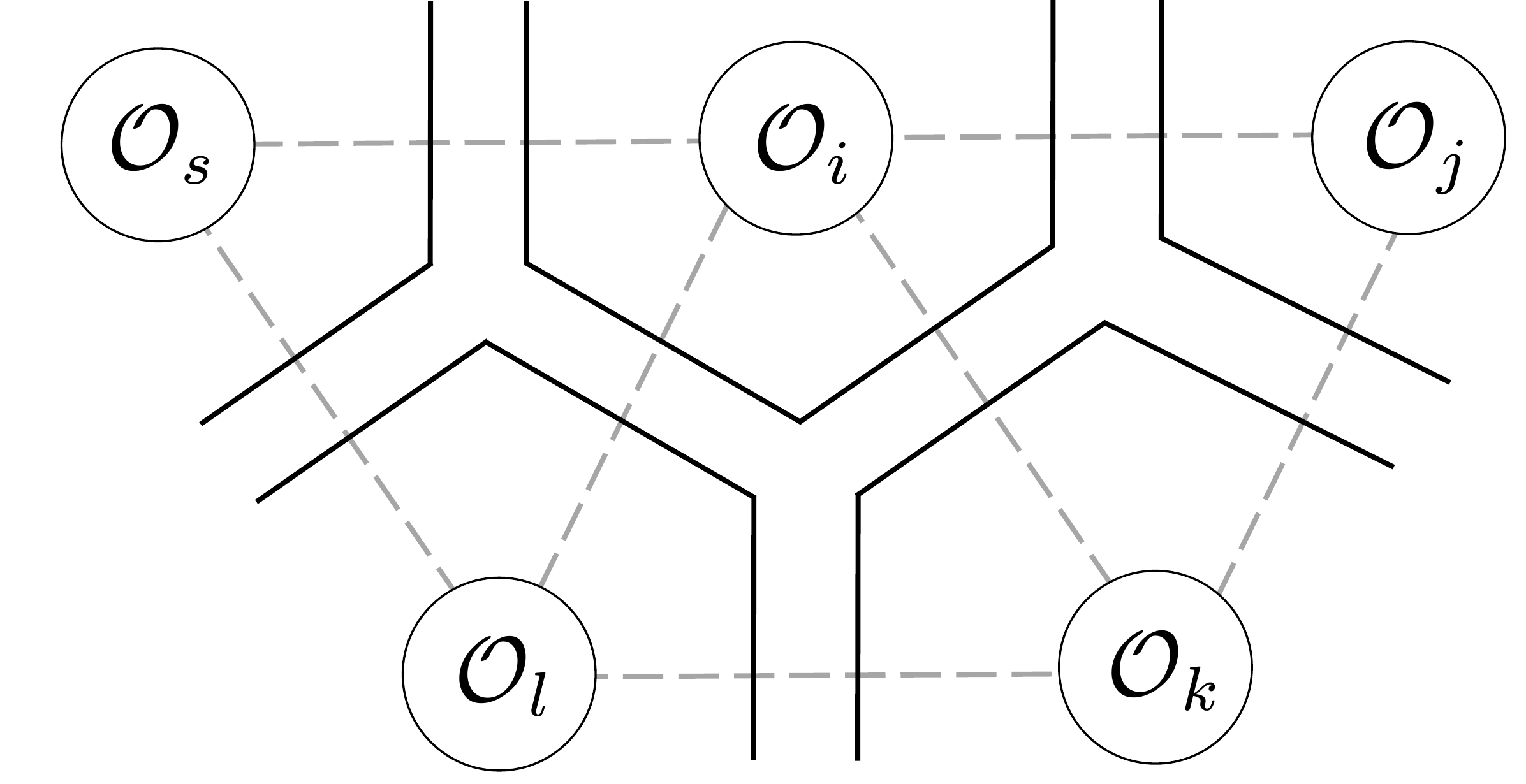}
\end{center}
\vspace{-0.5cm}
\caption{Hexagonalization and its dual graph. The dual graph of a hexagonalization is a ribbon graph, which can be interpreted as an open string diagram in the mirror channel. From this point of view, the hexagon form factor plays the role of a fundamental vertex of ``mirror open strings''.
\label{fig:sft}}
\vspace{-0.2cm}
\end{figure}
Our approach is perhaps pointing towards something like ``string field theory of mirror open strings'', in which the hexagon form factor plays the role of a fundamental vertex (see figure \ref{fig:sft}). In its current form, it appears quite different\footnote{The most notable difference is that our ``open strings'' are ending on physical closed-string states not on boundary states.} from conventional string field theories, but it might be rewarding to pursue this analogy and clarify the relation with the standard approaches. A step in this direction would be to understand flat-space amplitudes in a similar manner. 

 A related question is whether we can interpret the sum over graphs as the integration over the worldsheet moduli: For $n$-point functions, there are $2(n-3)$ unfixed bridge lengths and the number precisely matches the dimension of the moduli space. A crucial difference however is that the bridge lengths are all discrete whereas the moduli is continuous. It is generally believed that the discrete lightcone quantization (DLCQ) partially discretizes the moduli space \cite{GOPS,DVV}, but it would be desirable to understand how it works in our set-up. Similar but different ideas are proposed in \cite{Gopakumar,Razamat}, which are also quite inspirational.

The integrability approach to the AdS/CFT correspondence is sometimes regarded as a mere technical tool. It may be true that the progress so far has been technical, at least to some extent. However, with the method to study general correlation functions at hand, we truly believe that the time is ripe for asking more physical questions and using it to sharpen our understanding of quantum gravity and string theory.
\subsection*{Acknowledgement}
We acknowledge discussions with and comments by B.~Basso, J.~Caetano, 
V.~Goncalves, R.~Gopakumar, M.~Kruczenski, R.~Pius, R.~Suzuki, J.~Toledo and P. Vieira. We also thank V.~Kazakov, I.~Kostov, H.~Nastase and D.~Serban for comments on the manuscript. 
T.F. would like to thank J. Caetano for many 
inspiring discussions, for introducing him to
the four-point function problem and for pointing to
useful references. S.K. would like to thank D.~Honda for dicussions four years ago about the possibility of describing the string worldsheet based on a triangulation, which later led him to the idea discussed in this paper.
T.F. thanks Perimeter Institute for Theoretical Physics, where this work was initiated, and S.K. thanks ICTP-SAIFR in Brazil, where this work was completed. 
T.F. would like to thank FAPESP grants 13/12416-7 and 15/01135-2 for financial support. The research of S.K. is supported in part by Perimeter Institute for Theoretical Physics. Research at Perimeter Institute is supported by the Government of Canada through the Department of Innovation, Science and Economic Development Canada and by the Province of Ontario through the Ministry of Research, Innovation and Science.
\appendix
\section{Change of Conformal Frames and Crossing Rule\label{ap:crossing}}
Here we revisit  a crossing rule, first conjectured in \cite{BKV} and worked out more in detail in \cite{fermionic}, from the viewpoint of the symmetry. The basic strategy is to study the effect of the frame-changing transformation $r=e^{-\mathcal{K}}e^{-\mathcal{T}}$ on the magnon-symmetry group $\mathfrak{psu}(2|2)^2$.

Let us consider a three-point function with magnons in the operator $\mathcal{O}_1$. The frame suitable for describing this situation is the frame $1$ in figure \ref{fig:multi}, in which the $\mathfrak{psu}(2|2)^2$ symmetry of $\mathcal{O}_1$ is realized in a standard way. On the other hand, if we want to analyze this configuration from the point of view of $\mathcal{O}_2$, we should use the frame $2$ in figure \ref{fig:multi}, which is obtained from the frame $1$ by the transformation $r$. Under this transformation, the magnons in $\mathcal{O}_1$ transform as
\beq\label{defchi12}
|\chi^{A\dot{A}}(u)\rangle_1\quad  \to\quad  |\tilde{\chi}^{A\dot{A}}(u)\rangle_{2} \equiv r |\chi^{A\dot{A}}(u)\rangle_{1}
\eeq
where the subscripts $1$ and $2$ signifies the frame in which the state is defined. To understand the property of the state $|\tilde{\chi}^{A\dot{A}}\rangle_{2}$, we consider the action of the $\mathfrak{psu}(2|2)^2$ generators of the frame 2, which can be expressed using \eqref{defchi12} as
\beq\label{action2on1}
g|\tilde{\chi}^{A\dot{A}}(u)\rangle_{2} = gr |\chi^{A\dot{A}}(u)\rangle_{1}=r\left[(r^{-1}gr) |\chi^{A\dot{A}}(u)\rangle_{1}\right] \hspace{40pt} g\in \mathfrak{psu}(2|2)^2\period
\eeq
Under the conjugation $r$, the generators of $\mathfrak{psu}(2|2)$ transform as follows: 
\beq
\label{conjugationrule}
 \begin{aligned}
 &r^{-1} L^{\alpha}{}_{\beta}\,r= \dot{L}^{\alpha}{}_{\beta}\blue{+i( \epsilon^{\alpha\gamma}\delta^{\dot{\gamma}}_{\beta}+\frac{1}{2}\delta^{\alpha}{}_{\beta}\epsilon^{\gamma \dot{\gamma}})K_{\gamma\dot{\gamma}}}\comma &&r^{-1} \dot{L}^{\dot{\alpha}}{}_{\dot{\beta}}\,r=L^{\dot{\alpha}}{}_{\dot{\beta}}\blue{-i( \epsilon^{\dot{\alpha}\gamma}\delta^{\dot{\gamma}}_{\dot{\beta}}+\frac{1}{2}\delta^{\dot{\alpha}}{}_{\dot{\beta}}\epsilon^{\gamma \dot{\gamma}})K_{\gamma\dot{\gamma}}}\comma\\
 &r ^{-1}R^{a}{}_{b}\,r=\dot{R}^{\dot{a}}{}_{\dot{b}}\blue{-( \epsilon^{a\dot{c}}\delta^{c}_{b}+\frac{1}{2}\delta^{a}{}_{b}\epsilon^{\dot{c}c})R_{\dot{c}c}}\comma &&r^{-1} \dot{R}^{\dot{a}}{}_{\dot{b}}\,r=R^{a}{}_{b}\blue{+( \epsilon^{a\dot{c}}\delta^{c}_{b}+\frac{1}{2}\delta^{a}{}_{b}\epsilon^{\dot{c}c})R_{\dot{c}c}}\comma\\
 &r^{-1}Q^{\alpha}{}_{a}\,r=i\epsilon^{\alpha\dot{\beta}}\epsilon_{a\dot{b}}\dot{S}^{\dot{b}}{}_{\dot{\beta}}\blue{+i\epsilon^{\alpha\dot{\beta}}S_{\dot{\beta}a}}\comma &&r^{-1} \dot{S}^{\dot{a}}{}_{\dot{\alpha}}\,r=-i\epsilon^{\dot{a}b}\epsilon_{\dot{\alpha}\beta}Q^{\beta}{}_{b}\blue{-\epsilon^{\dot{a}b}S_{\dot{\alpha}b}}\comma\\
 &r^{-1} S^{a}{}_{\alpha}\,r= i\epsilon^{a\dot{b}}\epsilon_{\alpha\dot{\beta}}\dot{Q}^{\dot{\beta}}{}_{\dot{b}}\blue{-\epsilon^{a\dot{b}}S_{\dot{b}\alpha}}\comma &&r^{-1} \dot{Q}^{\dot{\alpha}}{}_{\dot{a}}\,r=-i\epsilon^{\dot{\alpha}\beta}\epsilon_{\dot{a}b}S^{b}{}_{\beta}\blue{-i\epsilon^{\dot{\alpha}\beta}S_{\dot{a}\beta}}\period
 \end{aligned}
 \eeq 
As can be seen from these relations, the conjugation by $r$ generates extra terms denoted in blue which are not inside $\mathfrak{psu}(2|2)^2$. However, these terms should be harmless in practice since they are all ``lowering'' generators of PSU$(2,2|4)$, and the net effect of their action on the Bethe states (with finite rapidities) is trivial owing to the highest weight property of the on-shell Bethe states.

As far as the bosonic generators are concerned, the conjugation simply swaps two $\mathfrak{psu}(2|2)$'s. On the other hand, the action of the fermionic generators is more complicated since the roles of $Q$'s and $S$'s are also swapped. For instance, from \eqref{conjugationrule} and \eqref{action2on1}, we can read off the action of $Q$ and $S$ as
\beq
\begin{aligned}
&Q^{\alpha}{}_{a}|\tilde{\chi}^{A\dot{b}}(u)\rangle=i(-1)^{|A|}c\delta^{\dot{b}}_{a}|\tilde{\chi}^{A\alpha}(u)\rangle\comma&&Q^{\alpha}{}_{a}|\tilde{\chi}^{A\dot{\beta}}(u)\rangle=i(-1)^{|A|}d\epsilon^{\alpha\dot{\beta}}\epsilon_{ab}|\tilde{\chi}^{Ab}(u)\rangle\comma\\
&S^{a}{}_{\alpha}|\tilde{\chi}^{A\dot{b}}(u)\rangle=i(-1)^{|A|}a\epsilon^{a\dot{b}}\epsilon_{\alpha\beta}|\tilde{\chi}^{A\beta}(u)\rangle\comma&&S^{a}{}_{\alpha}|\tilde{\chi}^{A\dot{\beta}}(u)\rangle=i(-1)^{|A|}b\delta^{\dot{\beta}}_{\alpha}|\tilde{\chi}^{Aa}(u)\rangle\period
\end{aligned}
\eeq
 Here $a$-$d$ are the ones in the string frame \cite{AF} and $|A|=0$ for bosonic indices ($a$'s and $b$'s) while $|A|=1$ for fermionic indices ($\alpha$'s and $\beta$'s). 
  By comparing these transformations (and the corresponding ones for $\mathfrak{psu}(2|2)_{R}$) with the standard transformation properties of a magnon in $\mathcal{O}_2$, we conclude that $|\tilde{\chi}^{A\dot{A}}(u)\rangle$ can be identified with an excitation in $\mathcal{O}_2$ as
 \beq
 \begin{aligned}
 &|\tilde{\chi}^{a\dot{a}}(u)\rangle =-|\chi^{\dot{a}a}(u^{-2\gamma})\rangle_{2}\comma\quad &&|\tilde{\chi}^{a\dot{\alpha}}(u)\rangle =-|\chi^{\dot{\alpha}a}(u^{-2\gamma})\rangle_{2}\comma\\
 &|\tilde{\chi}^{\alpha\dot{a}}(u)\rangle =|\chi^{\dot{a}\alpha}(u^{-2\gamma})\rangle_{2}\comma\quad &&|\tilde{\chi}^{\alpha\dot{\alpha}}(u)\rangle =-|\chi^{\dot{\alpha}\alpha}(u^{-2\gamma})\rangle_{2}\period
 \end{aligned}
 \eeq
 The relation physically means that having a magnon $\chi(u)$ in the operator $\mathcal{O}_1$ is equivalent to having a magnon $\chi (u^{-2\gamma})$ in the operator $\mathcal{O}_2$, up to signs and the change of indices. This nicely matches the crossing rule given in \cite{BKV,fermionic}. 
\section{Weak Coupling Expansions\label{ap:weak}}
In this appendix, we collect weak-coupling expressions of several quantities which are useful for the main text and future purposes. 

In what follows, $u=g(x+1/x)$, $v=g (y+1/y)$, $f^{\pm}=f (u\pm i/2)$ and $f^{[a]}=f(u+i a/2)$.
The energy and the momentum of a physical magnon are given by
\beq
\begin{aligned}
e^{i p(u)}&\equiv \frac{x^{+}}{x^{-}}=\frac{u+i/2}{u-i/2}\left(1+\frac{2 i u g^2}{\left(u^2+1/4\right)^2}\right)+O(g^4)\comma\\
E(u)&\equiv \frac{1}{2}+\frac{g}{i}\left(\frac{1}{x^{-}}-\frac{1}{x^{+}}\right)=\frac{1}{2}+\frac{g^2}{u^2+1/4}+O(g^4)\period
\end{aligned}
\eeq
On the other hand, the energy and the momentum in the mirror channel 
are given by
{{\small \beq
\begin{aligned}
e^{-\tilde{E}_{a}}&\equiv\frac{1}{x^{[+a]}x^{[-a]}}
=\frac{g^2}{u^2+\frac{a^2}{4}}\left[1+\frac{2g^2 \left(u^2-\frac{a^2}{4}\right)}{\left(u^2+\frac{a^2}{4}\right)^2}+g^4\frac{5\left(u^2-\frac{a^2}{4}\right)^2-3a^2u^2}{\left(u^2+\frac{a^2}{4}\right)^4}\right]\!\!+O(g^8)\comma\\
\tilde{p}_{a}&\equiv-i \left[\frac{a}{2}+\frac{g}{i}\left(\frac{1}{x^{[-a]}}-x^{[+a]}\right)\right]=u\left[1-\frac{2 g^2}{u^2+\frac{a^2}{4}}-\frac{2g^4\left(u^2-\frac{3a^2}{4}\right)}{\left(u^2+\frac{a^2}{4}\right)^3}\right]+O(g^6)\comma
\end{aligned}
\eeq}}
The measures for physical and mirror particles are given by
\beq
\begin{aligned}
\mu_{a}(u)&=\frac{1}{a}-\frac{a g^2}{\left(u^2+\frac{a^2}{4}\right)^2}+\frac{ag^4(a^2-8u^2)}{\left(u^2+\frac{a^2}{4}\right)^4}+O(g^6)\comma\\
\mu_{a}(u^{\gamma})&=\frac{a g^2}{\left(u^2+\frac{a^2}{4}\right)^2}-\frac{ag^4(a^2-8u^2)}{\left(u^2+\frac{a^2}{4}\right)^4}+\frac{a g^6(a^4-24a^2 u^2+48u^4)}{\left(u^2+\frac{a^2}{4}\right)^6}+O(g^8)\period
\end{aligned}
\eeq

The S-matrix in the SL(2) sector is given by
{\small
\beq
S(u,v)=\frac{u-v+i}{u-v-i}\left(\frac{1-\frac{1}{x^{-}y^{+}}}{1-\frac{1}{x^{+}y^{-}}}\right)^2\frac{1}{\sigma^2 (u,v)}=\frac{u-v+i}{u-v-i}\left[1+\frac{2i g^2(u-v)}{\left(u^2+\frac{1}{4}\right)\left(v^2+\frac{1}{4}\right)}\right]+O(g^4)\comma
\eeq
}where $\sigma (u,v)$ is the BES dressing phase \cite{BES}.
For the computation involving a Konishi operator, we use the 
following (fused) hexagon form factors:
{\small
\beq
\begin{aligned}
h(u,v)&= \frac{x^{-}-y^{-}}{x^{-}-y^{+}}\frac{1-\frac{1}{x^{-}y^{+}}}{1-\frac{1}{x^{+}y^{+}}}\frac{1}{\sigma (u,v)}=\frac{u-v}{u-v-i}\left(1+\frac{g^2 i(u-v+i)}{\left(u^2+\frac{1}{4}\right)\left(v^2+\frac{1}{4}\right)}\right) +O(g^4)\comma\\
h_{1a}(u,v^{-\gamma})&=\frac{u-\frac{i}{2}}{u-v-\frac{i (a+1)}{2}}+O(g^2)\comma\\
h_{1a}(u,v^{-3\gamma})&=\frac{1}{h_{a1}(v^{\gamma},u)}=\frac{u-v+\frac{i(a+1)}{2}}{u+\frac{i}{2}}+O(g^2)\comma\\
h_{1a}(u,v^{-5\gamma})&=\frac{1}{h_{a1}(v^{-\gamma},u)}=\frac{\left(u+\frac{i}{2}\right)^2}{u-\frac{i}{2}}\frac{u-v+\frac{i(a-1)}{2}}{\left(u-v-\frac{i(a-1)}{2}\right)\left(u-v+\frac{i(a+1)}{2}\right)}+O(g^2)
\end{aligned}
\eeq
}Here the subscripts $ab$ denote the bound-state indices of the first 
and the second particle and we used the 
important relation $h(u^{4\gamma},v)=1/h(v,u)$ when rewriting the results.

\section{OPE of Five-Point Functions\label{ap:fivept}}
Here we sketch how to obtain correlators with a Konishi operator by the OPE decomposition of higher-point functions. 

We start with a $(n+1)$-point function of BPS operators. Since we want a Konishi operator in the SL(2) sector, we place all the operators on the $x^2$-$x^3$ plane. Thus the positions of the operators are characterized by sets of holomorphic and anti-holomorphic coordinates $(x^{+}=x^{1}+ix^{2}, x^{-}=x^{1}-ix^{2})$:
\beq
\mathcal{O}_n (x^{+}_n,x^{-}_n)\period
\eeq
We furthermore take two of such operators to be
\beq
\mathcal{O}_1 \propto\Tr \big(ZX\big) \comma\qquad  \mathcal{O}_{n+1}\propto\Tr \big(Z\bar{X}\big)\period
\eeq
By bringing $\mathcal{O}_1$ and $\mathcal{O}_{n+1}$ close to each other, we obtain the  OPE series \cite{DolanOsborn},
{{\small
\beq
\begin{aligned}
&\mathcal{O}_1(0,0) \mathcal{O}_{n+1}(w^{+},w^{-})=\frac{c_{\rm BPS}}{w^{+}w^{-}}\left[\mathcal{O}_{\rm BPS}+\frac{1}{2}\left(w^{+}\del_{+}+w^{-}\del_{-}\right)\mathcal{O}_{\rm BPS}\right.\\
&\left.+\frac{1}{6}\left((w^{+})^2\del_{+}^2+w^{+}w^{-}\del_{+}\del_{-}+(w^{-})^2\del_{-}^{2}\right)\mathcal{O}_{\rm BPS}+\cdots\right]+c_{\rm Konishi} \frac{w^{+}}{w^{-}}\left(w^{+}w^{-}\right)^{\gamma/2}K +\cdots
\end{aligned}
\eeq
}}
where $\del_{\pm}=(\del_2\mp i\del_3)/2$, $K$ denotes the Konishi 
operator and $\gamma$ is the anomalous dimension of the Konishi 
operator. $c_{\rm BPS}$ is the structure constant for the BPS operator
\beq
\mathcal{O}_{\rm BPS}\propto \Tr \left(Z^2\right)\comma
\eeq
which (in appropriate normalization) is $c_{\rm BPS}=\sqrt{2}$ while 
$c_{\rm Konishi}$ is the structure constant for the Konishi operator, 
given by $c_{\rm Konishi}=1/(2\sqrt{3})-\sqrt{3}g^{2}$. 
Inserting this series into a correlation function, we obtain the relation
\beq
\begin{aligned}
\langle\mathcal{O}_1 \cdots \mathcal{O}_{n+1} \rangle=&\frac{c_{\rm BPS}}{w^{+}w^{-}}\langle \mathcal{O}_{\rm BPS}\cdots \mathcal{O}_n\rangle+\frac{c_{\rm BPS}}{6}\frac{w^{+}}{w^{-}}\langle \del_{+}^2\mathcal{O}_{\rm BPS}\cdots \mathcal{O}_{n}\rangle+\cdots\\
&+c_{\rm Konishi} \frac{w^{+}}{w^{-}}\left(w^{+}w^{-}\right)^{\gamma/2}\langle K \cdots \mathcal{O}_n\rangle+\cdots \period
\end{aligned}
\eeq
Since the anomalous dimension is treated as an infinitesimal quantity at weak coupling, $(w^{+}w^{-})^{\gamma/2}$ only appears as a series $1+\gamma \log (w^{+}w^{-})/2 +\cdots$.
Thus the correlator of a Konishi operator can be obtained by taking the limit of $(n+1)$-point function, reading off the term proportional to $w^{+}/w^{-}$, and subtracting a correlator  of a descendant of the BPS operator $\del^2_{+}\mathcal{O}_{\rm BPS}$. Applying this analysis to the five-point function of ${\bf 20}^{\prime}$'s given in \cite{DP}, we obtain the expression \eqref{dataKonishi}.
\section{Dressed Twisted Transfer Matrix\label{ap:transfer}}
In this appendix, we derive twisted transfer matrices and compute their weak-coupling expressions. To derive the results correctly, we have to take into account the effect of $Z$ markers.

\begin{figure}[t]
\begin{center}
\includegraphics[clip,height=4cm]{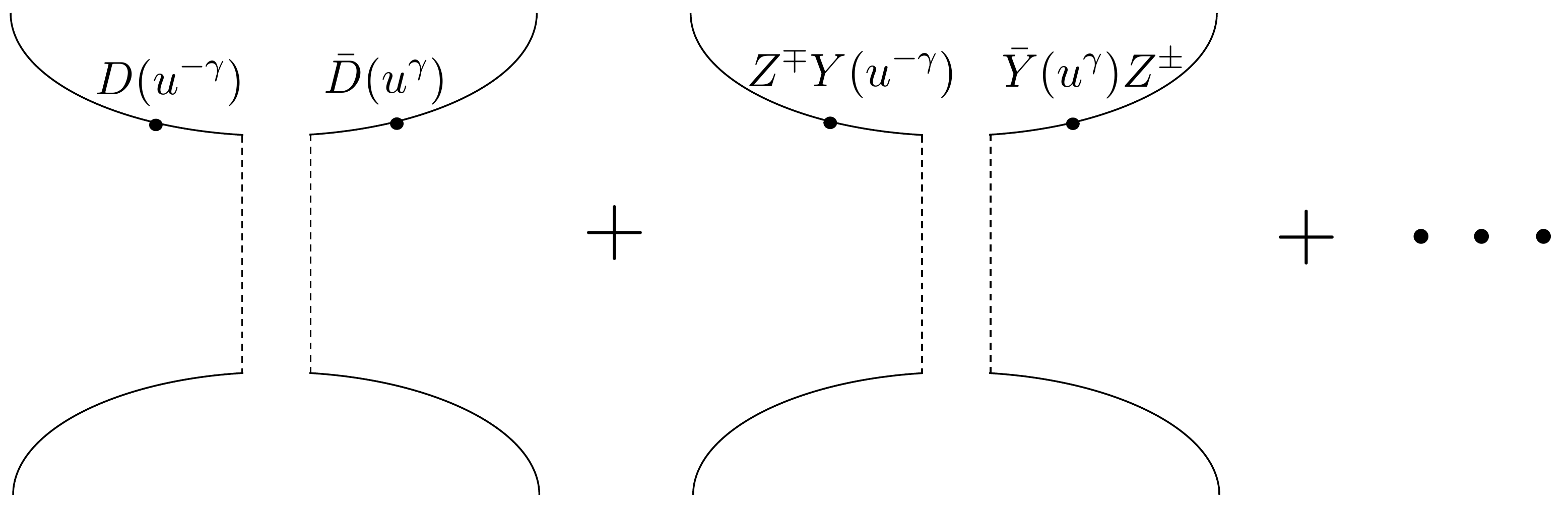}
\end{center}
\vspace{-0.5cm}
\caption{Dressed basis for the mirror channel. The dressing introduced in \eqref{modifiedstates} correspond to adding/subtracting $Z$ markers from scalar excitations as shown above.
\label{fig:basis}}
\end{figure}

For this pupose, it is convenient to take an alternative viewpoint on the finite-size correction explained in figure \ref{fig:mirror}. As we see below, the dressing by $Z$ markers given in \eqref{modifiedstates} corresponds in this viewpoint to adding and subtracting $Z$ markers appropriately from a virtual-particle pair whenever the excitations are scalar. To illustrate the idea, let us consider a fundamental mirror magnon. Were it not for the dressing, putting a one-particle mirror state on the mirror edge would correspond to adding a virtual-particle pair on two adjacent physical edges as 
\beq
|D(u^{-\gamma})\rangle\otimes |\bar{D}(u^{\gamma})\rangle+|Y(u^{-\gamma})\rangle\otimes |\bar{Y}(u^{\gamma})\rangle +\cdots\period
 \eeq 
As given in \eqref{modifiedstates}, we have to dress these states by $Z$ markers when the excitations are scalar. A priori, there is some ambiguity about where to insert $Z$ markers, but the correct one, which reproduces the weight factor \eqref{modifiedweight}, turns out to be\footnote{There are also terms with two fermions, and they are dressed with $Z^{\pm 1/2}$. However, because of the charge conservation, those terms do not contribute to the correlators we studied in this paper.} (see also figure \ref{fig:basis})
\beq
|D(u^{-\gamma})\rangle\otimes |\bar{D}(u^{\gamma})\rangle+|Z^{\mp}Y(u^{-\gamma})\rangle\otimes |\bar{Y}(u^{\gamma})Z^{\pm}\rangle +\cdots\period
\eeq
(As in \eqref{modifiedstates}, we need to average over two choices of signs.)
Note that here we have $Z^{\pm}$ instead of $Z^{\pm 1/2}$ as in \eqref{modifiedstates}. This is because the expression given in \eqref{modifiedstates} is for a single $\mathfrak{psu}(2|2)$ whereas the real excitation is made out of two copies of $\mathfrak{psu}(2|2)$'s.

When there are physical magnons, the $Z$ markers bring about two effects: One is the extra $e^{\varphi}$ dependence coming from $Z$ markers\footnote{The precise rule is to multiply $e^{\pm\varphi}$ when the {\it second} hexagon (or the right hexagon) contains $Z^{\pm}$. This is because, in the conformal frame used in figure \ref{fig:weightsym}, the cross ratios are associated with the second hexagon whereas the first hexagon is simply in the canonical configuration.}. The other is the phase factor derived in Appendix C of \cite{BKV}.
 
By carefully taking into account these two effects, we can derive transfer matrices for any mirror bound states. To write down the result, it is convenient to introduce generating functions $\mathcal{W}_{\pm}$ given by
\beq
\begin{aligned}
\mathcal{W}_{\pm}=\left(1-t_{1\pm}{\rm D}^2\right)^{-1}\left(1-t_{2\pm}{\rm D}^2\right)\left(1-t_{3\pm}{\rm D}^2\right)\left(1-t_{4\pm}{\rm D}^2\right)^{-1}
\end{aligned}
\eeq
with
\beq
\begin{aligned}
&t_{1\pm}=e^{i\phi}\frac{B^{(+)+}R^{(+)-}}{B^{(-)+}R^{(-)-}}\comma \quad &&t_{2\pm}=e^{\pm\left(i\frac{P}{2}+\varphi\right) }e^{i\left(\frac{P}{2}+\theta\right)}\frac{R^{(+)-}}{R^{(-)-}}\comma\\
&t_{3\pm}=e^{\mp(i\frac{P}{2}+\varphi) }e^{i\left(\frac{P}{2}-\theta\right)}\frac{R^{(+)-}}{R^{(-)-}}\comma \quad &&t_{4\pm}=e^{-i\phi}\comma
\end{aligned}
\eeq
where ${\rm D}\equiv e^{i\del_v /2}$, and $R^{(\pm)}$ and $B^{(\pm)}$ are given by
\beq
R^{(\pm)}(v)\equiv \prod_{i} (x(v)-x^{\mp}(u_i))\comma \quad B^{(\pm)}(v)\equiv \prod_{i} \left(1-\frac{1}{x(v)x^{\mp}(u_i)}\right).
\eeq
From these generating functions, we can define the {\it dressed twisted transfer matrix} (with a physical rapidity) as
\beq
\frac{1}{2}\left(\mathcal{W}_{+}+\mathcal{W}_{-}\right)=\sum_{a=1}^{\infty}(-1)^{a}\tilde{T}_{a,\{\phi,\theta,\varphi\}}^{[a-1]}(v){\rm D}^{2a}\period
\eeq
This yields the following explicit expression for $\tilde{T}_{a,\{\phi,\theta,\varphi\}}(v)$:
\beq\label{explicitT}
\tilde{T}_{a,\{\phi,\theta,\varphi\}}\!=(-1)^{a}\!\!\sum_{n=-1}^{1}f_n \prod_{m=0}^{n}e^{i\frac{P}{2}}\frac{R^{(+)[2m-a]}}
{R^{(-)[2m-a]}}\sum_{j=\frac{2-a}{2}}^{\frac{a-2n}{2}}e^{i(1-2j-n)\phi}\prod_{k=j+n}^{\frac{a-2}{2}}\frac{R^{(+)[2n-2k]}B^{(+)[-2k]}}{R^{(-)[2n-2k]}B^{(-)[-2k]}}
\eeq
with
\beq
f_{-1}=1\comma \quad f_{0}=-2\cos \theta \cosh\left(\varphi+iP/2\right) \comma \quad f_{1}=1\period
\eeq

For the computation performed in section \ref{sec:finite}, we need a transfer matrix $\tilde{T}_{a,\{\phi,\theta,\varphi\}}(v^{-3\gamma})$ since the mirror channel is distant from the physical magnons by the $3\gamma$ transformation. Using \eqref{explicitT}, we can compute the leading order expression of this object as\footnote{Because of the periodicity under $4\gamma$, we have $\tilde{T}_{a,\{\phi,\theta,\varphi\}}(v^{-3\gamma})=\tilde{T}_{a,\{\phi,\theta,\varphi\}}(v^{\gamma})$.}
\beq\label{oppositeT}
\begin{aligned}
\tilde{T}_{a,\{\phi,\theta,\varphi\}} (v^{-3\gamma})=&\frac{(-1)^{a}}{Q^{[-1-a]}}\left[e^{i(a\phi +P)}Q^{[a+1]}+\sum_{n=1-\frac{a}{2}}^{\frac{a}{2}}e^{-2i n\phi }Q^{[-2n-1]}\right.\\
&\left.\hspace{-20pt}-\cos \theta \left(e^{\varphi+iP}+e^{-\varphi}\right)\sum_{n=\frac{1-a}{2}}^{\frac{a-1}{2}}e^{-2in\phi }Q^{[-2n]}+e^{iP}\sum_{n=1-\frac{a}{2}}^{\frac{a}{2}-1}e^{-2in \phi }Q^{[-2n+1]}\right]\comma
\end{aligned}
\eeq
with $Q$ being the Baxter polynomial $Q(v)=\prod_{i=1}^{M}(v-u_i)$, and $e^{iP}$ being the total momentum of physical magnons.

In Appendix \ref{ap:flip}, we compute an octagon by cutting it in a different way. In that computation, we also need a transfer matrix $\tilde{T}_{a,\{\phi,\theta,\varphi\}}(v^{-\gamma})$ (see Appendix \ref{ap:flip} for more details). Its weak-coupling expression is given by
\beq
\begin{aligned}
\tilde{T}_{a,\{\phi,\theta,\varphi\}} (v^{-\gamma})=&\frac{(-1)^{a}}{Q^{[1-a]}}\left[e^{-ia\phi }Q^{[1-a]}+\sum_{n=-\frac{a}{2}}^{\frac{a}{2}-1}e^{-i (2n\phi +P)}Q^{[-2n-1]}\right.\\
&\left.\hspace{-20pt}-\cos \theta \left(e^{\varphi}+e^{-\varphi-iP}\right)\sum_{n=\frac{1-a}{2}}^{\frac{a-1}{2}}e^{-2in\phi }Q^{[-2n]}+\sum_{n=1-\frac{a}{2}}^{\frac{a}{2}-1}e^{-2in \phi }Q^{[-2n+1]}\right]\period
\end{aligned}
\eeq

\section{Correlators of Konishi and Longer BPS's\label{ap:longer}}
In \cite{DP}, explicit expressions for several five-point functions are written down. They are given in terms of correlators of length 2 BPS operators. For instance, the correlation function of three length 2 operators and two length 4 operators is given (in their normalization) by
\begin{align}\label{casestudy}
&\bar{G}_{22244}(x_0,x_1,x_2,x_3,x_4)= 
4 d_{34}^2 \bar{G}_{22222}(x_0,x_1,x_2,x_3,x_4)+
16d_{23}d_{24}d_{34}\bar{G}_{2222}(x_0,x_1,x_3,x_4)\nn\\
&\qquad \qquad +16 d_{13}d_{14}d_{34}\bar{G}_{2222}(x_0,x_2,x_3,x_4)+16d_{03}d_{04}d_{34}\bar{G}_{2222}(x_1,x_2,x_3,x_4)\period
\end{align}
By taking the OPE limit explained in Appendix \ref{ap:fivept}, we can compute correlators involving a Konishi operator. It turns out that the terms on the second line of \eqref{casestudy} do not contribute since they are non-singular in the limit $x_0\to x_1$. As a result, we obtain
\beq
\bar{G}_{K244}(x_1,x_2,x_3,x_4)=4 d_{34}^2 \bar{G}_{K222}(x_1,x_2,x_3,x_4)+16d_{23}d_{24}d_{34}\bar{G}_{K22}(x_1,x_3,x_4)\period
\eeq
The relation between correlators in \cite{DP} and the ones in this paper turns out to be
\beq
\bar{G}_{K,\{L_i\}}=\left(\prod_{i}L_i \right)G_{K,\{L_i\}}\period
\eeq
Therefore, we have
\beq\label{relationK244}
G_{K244}(x_1,x_2,x_3,x_4)= d_{34}^2 G_{K222}(x_1,x_2,x_3,x_4)+2d_{23}d_{24}d_{34}G_{K22}(x_1,x_3,x_4)\period
\eeq

\begin{figure}[t]
\begin{center}
\includegraphics[clip,height=5.3cm]{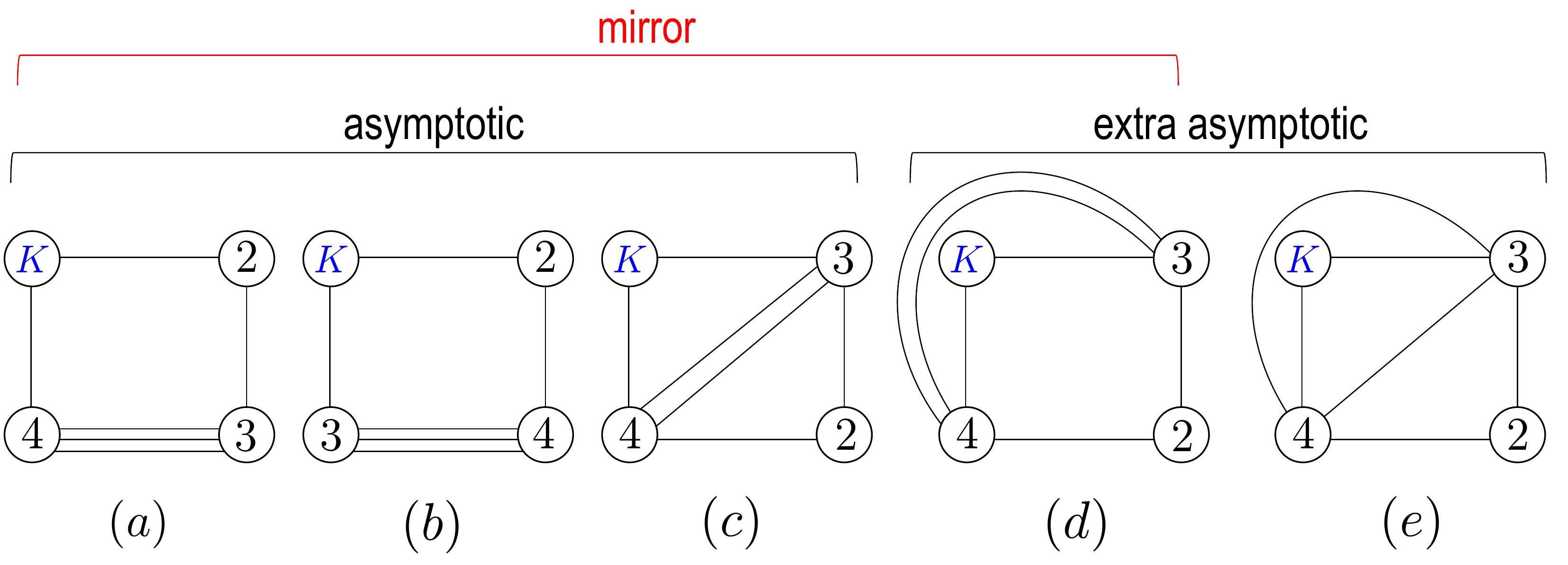}
\end{center}
\vspace{-0.5cm}
\caption{Graphs relevant for $G_{K244}$. The first three graphs give the asymptotic part of $G_{K222}$ and the last two give $G_{K22}$ (up to trivial propagation factors). The mirror correction comes from the first four graphs and it matches the one for $G_{K222}$.
\label{fig:longer}}
\end{figure}

Let us now show that the result \eqref{relationK244} is reproduced from integrability. Since \eqref{relationK244} is written in terms of correlators of ${\bf 20}^{\prime}$ operators, for which we have already seen a match between perturbation and integrability, it essentially boils down to checking the combinatorics: For this correlator, there are five distinct graphs as shown in figure \ref{fig:longer}. The first four receive the mirror correction: More precisely, the graphs $(a)$ and $(b)$ have two mirror channels whereas the graphs $(c)$ and $(d)$ have only one mirror channel. Dressing them by propagators and adding up contributions, we find that the mirror correction is the same as the one for $\langle K \mathcal{O}_{{\bf 20}^{\prime}}\mathcal{O}_{{\bf 20}^{\prime}}\mathcal{O}_{{\bf 20}^{\prime}}\rangle$ times an extra propagator factor $d_{34}^2$. On the other hand, as for the asymptotic part, the first {\it three} diagrams give the asymptotic part of $\langle K \mathcal{O}_{{\bf 20}^{\prime}}\mathcal{O}_{{\bf 20}^{\prime}}\mathcal{O}_{{\bf 20}^{\prime}}\rangle$ times  $d_{34}^2$. Combining these contributions we can reproduce the first term in \eqref{relationK244}. It is then easy to check that the remaining asymptotic part (coming from $(d)$ and $(e)$) matches the second term, since the asymptotic part is essentially given by the structure constant as shown in section \ref{sec:asympt}.

We performed such analysis also for other five-point functions given in Appendix A of \cite{DP} and we found an agreement in all the cases\footnote{For $G_{K345}$, the results are consistent if the coefficient of the second term in (76) of \cite{DP} is $15$ not $30$. We performed a perturbative computation for this correlator by ourselves and found that the coefficient is indeed $15$.} shown below:
\beq
\{G_{K244}\comma \quad \red{G_{K235}}\comma \quad G_{K334}\comma \quad \red{G_{K246}}\comma\quad G_{K336}\comma \quad G_{K255}\comma \quad G_{K345}\}\period
\eeq
The correlators denoted in red receive contributions from non-1EI graphs. We confirmed that our prescription given in section \ref{subsec:graph} correctly reproduces the data in such cases as well.
\section{Flip Invariance of Octagon\label{ap:flip}}
In section \ref{sec:4BPS}, we have seen that the hexagonalization of the BPS four-point function is flip-invariant; namely the result does not change if we cut the four-point function into hexagons in a different way. In this appendix, we will show that this property holds even for correlators involving a Konishi operator.

\begin{figure}[t]
\begin{center}
\includegraphics[clip,height=4.3cm]{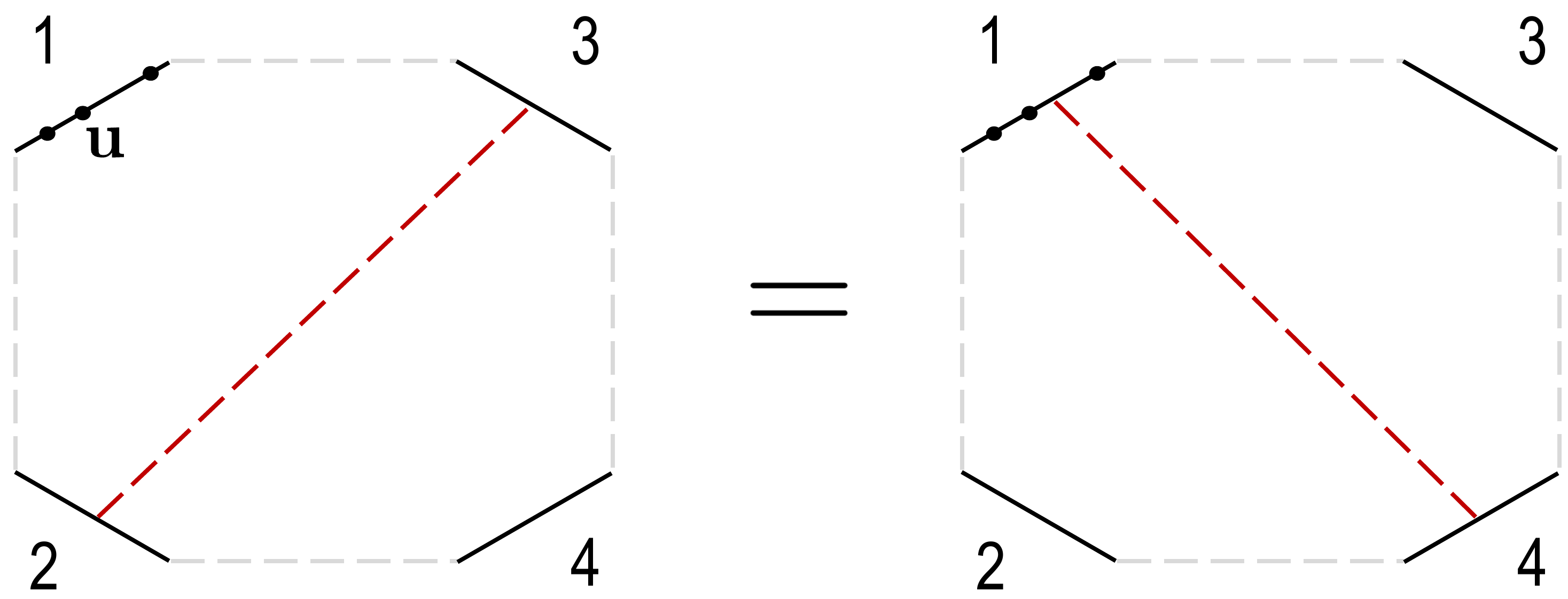}
\end{center}
\vspace{-0.5cm}
\caption{Flip invariance of an octagon with physical magnons. The red dashed lines denote zero-length bridges along which we stitch together two hexagons. The left octagon will be denoted by ${\sf O}_{23}$ whereas the right octagon will be denoted by ${\sf O}_{14}$.
\label{fig:oct}}
\end{figure}

Showing the flip-invariance of the four-point function boils down to proving the flip-invariance of an {\it octagon}\footnote{A similar object is discussed recently in the study of the lightcone string vertex \cite{stringvertex}.}, which can be obtained by gluing two hexagons along a zero-length bridge. In what follows, we study octagons with ${\rm SL}(2)$ magnons on a physical edge (see figure \ref{fig:oct}), since they are precisely the ones relevant for the computation of those correlators. 

The easiest way to compute such octagons is to cut them along the edge $23$. In this way of cutting, the asymptotic part is simply given by a product of hexagon form factors and the overall spacetime dependence,
\beq
{\sf O}^{(0)}_{23}=\left(\frac{x_{23}^{+}}{x_{12}^{+}x_{13}^{+}}\right)^{E_{\bf u}+\frac{M}{2}}\left(\frac{x_{23}^{-}}{x_{12}^{-}x_{13}^{-}}\right)^{E_{\bf u}-\frac{M}{2}}\prod_{i<j}h(u_i,u_j)\comma
\eeq
where $E_{{\bf u}}$ and $M$ are the total energy and the total number of physical magnons respectively. As discussed in the main text, the one-particle mirror correction ${\sf O}_{23}^{(1)}$ is given by
\beq
\begin{aligned}
\frac{{\sf O}^{(1)}_{23}}{{\sf O}^{(0)}_{23}}=&\int\frac{dv}{2\pi}(-1)^{a}\tilde{T}_{a,\{\phi,\theta,\varphi\}}(v^{-3\gamma})\mu_{a}(v^{\gamma})e^{-i \tilde{p}(v)\log z\bar{z}}\prod_{j}h_{1a}(u_j,v^{-3\gamma})\period
\end{aligned}
\eeq
Up to one loop, these two are the only relevant contributions.

Now, to show the flip invariance, we need to reproduce these results by cutting the edge $14$ instead. Since cutting the edge $14$ splits the edge on which magnons are living (see figure \ref{fig:oct}), the asymptotic part is now given by a sum over partition,
\beq\label{asympt14}
\begin{aligned}
&\frac{{\sf O}_{14}^{(0)}}{{\sf O}_{23}^{(0)}}=|1-z|^{-2E_{\bf u}}\left(\frac{1-\bar{z}}{1-z}\right)^{\frac{M}{2}}\sum_{\alpha\cup\bar{\alpha}={\bf u}}(-1)^{|\bar{\alpha}|}\prod_{k\in \bar{\alpha}}z^{E_k+\frac{1}{2}}\bar{z}^{E_k-\frac{1}{2}}\prod_{i\in \alpha,j\in\bar{\alpha}}\frac{1}{h(u_i,u_j)}\period
\end{aligned}
\eeq
with $E_k=E(u_k)$. 

To compute the one-particle mirror correction to \eqref{asympt14}, we need to compute the matrix part. Using the relation between the edges and the cross ratios depicted in figure \ref{fig:ratiorule}, one finds that the cross ratios for the edge $14$ are the inverses of the ones for the edge $23$. This however does not mean that the matrix part is given by $\tilde{T}_{a,\{-\phi,-\theta,-\varphi\}}$. This is because the crossing transformation ($v\to v^{2\gamma}$) changes the flavor indices as explained in Appendix \ref{ap:crossing} and it compensates the effect of the inversion of the cross ratios. As a result, we obtain the following expression for the one-particle correction for the channel  $14$:
\beq
\begin{aligned}
\frac{{\sf O}^{(1)}_{14}}{{\sf O}^{(0)}_{23}}=&|1-z|^{-2E_{\bf u}}\left(\frac{1-\bar{z}}{1-z}\right)^{\frac{M}{2}}\int\frac{dv}{2\pi}(-1)^{a}\tilde{T}_{a,\{\phi,\theta,\varphi\}}(v^{-\gamma})\mu_{a}(v^{\gamma})e^{i\tilde{p}(v)\log z\bar{z}}\\
&\times \prod_{j}h_{1a}(u_j,v^{-\gamma})\sum_{\alpha\cup\bar{\alpha}={\bf u}}(-1)^{|\bar{\alpha}|}\prod_{k\in \bar{\alpha}}\frac{z^{E_k+\frac{1}{2}}\bar{z}^{E_k-\frac{1}{2}}}{h_{1a}(u_k,v^{-\gamma})h_{a1}(v^{-\gamma},u_k)}\prod_{i\in \alpha,j\in\bar{\alpha}}\frac{1}{h(u_i,u_j)}\period
\end{aligned}
\eeq

Let us now check the flip invariance at tree level and one loop using the formulas above. At tree level, we only get contributions from the asymptotic part, which reads
\beq
\begin{aligned}
&\left.\frac{{\sf O}_{14}^{(0)}}{{\sf O}_{23}^{(0)}}\right|_{\rm tree}=\left(\frac{1}{1-z}\right)^{M}\underbrace{\sum_{\alpha\cup\bar{\alpha}={\bf u}}(-z)^{|\bar{\alpha}|}\prod_{i\in \alpha,j\in\bar{\alpha}}\frac{u_i-u_j-i}{u_i-u_j}}_{(\ast)}\period
\end{aligned}
\eeq
Such a summation was studied in \cite{TailoringIII} and as it was shown there, one can prove
\beq
(\ast)=(1-z)^{M}\period
\eeq
From this equality, the flip invariance follows immediately. 

At one loop, one has to consider both the asymptotic part and the mirror correction. As was done in the main text, the mirror correction for the channel $23$ can be easily computed by taking the residues at $v=ia/2$. On the other hand, the mirror integrand for the channel $14$ has extra poles at $v=u_i \pm i(a\pm1)/2$. In the case of the spectrum, these poles corresponded to the $\mu$ terms in the Luscher formula, which vanish\footnote{It is because the fundamental particle cannot decay into two physical particles at weak coupling. See section 7 of \cite{muterm}.} at weak coupling. Also here, we found that the contribution from these poles vanishes after the summation over the bound-state indices. Therefore, the mirror correction can be computed just by taking the residues at $v=ia/2$. We performed the computation explicitly for one and two physical magnons\footnote{When performing the computation, we did not impose any conditions on the rapidities.} and confirmed that the equality
\beq
\left.\frac{{\sf O}^{(0)}_{14}+{\sf O}^{(1)}_{14}}{{\sf O}_{23}^{(0)}}
\right|_{\text{one-loop}}=\left.
\frac{{\sf O}^{(0)}_{23}+{\sf O}^{(1)}_{23}}{{\sf O}_{23}^{(0)}}
\right|_{\text{one-loop}}\comma
\eeq
is indeed satisfied. This proves the flip invariance of correlators involving a Konishi state.

It would be an interesting future problem to show the invariance for an arbitrary number of physical magnons at one loop. An even more ambitious goal would be to show the invariance at finite coupling.

\end{document}